\documentclass[10pt,%
 twocolumn,
superscriptaddress,
bibnotes,
 amsmath,amssymb,
 aps,
 prl,
floatfix,
]{revtex4-2}

\usepackage{multibib}
\newcites{appendix}{Appendix References}

\usepackage{graphicx}
\usepackage[caption=false]{subfig}

\usepackage[percent]{overpic}

\usepackage[normalem]{ulem}

\usepackage{ragged2e}

\usepackage{amsmath}
\usepackage{bm}
\usepackage{graphicx}
\usepackage{dcolumn}
\usepackage{bm}

\usepackage{tikz-cd}

 \usepackage[colorinlistoftodos,prependcaption,textsize=tiny]{todonotes}

\newcommand\sbullet[1][.5]{\mathbin{\vcenter{\hbox{\scalebox{#1}{$\bullet$}}}}}

\usepackage{changes}
\definechangesauthor[name=XZ, color=blue]{x}
\definechangesauthor[name=Bin, color=red]{b}
\definechangesauthor[name=YF, color=yellow]{y}

\renewcommand{\d}{\ensuremath{\,\mathrm{d}}}

\usepackage{algorithm2e}
\usepackage{amsmath}
 \graphicspath{{./Newfigure/}}
 \usepackage{xcolor}

\usepackage[titletoc]{appendix}

\begin{document}

\preprint{APS/123-QED}

\title{Entropy Production in Non-Gaussian Active Matter: A Unified Fluctuation Theorem and Deep Learning Framework}

\author{Yuanfei Huang}
\affiliation{Asia Pacific Center for Theoretical Physics, Pohang-si, Gyeongsangbuk-do, 37673, Republic of Korea.}
\affiliation{Department of Mathematics, City University of Hong Kong, Kowloon, Hong Kong SAR.}

\author{Chengyu Liu}
\affiliation{ Department of Data Science, City University of Hong Kong, Kowloon, Hong Kong SAR. }

\author{Bing Miao*}
\affiliation{Center of Materials Science and Optoelectronics Engineering, College of Materials Science and Opto-Electronic Technology, University of Chinese Academy of Sciences (UCAS), Beijing 100049, China.}

\author{Xiang Zhou*}
\affiliation{Department of Mathematics, City University of Hong Kong, Kowloon, Hong Kong SAR.}


\begin{abstract}
We present a general framework for deriving entropy production rates (EPRs) in active matter systems driven by non-Gaussian active fluctuations. Employing the probability-flow equivalence technique, we rigorously obtain an entropy production (EP) decomposition formula. We demonstrate that the EP, $\Delta s_\mathrm{tot}$, satisfies a detailed fluctuation theorem, $\rho_{\mathcal{R}}(\Sigma)/\rho_{\mathcal{R}}(-\Sigma)=e^{\Sigma}$, which holds for the distribution $\rho_{\mathcal{R}}(\Sigma)$ defined as the probability of observing a value $\Sigma$ of the quantity $\mathcal{R}\equiv \Delta s_\mathrm{tot}-B_\mathrm{act}$, where $B_\mathrm{act}$ is a path-dependent random variable associated with active fluctuations. Moreover, an integral fluctuation theorem, $\langle e^{-\mathcal{R}} \rangle = 1$, and the generalized second law of thermodynamics, $\langle \Delta s_\mathrm{tot} \rangle \ge \langle B_\mathrm{act} \rangle$, follow directly. Our results hold under steady-state conditions and can be straightforwardly extended to arbitrary initial states. In the limiting case where active fluctuations vanish, these theorems reduce to the established results of stochastic thermodynamics. Building on this theoretical foundation, we introduce a deep-learning-based methodology for efficiently computing the EP, utilizing the L\'{e}vy score we propose. To illustrate the validity of our approach, we apply it to two representative systems: a Brownian particle in a periodic active bath and an active polymer composed of an active Brownian cross-linker interacting with passive Brownian beads. Our work provides a unified framework for analyzing EP in active matter and offers practical computational tools for investigating complex nonequilibrium behavior.

\end{abstract}

\maketitle

{\itshape Introduction---}Fluctuations are intrinsic to many microscopic systems with Gaussian thermal noise that forms the foundation of stochastic thermodynamics \cite{seifert2012stochastic,van2015ensemble,peliti2021stochastic,gaspard2022statistical}. However, recent advances in active matter systems have revealed the ubiquitous presence of non-Gaussian fluctuations, observed in phenomena such as stochastic navigation in eukaryotic microorganisms \cite{li2025biased}, bacterial dynamics \cite{lauga2012dance,matthaus2009coli}, self-propelling colloidal particles \cite{goswami2024anomalous,leptos2009dynamics,wu2000particle,chen2007fluctuations,palacci2010sedimentation}, molecular motor-driven transport \cite{ariga2021noise,kurihara2017non,park2020rapid,ezber2020dynein}, active living cells \cite{paneru2021transport,sang2022single,chen2015memoryless,song2018neuronal}, and active solids \cite{baconnier2022selective,hernandez2024model,caprini2023entropons}. These discoveries have inspired the development of generalized Langevin equations that incorporate both thermal and active noise \cite{barik2006langevin,demaerel2018active,um2019langevin,goerlich2022harvesting,bialas2023periodic,bialas2023mechanism,das2023enhanced,chaki2019enhanced,bialas2023control,ariga2024nonthermal,joo2020anomalous}. In theory, fluctuations in active matters are typically studied by different stochastic models. Such as active Ornstein--Uhlenbeck processes \cite{mandal2017entropy},  run-and-tumble processes \cite{goswami2024anomalous,bressloff2025stochastic}, and L\'{e}vy-type processes, including compound Poisson \cite{park2020rapid} and $\alpha$-stable L\'{e}vy processes \cite{ariga2021noise}.

Some key questions in stochastic thermodynamics are as follows: (i) how to rigorously define and compute EP, (ii) how to consistently extend stochastic thermodynamics to systems governed by active fluctuations, and (iii) how active fluctuations influence EP and dissipation mechanisms. Addressing these questions is essential to advance our understanding of the thermodynamics of active matter and to uncover  the fundamental principles underlying nonequilibrium processes in systems subject to complex fluctuations \cite{o2022time,fodor2022irreversibility}.
These questions have been extensively explored    for diffusion systems in steady states \cite{mandal2017entropy,dabelow2019irreversibility}   and  discrete-state Markov jump process  over finite-length trajectories \cite{andrieux2008temporal,polettini2014transient}. 

However, there are limited references addressing these issues in systems driven by L\'{e}vy-type active fluctuations. Previous studies on nonequilibrium heat transport \cite{kanazawa2013heat,kanazawa2015minimal,kanazawa2015asymptotic,kanazawa2017statistical} and fluctuations under L\'{e}vy and Poisson noise \cite{touchette2007fluctuation,touchette2009anomalous,baule2009fluctuation,budini2012generalized,lucente2023statistical,faria2025nonequilibrium} have provided valuable insights. However, these works focus predominantly on specific cases or using approximation techniques, and a unified theoretical framework for stochastic thermodynamics of active systems with non-Gaussian fluctuations remains largely undeveloped. 

In this Letter we address these challenges through proving unified fluctuation theorems in non-Gaussian active matter and providing a deep learning framework to compute the entropy production (EP). First, using the probability flow equivalence technique \cite{huang2024probability,huang2024vy}, we rigorously derive the EP decomposition formula:  
\begin{eqnarray}\label{eprdecomposition}
    \Delta s_\mathrm{tot} =&\ \Delta s_\mathrm{sys} + \Delta s_\mathrm{m} + \Delta s_\mathrm{act}.
\end{eqnarray}
The total EP, $\Delta s_\mathrm{tot}$, is decomposed into three parts: system $\Delta s_\mathrm{sys}$, medium  $\Delta s_\mathrm{m}$, and active fluctuation  $\Delta s_\mathrm{act}$.

After introducing $B_\mathrm{act}$ as  a path-dependent random variable induced by active fluctuations in Appendix A, of End Matter,  we define 
$\mathcal{R}\equiv\Delta s_\mathrm{tot} - B_\mathrm{act}$, and 
show  the following  detailed fluctuation theorem (DFT)
\begin{equation}\label{DFT}
        \rho_{\mathcal{R}}(\Sigma)/\rho_{\mathcal{R}}(-\Sigma) =e^{\Sigma},
\end{equation}
where $\rho_{\mathcal{R}}$ is the probability of observing a value $\Sigma$ for $\mathcal{R}$.

 We then can obtain an integral fluctuation theorem (IFT):
\begin{eqnarray}
       &&\langle e^{  -\mathcal{R}} \rangle = 1, \label{FT}
\end{eqnarray}
where $\langle\cdots\rangle$ describes the ensemble average over all microscopic trajectories,  This fluctuation relation bears a formal resemblance to the Sagawa--Ueda relation in information thermodynamics \cite{sagawa2012fluctuation}. Applying Jensen's inequality to \eqref{FT}, we derive:  
\begin{eqnarray}
     \langle \mathcal{R}\rangle \ge 0, ~\mbox{ i.e., }~  \langle\Delta s_\mathrm{tot}\rangle\geq&&  \langle B_\mathrm{act}\rangle.\label{SL}
\end{eqnarray}

\par 
These relations, \eqref{DFT}, \eqref{FT}, and \eqref{SL}, hold in steady-state conditions, but we also show that  under general nonequilibrium settings, the same relations remain valid provided that $\Delta s_\mathrm{tot}$ in $\mathcal{R}$ is replaced by its time-reversed counterpart $\Delta \tilde{s}_\mathrm{tot}$, which is defined via the reversed process in \eqref{eqn:SDEreversed}. In the absence of active fluctuations, $B_\mathrm{act}$ vanishes and Eqs. \eqref{DFT}, \eqref{FT} and inequality \eqref{SL} reduce to the conventional DFT, IFT, and second law of thermodynamics  (e.g., Ref. \cite{seifert2005entropy,boffi2024deep}), respectively.

We note that while previous studies on IFT and DFT in stochastic thermodynamics have provided valuable insights into nonequilibrium thermodynamics, they have primarily focused on Gaussian systems and their methodologies are often based on physical arguments. Few works have investigated in detail or explicitly highlighted the perspective of probability flow \cite{huang2024probability,huang2024vy} for time-reversed stochastic processes. In contrast, our results not only extend the applicability to systems with non-Gaussian active fluctuations but also provide a rigorous mathematical formulation that explicitly incorporates the time-reversed process and its corresponding path-space measure, thereby offering a new perspective on EP.

Finally, we introduce an efficient  method based on a L\'{e}vy score particle algorithm \cite{huang2024vy} for calculating EP in active systems. This approach allows us to compute EP components ($\Delta S_\mathrm{tot}$, $\Delta S_\mathrm{act}$, $\Delta S_\mathrm{sys}$, $\Delta S_\mathrm{m}$), thereby facilitating the analysis of stochastic thermodynamic properties in systems with non-Gaussian active fluctuations. Our framework distinguishes itself from the recent  model-agnostic numerical methods in ~\cite{kim2020learning,otsubo2020estimating,ro2022model}, which estimate the total EPR from a large number of  observed trajectories but are unable to quantify the distinct contributions from physically meaningful components arising from different types of fluctuations. We adapt our deep learning algorithm to a range of dynamical scenarios, demonstrating its robustness and effectiveness in capturing the nonequilibrium characteristics of active matter. 

{\itshape Model---}In Euclidean space $\mathbb{R}^d$, we consider the overdamped dynamics of a single particle driven by a force $\bm{F}(\bm{r})=-\nabla U(\bm{r}) + \bm{f}(\bm{r})$, where the first and second term denote conservative and non-conservative parts, respectively. The motion is described by Langevin equation:
\begin{equation}\label{eqn:SDE}
\d \bm{r}(t)/\d t = \bm{F}(\bm{r}(t))/\Gamma + \bm{\eta}_{\mathrm{th}}(t   ) + \bm{\eta}_{\mathrm{act}}(t),
\end{equation}
where $\Gamma$ is the friction coefficient, $\bm{\eta}_{\mathrm{th}}(t)$ and $\bm{\eta}_{\mathrm{act}}(t)$ are the thermal and active noise, respectively.
The thermal noise $\bm{\eta}_{\mathrm{th}}$ is a Gaussian white noise with zero mean and variance $\langle \eta_{\mathrm{th,i}}(t) \eta_{\mathrm{th,j}}(t') \rangle = 2D_{\mathrm{th}} \delta_\mathrm{i,j} \delta(t-t')$, 
where $D_{\mathrm{th}}$ is the diffusion coefficient and the indices $\mathrm{i,j} = {1, \ldots, d}$ denote spatial directions. The diffusion and friction coefficients satisfy the Einstein relation $D_{\mathrm{th}} =  k_B \mathcal{T}/\Gamma$, where $\mathcal{T}$ denotes the bath temperature. The active noise $\bm{\eta}_{\mathrm{act}}$ is modeled as the compound Poisson process which consists of discrete ``kicks'' occurring at a rate $\lambda_0$, and is expressed as: 
$\eta_{\mathrm{act,i}}(t) = \sum_{k=1}^{N_t} A_{k,\mathrm{i}} \delta(t-t_k)$, where the times $t_k$ are distributed according to the Poisson process with rate $\lambda_0$; the total number of kicks $N_t$ in the interval $[0, t]$ follows a Poisson distribution with mean $\lambda_0 t$; the kick amplitudes $A_{k,\mathrm{i}}$ are independent and identically distributed random variables drawn from a fixed probability distribution with intensity $\nu_A$. The associated L\'{e}vy measure $\nu$ of this active noise process is $\nu = \lambda_0 \nu_A$.

The L\'{e}vy--Fokker--Planck equation (LFPE) corresponding to Eq. \eqref{eqn:SDE} reads \cite{huang2024vy,huang2024probability}
\begin{equation}\label{eqn:LFP} 
\begin{aligned}
        \partial_t P(\bm{r},t) 
        =&-\nabla\cdot\bigg[\bigg( \bm{F}(\bm{r})/\Gamma - D_{\mathrm{th}} \nabla\log P(\bm{r},t) \\
        +&  \int_0^1\d\theta\int \nu(\d \bm{z})\frac{\bm{z} P(\bm{r} -\theta\bm{z},t)}{P(\bm{r},t)} \bigg)P(\bm{r},t)\bigg] \\
        \equiv & -\nabla\cdot \bm{J}(\bm{r},t) 
        \equiv -\nabla\cdot\left[  \bm{V}(\bm{r},t)P(\bm{r},t) \right],
\end{aligned}
\end{equation}
where we have defined the probability current $\bm{J}(\bm{r},t)$ and the velocity field $\bm{V}(\bm{r},t)$.
As analogy to referring to $\nabla \log P$ as the 
Stein score function, we call 
\begin{eqnarray}\label{levyscore}
    \bm{S}_\mathrm{L}(\bm{r},t)\equiv  -\frac{ \int_0^1\d\theta\int \nu(\d \bm{z}) \bm{z} P(\bm{r} -\theta\bm{z},t)  }{ P(\bm{r},t)}
\end{eqnarray}
the  \emph{L\'{e}vy score function} \cite{huang2024vy}. The Shannon entropy
\begin{eqnarray}\label{entropy:sys}
    S_{\mathrm{sys}}(t)\equiv -\int\d \bm{r} P( \bm{r},t)\log P(\bm{r},t)\equiv \langle s_{\mathrm{sys}}(t)\rangle,
\end{eqnarray}
suggests to define a trajectory-dependent entropy 
\begin{eqnarray}\label{stoentropy}
    s_{\mathrm{sys}}(t)= -\log P(\bm{r}(t),t),
\end{eqnarray}
where $P(\bm{r},t)$ obtained by solving the LFPE \eqref{eqn:LFP} is evaluated along a trajectory $\bm{r}(t)$. Obviously, for any trajectory $\bm{r}(t)$,
the stochastic entropy $s_\mathrm{sys}$ depends on the initial data $P_0(\bm{r})$ and
thus contains information on the whole ensemble. The definition \eqref{stoentropy} has been used previously by Crooks for
stochastic microscopically reversible dynamics \cite{crooks1999entropy}, by
Qian for stochastic dynamics of macromolecules \cite{qian2001mesoscopic}, and  by Seifert for stochastic nonequilibrium dynamics \cite{seifert2005entropy}. All of these works, however, only discussed Gaussian fluctuations for this stochastic entropy.

The rate of change of the system entropy \eqref{stoentropy} is
 \begin{subequations}
 \label{eqn:Ssys}
\begin{align}
        \dot{s}_{\mathrm{sys}}(t)=& -\frac{\partial_t P(\bm{r},t)}{P(\bm{r},t) }\bigg|_{\bm{r}(t)} - \frac{\nabla P(\bm{r},t)}{P(\bm{r},t) }\bigg|_{\bm{r}(t)}\diamond \dot{\bm{r}}\nonumber\\
        =& -\frac{\partial_t P(\bm{r},t)}{P(\bm{r},t) }\bigg|_{\bm{r}(t)} + \frac{\bm{J}(\bm{r},t)}{D_{\mathrm{th}} P(\bm{r},t)}\bigg|_{\bm{r}(t)} \diamond\dot{\bm{r}}\label{eqn:entropyderivationJ} \\
        & - \frac{\bm{F}(\bm{r})}{\Gamma D_{\mathrm{th}} }\bigg|_{\bm{r}(t)}\diamond \dot{\bm{r}} 
       + \frac{ \bm{S}_\mathrm{L}(\bm{r},t)   }{D_{\mathrm{th}}}
   \bigg|_{\bm{r}(t)}\diamond\dot{\bm{r}}, \label{eqn:entropyderivationF} 
\end{align}
 \end{subequations}
where $\diamond$ denotes the Marcus canonical integral that preserves the chain rule  for stochastic differential with jumps \cite{applebaum2009levy}. The Di Paola–Falsone calculus \cite{di1993stochastic,di1993ito} and the $(*)$-calculus \cite{kanazawa2012stochastic,fodor2018non} represent two additional stochastic calculi that preserve the chain rule for systems with jump noise. Although their definitions take different forms, both have been shown to be mathematically equivalent to the Marcus integral \cite{li2013marcus,falsone2018stochastic}. The first term in \eqref{eqn:entropyderivationF} captures the heat dissipation in the medium
\begin{eqnarray}
    \dot{q}(t)= \bm{F}(\bm{r}) \diamond\dot{\bm{r}} \equiv k_B\mathcal{T}\dot{s}_\mathrm{m},
\end{eqnarray}
where we identify the exchanged heat with an increase in entropy of the medium $s_\mathrm{m}$ at temperature $\mathcal{T}=D_{\mathrm{th}}\Gamma/k_B$. The second term in \eqref{eqn:entropyderivationF} corresponds to the entropy increase caused by the active fluctuation 
\begin{eqnarray}\label{actEPR}
    \dot{s}_{\mathrm{act}}(t)= - (\bm{S}_\mathrm{L}(\bm{r},t)  \diamond\dot{\bm{r}})/D_{\mathrm{th}} .
\end{eqnarray}
Then we have a dynamic balance equation for the trajectory-dependent total EP
\begin{eqnarray}\label{eqn:stotsmall}
    \dot{s}_{\mathrm{tot}}(t)=&& \dot{s}_{\mathrm{sys}}(t) + \dot{s}_{\mathrm{m}}(t) + \dot{s}_{\mathrm{act}}(t)\nonumber\\
    =&& -\frac{\partial_t P(\bm{r},t)}{P(\bm{r},t) }\bigg|_{\bm{r}(t)} + \frac{\bm{J}(\bm{r},t)}{D_{\mathrm{th}} P(\bm{r},t)}\bigg|_{\bm{r}(t)}\diamond \dot{\bm{r}},
\end{eqnarray}
which is our first central result. The first term on the right-hand-side signifies a change in $P(\bm{r},t)$, due to relaxation
from a nonstationary initial state $P(\bm{r},0)\neq P^\mathrm{s}(\bm{r})$.

Upon averaging, the EPR $\dot{s}_{\mathrm{tot}}$
has to be non-negative as required by the second law. This ensemble average proceeds in two steps. First, we average over all trajectories of \eqref{eqn:SDE} that pass a given position $\bm{r}$ at time $t$
leading to $\langle \dot{\bm{r}}\mid \bm{r},t\rangle = \bm{J}(\bm{r},t)/P(\bm{r},t)=\bm{V}(\bm{r},t)$ \cite[Appendix I]{supplemental_material}. Second, with $\int\d\bm{r}\partial_tP(\bm{r},t)=0$ due to total probability conservation, averaging over all $\bm{r}$ with $P(\bm{r},t)$ leads to
\begin{eqnarray}
    \dot{S}_{\mathrm{tot}}(t)\equiv&&\langle\dot{s}_{\mathrm{tot}}(t)\rangle
    = \int\d\bm{r} \frac{|\bm{J}(\bm{r},t)|^2}{D_{\mathrm{th}} P(\bm{r},t)}\nonumber\\
    =&& \int\d\bm{r}  |\bm{V}(\bm{r},t)|^2P(\bm{r},t) /D_{\mathrm{th}}.\label{entropy:tot}
\end{eqnarray}
The last  expression is always non-negative, and is only equal to zero in the case of  equilibrium. Similarly,
\begin{align}
    \dot{S}_{\mathrm{m}}(t)\equiv& \langle\dot{s}_{\mathrm{m}}(t)\rangle
    = \int \d\bm{r}\frac{ \bm{F}(\bm{r})\cdot\bm{J}(\bm{r},t) }{k_B\mathcal{T}},\label{entropy:m} \\
    \dot{S}_{\mathrm{act}}(t)\equiv&\langle\dot{s}_{\mathrm{act}}(t)\rangle
    =  \int\d\bm{r} \frac{ -\bm{S}_\mathrm{L}(\bm{r},t)   }{D_{\mathrm{th}} }\cdot\bm{J}(\bm{r},t).\label{entropy:act}
\end{align}
Hence, upon averaging, the increase in entropy of the
system itself becomes $\dot{S}_{\mathrm{sys}}(t)\equiv\langle \dot{s}_{\mathrm{sys}}(t)\rangle= \dot{S}_{\mathrm{tot}}(t) - \dot{S}_{\mathrm{m}}(t) - \dot{S}_{\mathrm{act}}(t)$. The foundation of this result is that we have obtained EP (or
annihilation) along a single stochastic trajectory in Eq. \eqref{eqn:stotsmall}, splitting it
into a medium part, a system part, and an active fluctuation part. These new concepts facilitate the discussion of fluctuation theorems.

{\itshape Fluctuation theorem---}For $P$ associated with \eqref{eqn:SDE}, we introduce a  
probability measure $\tilde{P}$  satisfying  the time reversal property $\tilde P(\bm{x}, T-t)=P(\bm{x}, t)$, associated with the following stochastic process on time interval $[0,T]$: 
\begin{equation}\label{eqn:SDEreversed}
\begin{split}
    \frac{\d\bm{\tilde{r}}(t)}{\d t} =& -\frac{\bm{F}(\bm{\tilde{r}}(t)) }{\Gamma} + 2D_\mathrm{th}\nabla\log P(\bm{\tilde{r}}(t),T-t )
    \\
     &   +\tilde{\bm{\eta}}_\mathrm{th}(t)+ \tilde{\bm{\eta}}_\mathrm{act}(t), \quad  
    \bm{\tilde{r}}(0)\sim    P(\sbullet,T)
\end{split}
\end{equation}
where $\tilde{\bm{\eta}}_\mathrm{th}$ is a Gaussian white noise with the same statistical property of $\bm{\eta}_\mathrm{th}$, and $\tilde{\bm{\eta}}_\mathrm{act}$ is a   compound Poisson process with the time- and state-dependent L\'{e}vy measure 
 \begin{equation}
\tilde{\nu}_{T-t}(\bm{r},\d\bm{z}) =\frac{ P( \bm{r} +\bm{z},T-t ) }{ P( \bm{r},T-t ) }\nu(-\d\bm{z}) .
 \end{equation}
Then $\bm{\tilde{r}}(t) = \bm{r}(T - t)$ for all $t$
\cite{privault2004markovian,conforti2022time} and is called the time-reversed stochastic process of Eq. \eqref{eqn:SDE}.  For this time-reversed non-Gaussian stochastic process $\bm{\tilde{r}}(t)$ and its distribution $\tilde{P}$,  we can define $\tilde{s}_\mathrm{tot}$  and
$\tilde{s}_\mathrm{sys}$, respectively, by  \eqref{stoentropy}  and \eqref{eqn:stotsmall} 
\cite[Eq. (SIII.7)]{supplemental_material}.
If the initial  distribution of Eq. \eqref{eqn:SDE} is the steady state, then Eq. \eqref{eqn:SDEreversed}
is also in steady state, and 
$\tilde{s}_\mathrm{tot}={s}_\mathrm{tot}$ and $\tilde{s}_\mathrm{sys}={s}_\mathrm{sys}$.

The explicit path integral representation for $\bm{r}(t)$ in \eqref{eqn:SDE} and $\bm{\tilde{r}}(t)$ in \eqref{eqn:SDEreversed} is generally infeasible to construct \cite{baule2023exponential}. 
However, the probability flow equivalence framework \cite{huang2024probability} enables us to bridge these processes through It\^{o} diffusion SDEs.
By discretizing an arbitrary trajectory $\{\bm{x}(t)\}_{0\leq t\leq T}$ into finite time intervals, we establish that

{\begin{equation}\label{eqn:Rn}
\begin{aligned}
    &
  \log \frac{P[\{\bm{x}(t_0),\cdots,\bm{x}(t_n) \} ] }{\tilde{P}[ \{ \bm{x}(t_0),\cdots,\bm{x}(t_n)\} ] } \\
  \simeq  &\sum_{i=0}^{n-1} \left(\dot{\tilde{s}}_\mathrm{sys}(t_i)+\dot{\tilde{s}}_\mathrm{m}(t_i)+\dot{\tilde{s}}_\mathrm{act}(t_i) \right)\Delta t -B_{\mathrm{act},n}( \{\bm{x}(t_i)\}_{i=0}^n) \\
   \rightarrow &  ~\Delta \tilde{s}_\mathrm{tot} - B_{\mathrm{act}}( \{\bm{x}(t)\}_{0\leq t\leq T}),\quad \mbox{as}\quad n\to\infty.
\end{aligned}
\end{equation}
}The quantity \( B_{\mathrm{act}}(\{\bm{x}(t)\}_{0\leq t\leq T}) \), which represents a random variable dependent on the trajectory \( \{\bm{x}(t)\}_{0\leq t\leq T}\), emerges from the effect mainly caused by symmetry-breaking when reversing the non-local active fluctuation $\bm{\eta}_\mathrm{act}$, manifested by the distinctive L\'{e}vy measure $\nu$ and $\tilde{\nu}$; see   End Matter Appendix A for more details.

We denote by $\rho_{\mathcal{R}}$ the distribution of $\mathcal{R} = \Delta \tilde{s}_\mathrm{tot} - B_\mathrm{act}$ in \eqref{eqn:Rn}, and prove the following DFT in End Matter:
\begin{align}
     &\rho_{\mathcal{R}}(\Sigma)/\rho_{\mathcal{R}}(-\Sigma) =e^{\Sigma}. \label{DFT1} 
\end{align}
Then  we easily derive the IFT:
\begin{align}
    \langle e^{-\mathcal{R}} 
    \rangle 
    =&\int_{\mathbb{R}}  \rho_{\mathcal{R}}(\Sigma)    e^{-\Sigma} \d \Sigma  =  \int_{\mathbb{R}}    ~\rho_{\mathcal{R}}(-\Sigma)   \d \Sigma=1.\label{eR=1}
\end{align} Here, the average is taken over the trajectories 
of the SDE \eqref{eqn:SDE} with 
the noise histories $\bm{\eta}_\mathrm{th}(t)$ and $\bm{\eta}_\mathrm{act}(t)$. It is easy to see that $\langle \mathcal{R} \rangle \geq 0$ from convexity.

These two fluctuation theorems, \eqref{DFT1} and \eqref{eR=1}, for $\tilde{s}_\mathrm{tot}$, are remarkably universal, as they hold for any initial condition (not just for $P_0(\bm{r}_0) = P^\mathrm{s}(\bm{r}_0)$), with ($\bm{f} \neq 0$) or without ($\bm{f} = 0$) external driving, and for any trajectory length $T$. When the system starts from its steady state, \eqref{DFT1} and \eqref{eR=1} become \eqref{DFT} and \eqref{FT}, respectively.

{\itshape Deep learning algorithm for EPRs---}Since the EP depends on the density-dependent velocity $\bm{V}(\bm{r}, t)$, directly calculating these values is challenging. We present a deep learning-based approach. Recall \eqref{eqn:LFP}, $P(\bm{r}, t)$ can be interpreted as a pushforward of $P_0(\bm{r})$ under the flow map $\bm{X}_{s,t}$ (for $t\geq s\geq0$):
\begin{align}
        \d \bm{X}_{s,t}(\bm{x})/\d t=\ \bm{V}(\bm{X}_{s,t}(\bm{x}),t),\quad \bm{X}_{s,s}(\bm{x})=\bm{x}.\label{eqn:interactingODE}
\end{align}In our previous work \cite{huang2024vy}, a score-based deep learning approach was proposed to learn the velocity field $\bm{V}(\bm{r}, t) = \bm{F}(\bm{r})/\Gamma - \bm{S}^\mathrm{NN}(\bm{r}, t)$ and the probability flow $P(\bm{r},t)$, using a single neural network $\bm{S}^\mathrm{NN}(\bm{r}, t)$. This neural network approximates the sum of the conventional score, $\nabla \log P(\bm{r}, t)$, and the L\'{e}vy score, $\bm{S}_\mathrm{L}(\bm{r}, t)$. However, since the EP of active fluctuations \eqref{entropy:act} explicitly involves the L\'{e}vy score, it becomes necessary to use two neural networks to learn the velocity field as:
\begin{align}
\bm{V}^\mathrm{NN}(\bm{r}, t) = \bm{F}(\bm{r})/\Gamma - D_\mathrm{th}\bm{S}_\mathrm{B}^\mathrm{NN}(\bm{r}, t) - \bm{S}_\mathrm{L}^\mathrm{NN}(\bm{r}, t),
\end{align}
where $\bm{S}_\mathrm{B}^\mathrm{NN}(\bm{r}, t)$ approximates the conventional score $\nabla \log P(\bm{r}, t)$, and $\bm{S}_\mathrm{L}^\mathrm{NN}(\bm{r}, t)$ approximates the L\'{e}vy score. To train them, we solve the following two score-matching optimization problems:
   {\small \begin{align}
        &\min_{\bm{S}_\mathrm{B}^\mathrm{NN} }\left\langle  \left|\bm{S}_\mathrm{B}^\mathrm{NN}  - \nabla\log P^\mathrm{NN}(\bm{r},t)  \right|^2 \right\rangle,\label{minimization}\\
        &\min_{  \bm{S}_\mathrm{L}^\mathrm{NN} }\left\langle  \left| \bm{S}_\mathrm{L}^\mathrm{NN}  + \int_0^1\d\theta\int \nu(\d \bm{z}) \frac{\bm{z} P^\mathrm{NN}(\bm{r} -\theta\bm{z},t)}{P^\mathrm{NN}(\bm{r},t)}\right|^2\right\rangle,\label{minimization2}
    \end{align}
}where $\langle \cdots \rangle$ denotes averaging over all trajectories $\{\bm{X}_{0,t}(\bm{x})\}_{0 \leq t \leq T}$ obtained from \eqref{eqn:interactingODE}, with $\bm{x}$ drawn from the initial distribution $P_0(\bm{r})$. The term $P^\mathrm{NN}(\bm{r}, t)$ represents the probability flow derived from \eqref{eqn:interactingODE} by replacing $\bm{V}$ with $\bm{V}^\mathrm{NN}$. In practice, $P^\mathrm{NN}(\bm{r}, t)$ is approximated by $ \frac{1}{N} \sum_{i=1}^N \delta(\bm{r} - \bm{X}_{0,t}(\bm{x}_i))$.
The primary challenge in solving \eqref{minimization} and \eqref{minimization2} lies in the fact that $P^\mathrm{NN}(\bm{r}, t)$ depends on both $\bm{S}_\mathrm{B}^\mathrm{NN}$ and $\bm{S}_\mathrm{L}^\mathrm{NN}$ in a self-consistent manner. To make the minimization problems   practical, we   train  $\bm{S}_\mathrm{B}^\mathrm{NN}$ and $\bm{S}_\mathrm{L}^\mathrm{NN}$   within each time sub-interval $[t,t+\Delta t]$ in a time-discrete framework with a given $P^\mathrm{NN}(\bm{r}, t)$. This approach freezes $P^\mathrm{NN}(\bm{r}, t)$   within a short time sub-interval, and then update it at the next time sub-interval from the learned scores.
 The loss functions are constructed as below.  At each $t \in [0, T]$ with a given $P^\mathrm{NN}(\bm{r}, t)$, we   expand the squared terms in \eqref{minimization} and \eqref{minimization2}. After discarding the terms that does not involve $\bm{S}_\mathrm{B}^\mathrm{NN}$ and $\bm{S}_\mathrm{L}^\mathrm{NN}$—as these remain constant during optimization—the loss functions at time $t$ are given by:
\begin{eqnarray}
    &&Loss_\mathrm{B}(t)\equiv \mathbb{E}_{\bm{X}_{t}\sim P^\mathrm{NN}(\cdot,t)}\left(  \left|\bm{S}_\mathrm{B}^\mathrm{NN}(\bm{X}_{t})\right|^2 \right) \label{opt:B}\\
    &&\  \qquad +2\mathbb{E}_{\bm{X}_{t}\sim P^\mathrm{NN}(\cdot,t)}\left[\nabla\cdot \bm{S}_\mathrm{B}^\mathrm{NN}(\bm{X}_{t}) \right],\nonumber\\
    && Loss_\mathrm{L}(t)\equiv \mathbb{E}_{\bm{X}_{t}\sim P^\mathrm{NN}(\cdot,t)}\left(\left|\bm{S}_\mathrm{L}^\mathrm{NN}(\bm{X}_{t}) \right|^2\right) \label{opt:L}\\
    &&\ {\color{blue}{+}} 2 \mathbb{E}_{\bm{X}_{t}\sim P^\mathrm{NN}(\cdot,t)}\left[\int\nu(\d\bm{z}) \int_0^1 \bm{S}_\mathrm{L}^\mathrm{NN}(\bm{X}_{t}+\theta\bm{z})\cdot \bm{z}\d\theta \right]\nonumber.
\end{eqnarray} 
Once the optimal $\bm{S}^\mathrm{NN}_\mathrm{B}(\bm{r}, t)$ and $\bm{S}^\mathrm{NN}_\mathrm{L}(\bm{r}, t)$ are obtained, the velocity field $\bm{V}^\mathrm{NN}(\bm{r}, t)$ is updated accordingly to generate the new samples $\bm{X}^\mathrm{NN}_{0, t+\Delta t}(\bm{x})$ for the next time step $t + \Delta t$ \cite{code}.
Under a similar argument of \cite{huang2024vy} we can show the error:  $  \mathbb{E}  \left|\bm{X}_{0,n \Delta t}-\bm{X}^\mathrm{NN}_{0,n \Delta t}\right|/T \leq   O(\varepsilon)+O(\Delta t) ,$ with $\varepsilon$ the error bounds of scores, and $ n\Delta t \le T$.

\begin{figure}[htbp]
  \centering
  \begin{minipage}[t]{\linewidth}
    \raggedright (a)\\[-0.5ex]
    \includegraphics[width=.95\linewidth]{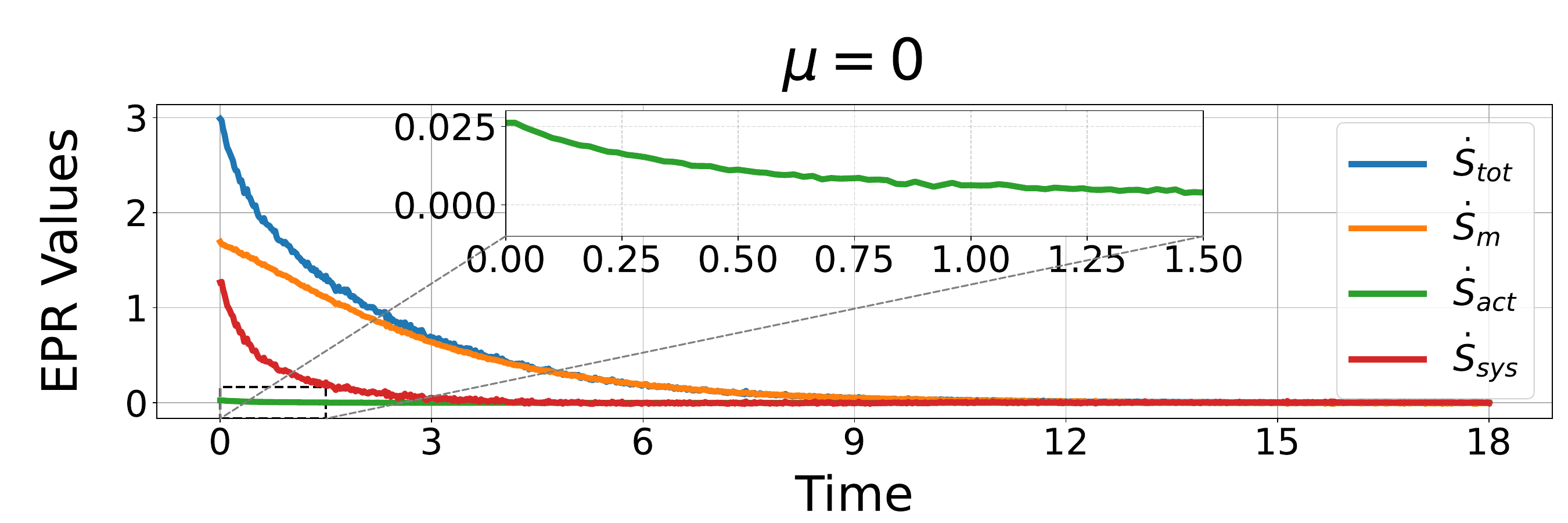}
  \end{minipage}\hfill
  \begin{minipage}[t]{\linewidth}
    \raggedright (b)\\[-0.5ex]
    \includegraphics[width=.95\linewidth]{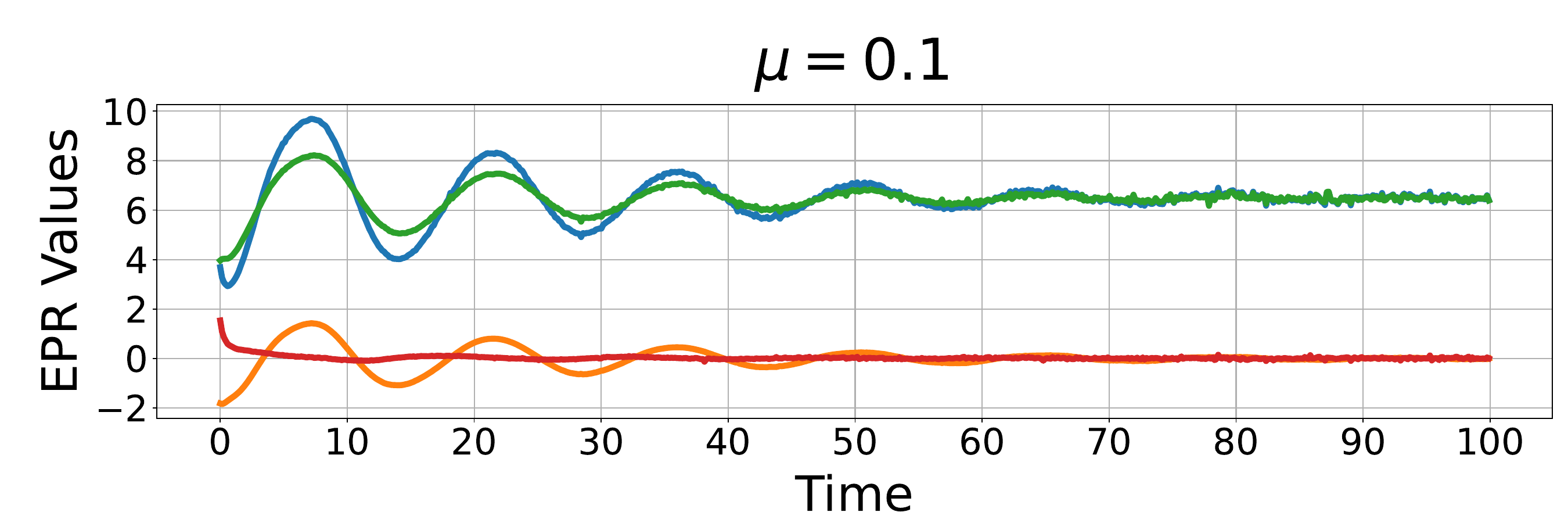}
  \end{minipage}\hfill
  \caption{EPRs of a Brownian particle in an active bath. Shown are $\dot{S}_\mathrm{tot}$ (blue) and its three components: $\dot{S}_\mathrm{m}$ (orange), $\dot{S}_\mathrm{act}$ (green), and $\dot{S}_\mathrm{sys}$ (red) in both plots. (a)  $\mu=0$. All EPRs decay to 0, showing equilibrium-like behavior. (b)  $\mu=0.1$. $\dot{S}_\mathrm{tot}$ saturates at a positive value due to active fluctuation, showing a nonequilibrium steady state.
}
  \label{fig:1DEPR}
\end{figure}

{\itshape Examples}---We first consider a Brownian particle starting from a standard Gaussian distribution immersed in an active bath \cite{paneru2021transport}, diﬀusing in a spatially asymmetric periodic potential of period $L$ and barrier height $2V_0$, $V(r)= V_0\left[ \sin(2\pi r/L) + 0.25\sin(4\pi r/L) \right]$:
\begin{eqnarray}
     \frac{\d r}{\d t}=&& -\frac{1}{\Gamma}\frac{\partial V(r)}{\partial r} + \eta_\mathrm{th}(t)+   \sum_{i=1}^{N(t)}A_i\delta(t-t_i),
\end{eqnarray}
where $N(t)$ is a Poisson counting process with parameters $\lambda_0$ and $A_i$ are Gaussian random variables with mean $\mu$ and standard deviation $\sigma$. The numerical parameters can be found in \cite[Appendix VII]{supplemental_material}. Using our proposed deep learning framework, we perform two numerical experiments: (1) \(\mu = 0\) and \(\sigma = 1/24\), and (2) \(\mu = 0.1\) and \(\sigma = 1/24\), respectively. FIG.~\ref{fig:1DEPR}  illustrates the EPRs for this example.   FIG.~\ref{fig:1DEPR}(a) is for  zero mean $\mu=0$. In this case, since the noise does not have net drift, its impact on the system is minimal, and all EPRs eventually decay to zero, indicating that the system   evolves towards equilibrium. This is similar to the case where there is only Gaussian thermal noise. That is, active non-Gaussian systems can still admit equilibrium steady states, as demonstrated here. Interestingly, \cite{faria2025nonequilibrium} shows that for a different system with non-Gaussian fluctuations, increasing non-Gaussianity in fact can reduce the total EP, indicating the system effectively approaches equilibrium. Similarly, \cite{fodor2016far} shows the active Ornstein--Uhlenbeck process still satisfies an effective fluctuation-dissipation theorem (FDT) of equilibrium fluctuations. FIG.~\ref{fig:1DEPR}(b) shows that increasing   $\mu$ to \(0.1\) drives the system away from equilibrium. As a result, the EPR (\(\dot{S}_{\text{tot}}\)) stabilizes to a nonzero steady-state value due to the contribution (\(\dot{S}_{\text{act}}\)) from active fluctuations, which defines the non-equilibrium steady state with active fluctuations breaking the detailed balance. These results highlight how the statistical properties of active noise, particularly the mean jump height, play a crucial role in determining the system's dynamical and steady-state EP. Moreover, we validate these numerical results and perform an examination of the influence of various $\mu$ parameters in \cite{supplemental_material}. 

\begin{figure*}[htbp]
  \begin{minipage}[l]{0.3\linewidth}
    \centering
    \begin{overpic}[width=.55\linewidth]{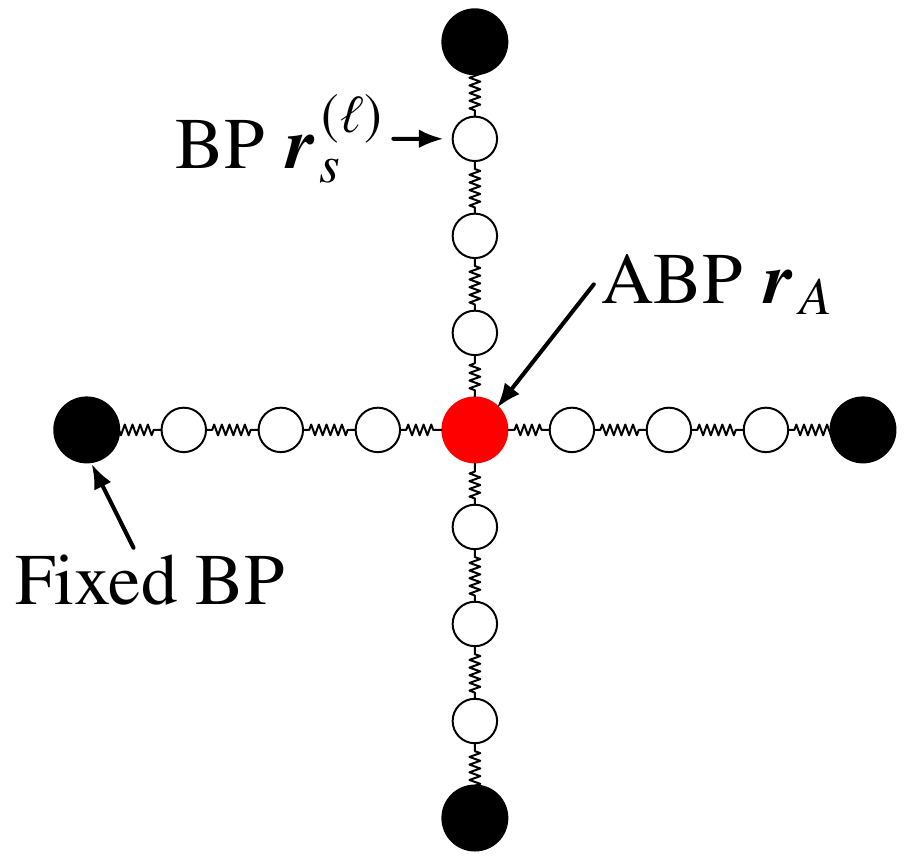}
      \put(1,90){(a)}
    \end{overpic}
  \end{minipage}
  \begin{minipage}[l]{0.344\linewidth} 
    \begin{overpic}[width=\linewidth]{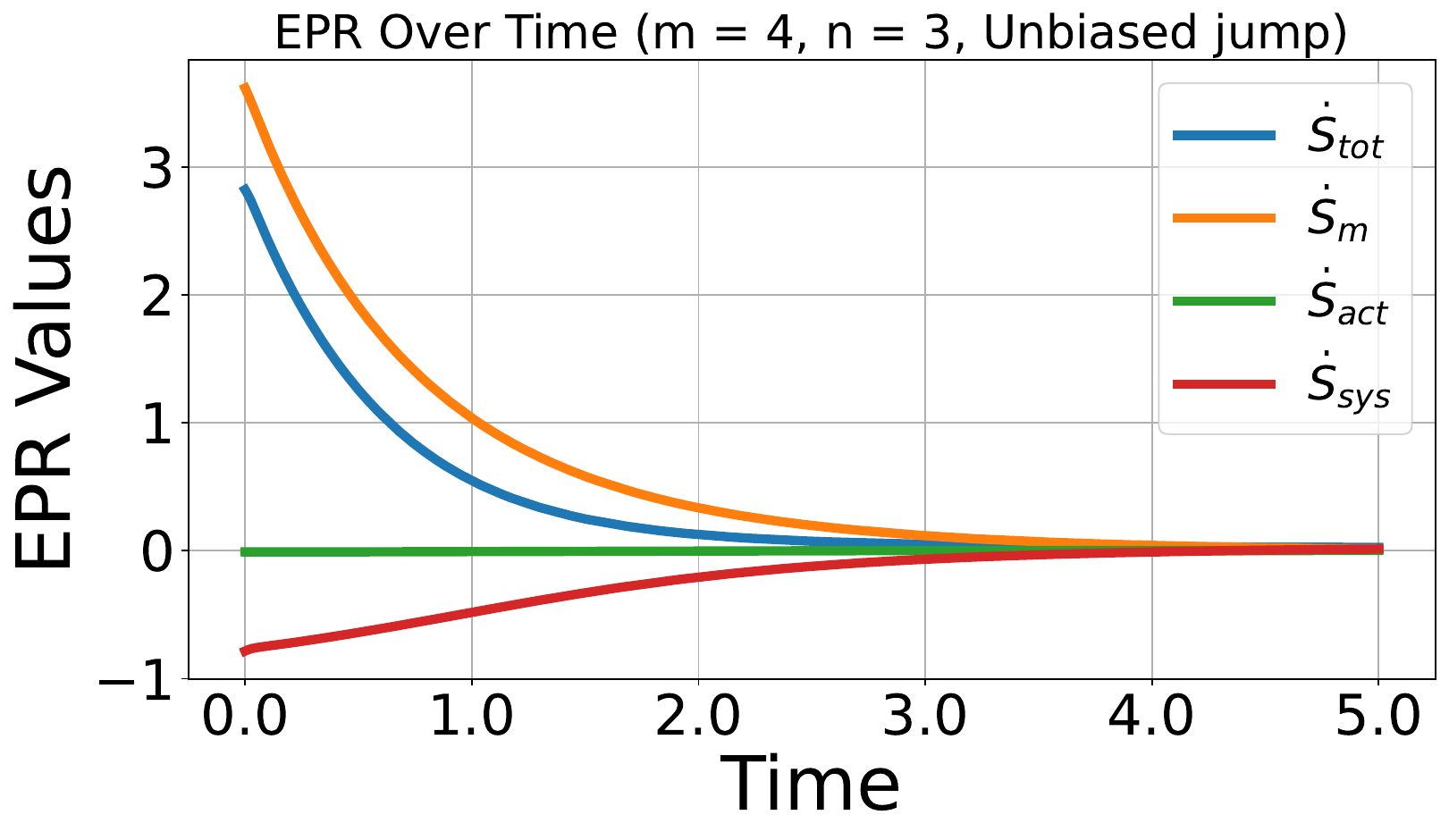}
      \put(1,55){(b)}
    \end{overpic}
  \end{minipage}
   \begin{minipage}[c]{0.344\linewidth} 
    \begin{overpic}[width=\linewidth]{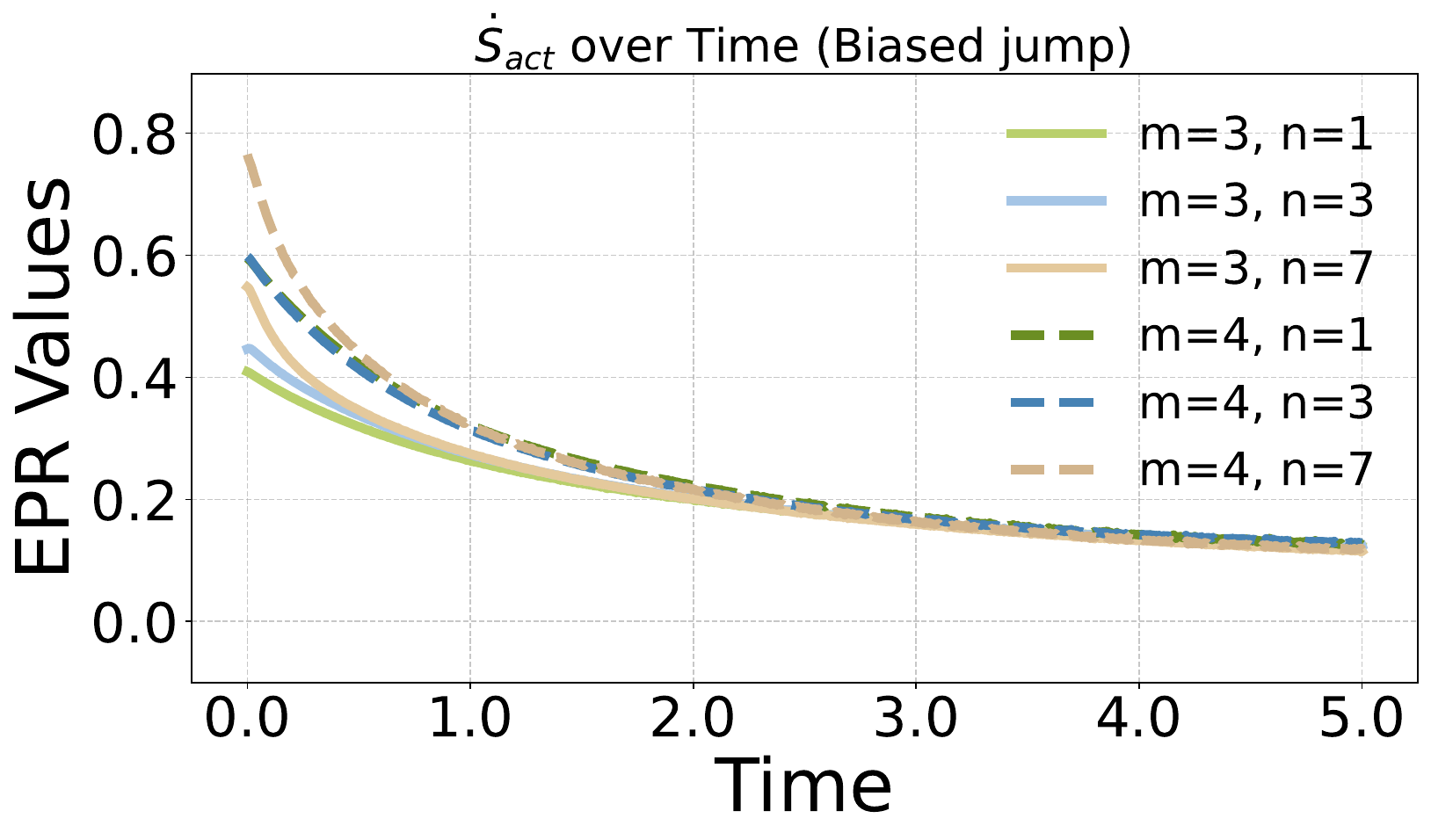}
      \put(1,55){(c)}
    \end{overpic}
  \end{minipage}
\caption{(a)  Schematic of the ABP system:  A central ABP (red) is connected to $m$ chains (illustrated with $m=4$) consisting of $n$ BPs (white) each (with $n=3$ shown) and a fixed BP (black) at the end of each chain. (b) EPRs of the ABP under unbiased jump noise with $m = 4$ and $n = 3$. Shown are $\dot{S}_\mathrm{tot}$ (blue) and its three components: $\dot{S}_\mathrm{m}$ (orange), $\dot{S}_\mathrm{act}$ (green), and $\dot{S}_\mathrm{sys}$ (red).
(c) Active EPR $\dot{S}_\mathrm{act}$ of the ABP under biased jump noise for various combinations of $(m, n)$.
Solid and dashed lines correspond to $m = 3$ and $m = 4$, respectively, with colors indicating different $n$ values: green ($n = 1$), blue ($n = 3$), and beige ($n = 7$).
}
\label{fig:beads&EPR}
\end{figure*}

To demonstrate that our numerical method is applicable to high dimensional problems, we next consider an active polymer system consisting of an active Brownian particle (ABP) cross-linker and $(n+1)m$ ordinary Brownian beads, as illustrated in FIG.~\ref{fig:beads&EPR}(a) and described   in End Matter Appendix C. The active fluctuations  examined include   both unbiased   
and   biased types, shown in FIG. \ref{fig:beads&EPR} (b) and (c) respectively.  FIG. \ref{fig:beads&EPR}(b) shows 
the numerical value of the EPR for the active polymer system under 
symmetric (unbiased) jump  applied to the   ABP. The EPR converges to zero, indicating that the system reaches an effective equilibrium state despite the presence of active fluctuations.
For the ABP subject to biased jump noise, FIG. \ref{fig:beads&EPR}(c) displays the time evolution of the active EPR  $\dot{S}_\mathrm{act}$. In this case, $\dot{S}_\mathrm{act}$ approaches a small but nonzero steady-state, reflecting a sustained non-equilibrium driving due to the bias jumps.  Consequentially, the total EPR $\dot{S}_\mathrm{tot}$ remains small yet finite. This result  demonstrates the distinct influence of non-Gaussian active noise on the non-equilibrium character of the polymer system. We remark that the small magnitude of the long time  entropy production in  FIG. \ref{fig:beads&EPR}(c) is attributed to the system's constraints—namely, the fixed end beads in each arm and the localization of active driving solely at the central particle—which restrict the overall dissipation.
More numerical validation and detailed results for this biased case are provided  in  \cite[Appendix VIII]{supplemental_material}.

{\itshape Conclusions and outlook---}Via the probability flow equivalence technique \cite{huang2024probability,huang2024vy}, we rigorously formulate the EP in active matter systems along a single trajectory as the sum of three contributions: the system, the medium, and the active fluctuations in Eq.\eqref{eprdecomposition}. 
This framework is particularly applicable to systems exhibiting non-Gaussian fluctuations like \eqref{eqn:SDE}. The total EP, combined with a random variable induced by active fluctuation, satisfies an integral fluctuation theorem  and a detailed fluctuation theorem  that hold universally—regardless of the initial conditions or the length of the trajectories.

Importantly, with our definition of entropy, the integral
fluctuation theorem for the total EP remains
valid not only in steady states but also for finite-length
trajectories.
Experimentally, the trajectory-dependent entropy of a particle could be measured under a time-dependent protocol by recording the probability distribution across many trajectories. From this data, the entropy $s_\mathrm{sys}$ of individual trajectories can be inferred, which may help enabling direct experimental validation of the  fluctuation theorems in Eqs.~\eqref{FT} and \eqref{DFT} and generalized second law in Eq.~\eqref{SL}.

Finally, our deep learning-based numerical method efficiently and accurately computes each contribution term of EP in active systems. This powerful computational tool explores nonequilibrium thermodynamics in complex systems, enabling insights into the role of active fluctuations in EP and dissipation.

As a perspective for future work, it would be intriguing to investigate whether the fluctuation theorem derived here for non-Gaussian active noise  can still be used to establish relations between response and correlation functions analogous to the equilibrium fluctuation–dissipation theorem.

\hfill\\
{\itshape Acknowledgments---} X.Z. acknowledges the support from 
 Hong Kong General Research Funds  (11318522, 11308323). B.M. acknowledges support from the National Natural Science Foundation of China (NSFC) (Grant Nos. 12575045 and 12034019). The authors acknowledge helpful discussions with Prof. Jong-Min Park. We also thank Prof. Tiezheng Qian and Prof. Dahai He for helpful comments on our draft of the manuscript.
 We are grateful to the anonymous reviewers for their insightful comments and detailed feedback, which helped us improve the clarity and robustness of this Letter.

Corresponding authors*:

Bing Miao: bmiao@ucas.ac.cn

Xiang Zhou: xiang.zhou@cityu.edu.hk

\bibliographystyle{apsrev4-2}
\bibliography{apssamp}

\begin{thebibliography}{75}%
\makeatletter
\providecommand \@ifxundefined [1]{%
 \@ifx{#1\undefined}
}%
\providecommand \@ifnum [1]{%
 \ifnum #1\expandafter \@firstoftwo
 \else \expandafter \@secondoftwo
 \fi
}%
\providecommand \@ifx [1]{%
 \ifx #1\expandafter \@firstoftwo
 \else \expandafter \@secondoftwo
 \fi
}%
\providecommand \natexlab [1]{#1}%
\providecommand \enquote  [1]{``#1''}%
\providecommand \bibnamefont  [1]{#1}%
\providecommand \bibfnamefont [1]{#1}%
\providecommand \citenamefont [1]{#1}%
\providecommand \href@noop [0]{\@secondoftwo}%
\providecommand \href [0]{\begingroup \@sanitize@url \@href}%
\providecommand \@href[1]{\@@startlink{#1}\@@href}%
\providecommand \@@href[1]{\endgroup#1\@@endlink}%
\providecommand \@sanitize@url [0]{\catcode `\\12\catcode `\$12\catcode `\&12\catcode `\#12\catcode `\^12\catcode `\_12\catcode `\%12\relax}%
\providecommand \@@startlink[1]{}%
\providecommand \@@endlink[0]{}%
\providecommand \url  [0]{\begingroup\@sanitize@url \@url }%
\providecommand \@url [1]{\endgroup\@href {#1}{\urlprefix }}%
\providecommand \urlprefix  [0]{URL }%
\providecommand \Eprint [0]{\href }%
\providecommand \doibase [0]{https://doi.org/}%
\providecommand \selectlanguage [0]{\@gobble}%
\providecommand \bibinfo  [0]{\@secondoftwo}%
\providecommand \bibfield  [0]{\@secondoftwo}%
\providecommand \translation [1]{[#1]}%
\providecommand \BibitemOpen [0]{}%
\providecommand \bibitemStop [0]{}%
\providecommand \bibitemNoStop [0]{.\EOS\space}%
\providecommand \EOS [0]{\spacefactor3000\relax}%
\providecommand \BibitemShut  [1]{\csname bibitem#1\endcsname}%
\let\auto@bib@innerbib\@empty
\bibitem [{\citenamefont {Seifert}(2012)}]{seifert2012stochastic}%
  \BibitemOpen
  \bibfield  {author} {\bibinfo {author} {\bibfnamefont {U.}~\bibnamefont {Seifert}},\ }\href@noop {} {\bibfield  {journal} {\bibinfo  {journal} {Rep. Prog. Phys.}\ }\textbf {\bibinfo {volume} {75}},\ \bibinfo {pages} {126001} (\bibinfo {year} {2012})}\BibitemShut {NoStop}%
\bibitem [{\citenamefont {Van~den Broeck}\ and\ \citenamefont {Esposito}(2015)}]{van2015ensemble}%
  \BibitemOpen
  \bibfield  {author} {\bibinfo {author} {\bibfnamefont {C.}~\bibnamefont {Van~den Broeck}}\ and\ \bibinfo {author} {\bibfnamefont {M.}~\bibnamefont {Esposito}},\ }\href@noop {} {\bibfield  {journal} {\bibinfo  {journal} {Physica A}\ }\textbf {\bibinfo {volume} {418}},\ \bibinfo {pages} {6} (\bibinfo {year} {2015})}\BibitemShut {NoStop}%
\bibitem [{\citenamefont {Peliti}\ and\ \citenamefont {Pigolotti}(2021)}]{peliti2021stochastic}%
  \BibitemOpen
  \bibfield  {author} {\bibinfo {author} {\bibfnamefont {L.}~\bibnamefont {Peliti}}\ and\ \bibinfo {author} {\bibfnamefont {S.}~\bibnamefont {Pigolotti}},\ }\href@noop {} {\emph {\bibinfo {title} {Stochastic thermodynamics: an introduction}}}\ (\bibinfo  {publisher} {Princeton University Press},\ \bibinfo {year} {2021})\BibitemShut {NoStop}%
\bibitem [{\citenamefont {Gaspard}(2022)}]{gaspard2022statistical}%
  \BibitemOpen
  \bibfield  {author} {\bibinfo {author} {\bibfnamefont {P.}~\bibnamefont {Gaspard}},\ }\href@noop {} {\emph {\bibinfo {title} {The Statistical Mechanics of Irreversible Phenomena}}}\ (\bibinfo  {publisher} {Cambridge University Press},\ \bibinfo {year} {2022})\BibitemShut {NoStop}%
\bibitem [{\citenamefont {Li}\ \emph {et~al.}(2025)\citenamefont {Li}, \citenamefont {Zhao}, \citenamefont {Yang}, \citenamefont {Tang},\ and\ \citenamefont {Zhang}}]{li2025biased}%
  \BibitemOpen
  \bibfield  {author} {\bibinfo {author} {\bibfnamefont {Y.}~\bibnamefont {Li}}, \bibinfo {author} {\bibfnamefont {Y.}~\bibnamefont {Zhao}}, \bibinfo {author} {\bibfnamefont {S.}~\bibnamefont {Yang}}, \bibinfo {author} {\bibfnamefont {M.}~\bibnamefont {Tang}},\ and\ \bibinfo {author} {\bibfnamefont {H.}~\bibnamefont {Zhang}},\ }\href@noop {} {\bibfield  {journal} {\bibinfo  {journal} {Phys. Rev. Lett.}\ }\textbf {\bibinfo {volume} {134}},\ \bibinfo {pages} {108301} (\bibinfo {year} {2025})}\BibitemShut {NoStop}%
\bibitem [{\citenamefont {Lauga}\ and\ \citenamefont {Goldstein}(2012)}]{lauga2012dance}%
  \BibitemOpen
  \bibfield  {author} {\bibinfo {author} {\bibfnamefont {E.}~\bibnamefont {Lauga}}\ and\ \bibinfo {author} {\bibfnamefont {R.~E.}\ \bibnamefont {Goldstein}},\ }\href@noop {} {\bibfield  {journal} {\bibinfo  {journal} {Phys. Today}\ }\textbf {\bibinfo {volume} {65}},\ \bibinfo {pages} {30} (\bibinfo {year} {2012})}\BibitemShut {NoStop}%
\bibitem [{\citenamefont {Matth{\"a}us}\ \emph {et~al.}(2009)\citenamefont {Matth{\"a}us}, \citenamefont {Jagodi{\v{c}}},\ and\ \citenamefont {Dobnikar}}]{matthaus2009coli}%
  \BibitemOpen
  \bibfield  {author} {\bibinfo {author} {\bibfnamefont {F.}~\bibnamefont {Matth{\"a}us}}, \bibinfo {author} {\bibfnamefont {M.}~\bibnamefont {Jagodi{\v{c}}}},\ and\ \bibinfo {author} {\bibfnamefont {J.}~\bibnamefont {Dobnikar}},\ }\href@noop {} {\bibfield  {journal} {\bibinfo  {journal} {Biophys. J.}\ }\textbf {\bibinfo {volume} {97}},\ \bibinfo {pages} {946} (\bibinfo {year} {2009})}\BibitemShut {NoStop}%
\bibitem [{\citenamefont {Goswami}\ \emph {et~al.}(2024)\citenamefont {Goswami}, \citenamefont {Cherstvy}, \citenamefont {Godec},\ and\ \citenamefont {Metzler}}]{goswami2024anomalous}%
  \BibitemOpen
  \bibfield  {author} {\bibinfo {author} {\bibfnamefont {K.}~\bibnamefont {Goswami}}, \bibinfo {author} {\bibfnamefont {A.~G.}\ \bibnamefont {Cherstvy}}, \bibinfo {author} {\bibfnamefont {A.}~\bibnamefont {Godec}},\ and\ \bibinfo {author} {\bibfnamefont {R.}~\bibnamefont {Metzler}},\ }\href@noop {} {\bibfield  {journal} {\bibinfo  {journal} {Phys. Rev. E}\ }\textbf {\bibinfo {volume} {110}},\ \bibinfo {pages} {044609} (\bibinfo {year} {2024})}\BibitemShut {NoStop}%
\bibitem [{\citenamefont {Leptos}\ \emph {et~al.}(2009)\citenamefont {Leptos}, \citenamefont {Guasto}, \citenamefont {Gollub}, \citenamefont {Pesci},\ and\ \citenamefont {Goldstein}}]{leptos2009dynamics}%
  \BibitemOpen
  \bibfield  {author} {\bibinfo {author} {\bibfnamefont {K.~C.}\ \bibnamefont {Leptos}}, \bibinfo {author} {\bibfnamefont {J.~S.}\ \bibnamefont {Guasto}}, \bibinfo {author} {\bibfnamefont {J.~P.}\ \bibnamefont {Gollub}}, \bibinfo {author} {\bibfnamefont {A.~I.}\ \bibnamefont {Pesci}},\ and\ \bibinfo {author} {\bibfnamefont {R.~E.}\ \bibnamefont {Goldstein}},\ }\href@noop {} {\bibfield  {journal} {\bibinfo  {journal} {Phys. Rev. Lett.}\ }\textbf {\bibinfo {volume} {103}},\ \bibinfo {pages} {198103} (\bibinfo {year} {2009})}\BibitemShut {NoStop}%
\bibitem [{\citenamefont {Wu}\ and\ \citenamefont {Libchaber}(2000)}]{wu2000particle}%
  \BibitemOpen
  \bibfield  {author} {\bibinfo {author} {\bibfnamefont {X.-L.}\ \bibnamefont {Wu}}\ and\ \bibinfo {author} {\bibfnamefont {A.}~\bibnamefont {Libchaber}},\ }\href@noop {} {\bibfield  {journal} {\bibinfo  {journal} {Phys. Rev. Lett.}\ }\textbf {\bibinfo {volume} {84}},\ \bibinfo {pages} {3017} (\bibinfo {year} {2000})}\BibitemShut {NoStop}%
\bibitem [{\citenamefont {Chen}\ \emph {et~al.}(2007)\citenamefont {Chen}, \citenamefont {Lau}, \citenamefont {Hough}, \citenamefont {Islam}, \citenamefont {Goulian}, \citenamefont {Lubensky},\ and\ \citenamefont {Yodh}}]{chen2007fluctuations}%
  \BibitemOpen
  \bibfield  {author} {\bibinfo {author} {\bibfnamefont {D.~T.}\ \bibnamefont {Chen}}, \bibinfo {author} {\bibfnamefont {A.}~\bibnamefont {Lau}}, \bibinfo {author} {\bibfnamefont {L.~A.}\ \bibnamefont {Hough}}, \bibinfo {author} {\bibfnamefont {M.~F.}\ \bibnamefont {Islam}}, \bibinfo {author} {\bibfnamefont {M.}~\bibnamefont {Goulian}}, \bibinfo {author} {\bibfnamefont {T.~C.}\ \bibnamefont {Lubensky}},\ and\ \bibinfo {author} {\bibfnamefont {A.~G.}\ \bibnamefont {Yodh}},\ }\href@noop {} {\bibfield  {journal} {\bibinfo  {journal} {Phys. Rev. Lett.}\ }\textbf {\bibinfo {volume} {99}},\ \bibinfo {pages} {148302} (\bibinfo {year} {2007})}\BibitemShut {NoStop}%
\bibitem [{\citenamefont {Palacci}\ \emph {et~al.}(2010)\citenamefont {Palacci}, \citenamefont {Cottin-Bizonne}, \citenamefont {Ybert},\ and\ \citenamefont {Bocquet}}]{palacci2010sedimentation}%
  \BibitemOpen
  \bibfield  {author} {\bibinfo {author} {\bibfnamefont {J.}~\bibnamefont {Palacci}}, \bibinfo {author} {\bibfnamefont {C.}~\bibnamefont {Cottin-Bizonne}}, \bibinfo {author} {\bibfnamefont {C.}~\bibnamefont {Ybert}},\ and\ \bibinfo {author} {\bibfnamefont {L.}~\bibnamefont {Bocquet}},\ }\href@noop {} {\bibfield  {journal} {\bibinfo  {journal} {Phys. Rev. Lett.}\ }\textbf {\bibinfo {volume} {105}},\ \bibinfo {pages} {088304} (\bibinfo {year} {2010})}\BibitemShut {NoStop}%
\bibitem [{\citenamefont {Ariga}\ \emph {et~al.}(2021)\citenamefont {Ariga}, \citenamefont {Tateishi}, \citenamefont {Tomishige},\ and\ \citenamefont {Mizuno}}]{ariga2021noise}%
  \BibitemOpen
  \bibfield  {author} {\bibinfo {author} {\bibfnamefont {T.}~\bibnamefont {Ariga}}, \bibinfo {author} {\bibfnamefont {K.}~\bibnamefont {Tateishi}}, \bibinfo {author} {\bibfnamefont {M.}~\bibnamefont {Tomishige}},\ and\ \bibinfo {author} {\bibfnamefont {D.}~\bibnamefont {Mizuno}},\ }\href@noop {} {\bibfield  {journal} {\bibinfo  {journal} {Phys. Rev. Lett.}\ }\textbf {\bibinfo {volume} {127}},\ \bibinfo {pages} {178101} (\bibinfo {year} {2021})}\BibitemShut {NoStop}%
\bibitem [{\citenamefont {Kurihara}\ \emph {et~al.}(2017)\citenamefont {Kurihara}, \citenamefont {Aridome}, \citenamefont {Ayade}, \citenamefont {Zaid},\ and\ \citenamefont {Mizuno}}]{kurihara2017non}%
  \BibitemOpen
  \bibfield  {author} {\bibinfo {author} {\bibfnamefont {T.}~\bibnamefont {Kurihara}}, \bibinfo {author} {\bibfnamefont {M.}~\bibnamefont {Aridome}}, \bibinfo {author} {\bibfnamefont {H.}~\bibnamefont {Ayade}}, \bibinfo {author} {\bibfnamefont {I.}~\bibnamefont {Zaid}},\ and\ \bibinfo {author} {\bibfnamefont {D.}~\bibnamefont {Mizuno}},\ }\href@noop {} {\bibfield  {journal} {\bibinfo  {journal} {Phys. Rev. E}\ }\textbf {\bibinfo {volume} {95}},\ \bibinfo {pages} {030601} (\bibinfo {year} {2017})}\BibitemShut {NoStop}%
\bibitem [{\citenamefont {Park}\ \emph {et~al.}(2020)\citenamefont {Park}, \citenamefont {Paneru}, \citenamefont {Kwon}, \citenamefont {Granick},\ and\ \citenamefont {Pak}}]{park2020rapid}%
  \BibitemOpen
  \bibfield  {author} {\bibinfo {author} {\bibfnamefont {J.~T.}\ \bibnamefont {Park}}, \bibinfo {author} {\bibfnamefont {G.}~\bibnamefont {Paneru}}, \bibinfo {author} {\bibfnamefont {C.}~\bibnamefont {Kwon}}, \bibinfo {author} {\bibfnamefont {S.}~\bibnamefont {Granick}},\ and\ \bibinfo {author} {\bibfnamefont {H.~K.}\ \bibnamefont {Pak}},\ }\href@noop {} {\bibfield  {journal} {\bibinfo  {journal} {Soft Matter}\ }\textbf {\bibinfo {volume} {16}},\ \bibinfo {pages} {8122} (\bibinfo {year} {2020})}\BibitemShut {NoStop}%
\bibitem [{\citenamefont {Ezber}\ \emph {et~al.}(2020)\citenamefont {Ezber}, \citenamefont {Belyy}, \citenamefont {Can},\ and\ \citenamefont {Yildiz}}]{ezber2020dynein}%
  \BibitemOpen
  \bibfield  {author} {\bibinfo {author} {\bibfnamefont {Y.}~\bibnamefont {Ezber}}, \bibinfo {author} {\bibfnamefont {V.}~\bibnamefont {Belyy}}, \bibinfo {author} {\bibfnamefont {S.}~\bibnamefont {Can}},\ and\ \bibinfo {author} {\bibfnamefont {A.}~\bibnamefont {Yildiz}},\ }\href@noop {} {\bibfield  {journal} {\bibinfo  {journal} {Nat. Phys.}\ }\textbf {\bibinfo {volume} {16}},\ \bibinfo {pages} {312} (\bibinfo {year} {2020})}\BibitemShut {NoStop}%
\bibitem [{\citenamefont {Paneru}\ \emph {et~al.}(2021)\citenamefont {Paneru}, \citenamefont {Park},\ and\ \citenamefont {Pak}}]{paneru2021transport}%
  \BibitemOpen
  \bibfield  {author} {\bibinfo {author} {\bibfnamefont {G.}~\bibnamefont {Paneru}}, \bibinfo {author} {\bibfnamefont {J.~T.}\ \bibnamefont {Park}},\ and\ \bibinfo {author} {\bibfnamefont {H.~K.}\ \bibnamefont {Pak}},\ }\href@noop {} {\bibfield  {journal} {\bibinfo  {journal} {J. Phys. Chem. Lett.}\ }\textbf {\bibinfo {volume} {12}},\ \bibinfo {pages} {11078} (\bibinfo {year} {2021})}\BibitemShut {NoStop}%
\bibitem [{\citenamefont {Sang}\ \emph {et~al.}(2022)\citenamefont {Sang}, \citenamefont {Wen},\ and\ \citenamefont {He}}]{sang2022single}%
  \BibitemOpen
  \bibfield  {author} {\bibinfo {author} {\bibfnamefont {Y.}~\bibnamefont {Sang}}, \bibinfo {author} {\bibfnamefont {X.}~\bibnamefont {Wen}},\ and\ \bibinfo {author} {\bibfnamefont {Y.}~\bibnamefont {He}},\ }\href@noop {} {\bibfield  {journal} {\bibinfo  {journal} {View}\ }\textbf {\bibinfo {volume} {3}},\ \bibinfo {pages} {20220047} (\bibinfo {year} {2022})}\BibitemShut {NoStop}%
\bibitem [{\citenamefont {Chen}\ \emph {et~al.}(2015)\citenamefont {Chen}, \citenamefont {Wang},\ and\ \citenamefont {Granick}}]{chen2015memoryless}%
  \BibitemOpen
  \bibfield  {author} {\bibinfo {author} {\bibfnamefont {K.}~\bibnamefont {Chen}}, \bibinfo {author} {\bibfnamefont {B.}~\bibnamefont {Wang}},\ and\ \bibinfo {author} {\bibfnamefont {S.}~\bibnamefont {Granick}},\ }\href@noop {} {\bibfield  {journal} {\bibinfo  {journal} {Nat. Mater.}\ }\textbf {\bibinfo {volume} {14}},\ \bibinfo {pages} {589} (\bibinfo {year} {2015})}\BibitemShut {NoStop}%
\bibitem [{\citenamefont {Song}\ \emph {et~al.}(2018)\citenamefont {Song}, \citenamefont {Moon}, \citenamefont {Jeon},\ and\ \citenamefont {Park}}]{song2018neuronal}%
  \BibitemOpen
  \bibfield  {author} {\bibinfo {author} {\bibfnamefont {M.~S.}\ \bibnamefont {Song}}, \bibinfo {author} {\bibfnamefont {H.~C.}\ \bibnamefont {Moon}}, \bibinfo {author} {\bibfnamefont {J.-H.}\ \bibnamefont {Jeon}},\ and\ \bibinfo {author} {\bibfnamefont {H.~Y.}\ \bibnamefont {Park}},\ }\href@noop {} {\bibfield  {journal} {\bibinfo  {journal} {Nat. Commun.}\ }\textbf {\bibinfo {volume} {9}},\ \bibinfo {pages} {1} (\bibinfo {year} {2018})}\BibitemShut {NoStop}%
\bibitem [{\citenamefont {Baconnier}\ \emph {et~al.}(2022)\citenamefont {Baconnier}, \citenamefont {Shohat}, \citenamefont {L{\'o}pez}, \citenamefont {Coulais}, \citenamefont {D{\'e}mery}, \citenamefont {D{\"u}ring},\ and\ \citenamefont {Dauchot}}]{baconnier2022selective}%
  \BibitemOpen
  \bibfield  {author} {\bibinfo {author} {\bibfnamefont {P.}~\bibnamefont {Baconnier}}, \bibinfo {author} {\bibfnamefont {D.}~\bibnamefont {Shohat}}, \bibinfo {author} {\bibfnamefont {C.~H.}\ \bibnamefont {L{\'o}pez}}, \bibinfo {author} {\bibfnamefont {C.}~\bibnamefont {Coulais}}, \bibinfo {author} {\bibfnamefont {V.}~\bibnamefont {D{\'e}mery}}, \bibinfo {author} {\bibfnamefont {G.}~\bibnamefont {D{\"u}ring}},\ and\ \bibinfo {author} {\bibfnamefont {O.}~\bibnamefont {Dauchot}},\ }\href@noop {} {\bibfield  {journal} {\bibinfo  {journal} {Nature Physics}\ }\textbf {\bibinfo {volume} {18}},\ \bibinfo {pages} {1234} (\bibinfo {year} {2022})}\BibitemShut {NoStop}%
\bibitem [{\citenamefont {Hern{\'a}ndez-L{\'o}pez}\ \emph {et~al.}(2024)\citenamefont {Hern{\'a}ndez-L{\'o}pez}, \citenamefont {Baconnier}, \citenamefont {Coulais}, \citenamefont {Dauchot},\ and\ \citenamefont {D{\"u}ring}}]{hernandez2024model}%
  \BibitemOpen
  \bibfield  {author} {\bibinfo {author} {\bibfnamefont {C.}~\bibnamefont {Hern{\'a}ndez-L{\'o}pez}}, \bibinfo {author} {\bibfnamefont {P.}~\bibnamefont {Baconnier}}, \bibinfo {author} {\bibfnamefont {C.}~\bibnamefont {Coulais}}, \bibinfo {author} {\bibfnamefont {O.}~\bibnamefont {Dauchot}},\ and\ \bibinfo {author} {\bibfnamefont {G.}~\bibnamefont {D{\"u}ring}},\ }\href@noop {} {\bibfield  {journal} {\bibinfo  {journal} {Physical Review Letters}\ }\textbf {\bibinfo {volume} {132}},\ \bibinfo {pages} {238303} (\bibinfo {year} {2024})}\BibitemShut {NoStop}%
\bibitem [{\citenamefont {Caprini}\ \emph {et~al.}(2023)\citenamefont {Caprini}, \citenamefont {Marini Bettolo~Marconi}, \citenamefont {Puglisi},\ and\ \citenamefont {L{\"o}wen}}]{caprini2023entropons}%
  \BibitemOpen
  \bibfield  {author} {\bibinfo {author} {\bibfnamefont {L.}~\bibnamefont {Caprini}}, \bibinfo {author} {\bibfnamefont {U.}~\bibnamefont {Marini Bettolo~Marconi}}, \bibinfo {author} {\bibfnamefont {A.}~\bibnamefont {Puglisi}},\ and\ \bibinfo {author} {\bibfnamefont {H.}~\bibnamefont {L{\"o}wen}},\ }\href@noop {} {\bibfield  {journal} {\bibinfo  {journal} {The Journal of Chemical Physics}\ }\textbf {\bibinfo {volume} {159}} (\bibinfo {year} {2023})}\BibitemShut {NoStop}%
\bibitem [{\citenamefont {Barik}\ \emph {et~al.}(2006)\citenamefont {Barik}, \citenamefont {Ghosh},\ and\ \citenamefont {Ray}}]{barik2006langevin}%
  \BibitemOpen
  \bibfield  {author} {\bibinfo {author} {\bibfnamefont {D.}~\bibnamefont {Barik}}, \bibinfo {author} {\bibfnamefont {P.~K.}\ \bibnamefont {Ghosh}},\ and\ \bibinfo {author} {\bibfnamefont {D.~S.}\ \bibnamefont {Ray}},\ }\href@noop {} {\bibfield  {journal} {\bibinfo  {journal} {J. Stat. Mech.-Theory Exp.}\ }\textbf {\bibinfo {volume} {2006}},\ \bibinfo {pages} {P03010} (\bibinfo {year} {2006})}\BibitemShut {NoStop}%
\bibitem [{\citenamefont {Demaerel}\ and\ \citenamefont {Maes}(2018)}]{demaerel2018active}%
  \BibitemOpen
  \bibfield  {author} {\bibinfo {author} {\bibfnamefont {T.}~\bibnamefont {Demaerel}}\ and\ \bibinfo {author} {\bibfnamefont {C.}~\bibnamefont {Maes}},\ }\href@noop {} {\bibfield  {journal} {\bibinfo  {journal} {Phys. Rev. E}\ }\textbf {\bibinfo {volume} {97}},\ \bibinfo {pages} {032604} (\bibinfo {year} {2018})}\BibitemShut {NoStop}%
\bibitem [{\citenamefont {Um}\ \emph {et~al.}(2019)\citenamefont {Um}, \citenamefont {Song},\ and\ \citenamefont {Jeon}}]{um2019langevin}%
  \BibitemOpen
  \bibfield  {author} {\bibinfo {author} {\bibfnamefont {J.}~\bibnamefont {Um}}, \bibinfo {author} {\bibfnamefont {T.}~\bibnamefont {Song}},\ and\ \bibinfo {author} {\bibfnamefont {J.-H.}\ \bibnamefont {Jeon}},\ }\href@noop {} {\bibfield  {journal} {\bibinfo  {journal} {Front. Physics}\ }\textbf {\bibinfo {volume} {7}},\ \bibinfo {pages} {143} (\bibinfo {year} {2019})}\BibitemShut {NoStop}%
\bibitem [{\citenamefont {Goerlich}\ \emph {et~al.}(2022)\citenamefont {Goerlich}, \citenamefont {Pires}, \citenamefont {Manfredi}, \citenamefont {Hervieux},\ and\ \citenamefont {Genet}}]{goerlich2022harvesting}%
  \BibitemOpen
  \bibfield  {author} {\bibinfo {author} {\bibfnamefont {R.}~\bibnamefont {Goerlich}}, \bibinfo {author} {\bibfnamefont {L.~B.}\ \bibnamefont {Pires}}, \bibinfo {author} {\bibfnamefont {G.}~\bibnamefont {Manfredi}}, \bibinfo {author} {\bibfnamefont {P.-A.}\ \bibnamefont {Hervieux}},\ and\ \bibinfo {author} {\bibfnamefont {C.}~\bibnamefont {Genet}},\ }\href@noop {} {\bibfield  {journal} {\bibinfo  {journal} {Phys. Rev. E}\ }\textbf {\bibinfo {volume} {106}},\ \bibinfo {pages} {054617} (\bibinfo {year} {2022})}\BibitemShut {NoStop}%
\bibitem [{\citenamefont {Bia{\l}as}\ \emph {et~al.}(2023{\natexlab{a}})\citenamefont {Bia{\l}as}, \citenamefont {{\L}uczka},\ and\ \citenamefont {Spiechowicz}}]{bialas2023periodic}%
  \BibitemOpen
  \bibfield  {author} {\bibinfo {author} {\bibfnamefont {K.}~\bibnamefont {Bia{\l}as}}, \bibinfo {author} {\bibfnamefont {J.}~\bibnamefont {{\L}uczka}},\ and\ \bibinfo {author} {\bibfnamefont {J.}~\bibnamefont {Spiechowicz}},\ }\href@noop {} {\bibfield  {journal} {\bibinfo  {journal} {Phys. Rev. E}\ }\textbf {\bibinfo {volume} {107}},\ \bibinfo {pages} {024107} (\bibinfo {year} {2023}{\natexlab{a}})}\BibitemShut {NoStop}%
\bibitem [{\citenamefont {Bia{\l}as}\ and\ \citenamefont {Spiechowicz}(2023)}]{bialas2023mechanism}%
  \BibitemOpen
  \bibfield  {author} {\bibinfo {author} {\bibfnamefont {K.}~\bibnamefont {Bia{\l}as}}\ and\ \bibinfo {author} {\bibfnamefont {J.}~\bibnamefont {Spiechowicz}},\ }\href@noop {} {\bibfield  {journal} {\bibinfo  {journal} {Phys. Rev. E}\ }\textbf {\bibinfo {volume} {107}},\ \bibinfo {pages} {064120} (\bibinfo {year} {2023})}\BibitemShut {NoStop}%
\bibitem [{\citenamefont {Das}\ \emph {et~al.}(2023)\citenamefont {Das}, \citenamefont {Paul}, \citenamefont {Manikandan},\ and\ \citenamefont {Banerjee}}]{das2023enhanced}%
  \BibitemOpen
  \bibfield  {author} {\bibinfo {author} {\bibfnamefont {B.}~\bibnamefont {Das}}, \bibinfo {author} {\bibfnamefont {S.}~\bibnamefont {Paul}}, \bibinfo {author} {\bibfnamefont {S.~K.}\ \bibnamefont {Manikandan}},\ and\ \bibinfo {author} {\bibfnamefont {A.}~\bibnamefont {Banerjee}},\ }\href@noop {} {\bibfield  {journal} {\bibinfo  {journal} {New J. Phys.}\ }\textbf {\bibinfo {volume} {25}},\ \bibinfo {pages} {093051} (\bibinfo {year} {2023})}\BibitemShut {NoStop}%
\bibitem [{\citenamefont {Chaki}\ and\ \citenamefont {Chakrabarti}(2019)}]{chaki2019enhanced}%
  \BibitemOpen
  \bibfield  {author} {\bibinfo {author} {\bibfnamefont {S.}~\bibnamefont {Chaki}}\ and\ \bibinfo {author} {\bibfnamefont {R.}~\bibnamefont {Chakrabarti}},\ }\href@noop {} {\bibfield  {journal} {\bibinfo  {journal} {J. Chem. Phys.}\ }\textbf {\bibinfo {volume} {150}} (\bibinfo {year} {2019})}\BibitemShut {NoStop}%
\bibitem [{\citenamefont {Bia{\l}as}\ \emph {et~al.}(2023{\natexlab{b}})\citenamefont {Bia{\l}as}, \citenamefont {{\L}uczka},\ and\ \citenamefont {Spiechowicz}}]{bialas2023control}%
  \BibitemOpen
  \bibfield  {author} {\bibinfo {author} {\bibfnamefont {K.}~\bibnamefont {Bia{\l}as}}, \bibinfo {author} {\bibfnamefont {J.}~\bibnamefont {{\L}uczka}},\ and\ \bibinfo {author} {\bibfnamefont {J.}~\bibnamefont {Spiechowicz}},\ }\href@noop {} {\bibfield  {journal} {\bibinfo  {journal} {The European Physical Journal Special Topics}\ }\textbf {\bibinfo {volume} {232}},\ \bibinfo {pages} {3191} (\bibinfo {year} {2023}{\natexlab{b}})}\BibitemShut {NoStop}%
\bibitem [{\citenamefont {Ariga}(2024)}]{ariga2024nonthermal}%
  \BibitemOpen
  \bibfield  {author} {\bibinfo {author} {\bibfnamefont {T.}~\bibnamefont {Ariga}},\ }\href@noop {} {\bibfield  {journal} {\bibinfo  {journal} {Biophys. Rev.}\ ,\ \bibinfo {pages} {1}} (\bibinfo {year} {2024})}\BibitemShut {NoStop}%
\bibitem [{\citenamefont {Joo}\ \emph {et~al.}(2020)\citenamefont {Joo}, \citenamefont {Durang}, \citenamefont {Lee},\ and\ \citenamefont {Jeon}}]{joo2020anomalous}%
  \BibitemOpen
  \bibfield  {author} {\bibinfo {author} {\bibfnamefont {S.}~\bibnamefont {Joo}}, \bibinfo {author} {\bibfnamefont {X.}~\bibnamefont {Durang}}, \bibinfo {author} {\bibfnamefont {O.-c.}\ \bibnamefont {Lee}},\ and\ \bibinfo {author} {\bibfnamefont {J.-H.}\ \bibnamefont {Jeon}},\ }\href@noop {} {\bibfield  {journal} {\bibinfo  {journal} {Soft Matter}\ }\textbf {\bibinfo {volume} {16}},\ \bibinfo {pages} {9188} (\bibinfo {year} {2020})}\BibitemShut {NoStop}%
\bibitem [{\citenamefont {Mandal}\ \emph {et~al.}(2017)\citenamefont {Mandal}, \citenamefont {Klymko},\ and\ \citenamefont {DeWeese}}]{mandal2017entropy}%
  \BibitemOpen
  \bibfield  {author} {\bibinfo {author} {\bibfnamefont {D.}~\bibnamefont {Mandal}}, \bibinfo {author} {\bibfnamefont {K.}~\bibnamefont {Klymko}},\ and\ \bibinfo {author} {\bibfnamefont {M.~R.}\ \bibnamefont {DeWeese}},\ }\href@noop {} {\bibfield  {journal} {\bibinfo  {journal} {Physical Review Letters}\ }\textbf {\bibinfo {volume} {119}},\ \bibinfo {pages} {258001} (\bibinfo {year} {2017})}\BibitemShut {NoStop}%
\bibitem [{\citenamefont {Bressloff}(2025)}]{bressloff2025stochastic}%
  \BibitemOpen
  \bibfield  {author} {\bibinfo {author} {\bibfnamefont {P.~C.}\ \bibnamefont {Bressloff}},\ }\href@noop {} {\bibfield  {journal} {\bibinfo  {journal} {Proceedings of the Royal Society A}\ }\textbf {\bibinfo {volume} {481}},\ \bibinfo {pages} {20240815} (\bibinfo {year} {2025})}\BibitemShut {NoStop}%
\bibitem [{\citenamefont {O’Byrne}\ \emph {et~al.}(2022)\citenamefont {O’Byrne}, \citenamefont {Kafri}, \citenamefont {Tailleur},\ and\ \citenamefont {van Wijland}}]{o2022time}%
  \BibitemOpen
  \bibfield  {author} {\bibinfo {author} {\bibfnamefont {J.}~\bibnamefont {O’Byrne}}, \bibinfo {author} {\bibfnamefont {Y.}~\bibnamefont {Kafri}}, \bibinfo {author} {\bibfnamefont {J.}~\bibnamefont {Tailleur}},\ and\ \bibinfo {author} {\bibfnamefont {F.}~\bibnamefont {van Wijland}},\ }\href@noop {} {\bibfield  {journal} {\bibinfo  {journal} {Nature Reviews Physics}\ }\textbf {\bibinfo {volume} {4}},\ \bibinfo {pages} {167} (\bibinfo {year} {2022})}\BibitemShut {NoStop}%
\bibitem [{\citenamefont {Fodor}\ \emph {et~al.}(2022)\citenamefont {Fodor}, \citenamefont {Jack},\ and\ \citenamefont {Cates}}]{fodor2022irreversibility}%
  \BibitemOpen
  \bibfield  {author} {\bibinfo {author} {\bibfnamefont {{\'E}.}~\bibnamefont {Fodor}}, \bibinfo {author} {\bibfnamefont {R.~L.}\ \bibnamefont {Jack}},\ and\ \bibinfo {author} {\bibfnamefont {M.~E.}\ \bibnamefont {Cates}},\ }\href@noop {} {\bibfield  {journal} {\bibinfo  {journal} {Annual Review of Condensed Matter Physics}\ }\textbf {\bibinfo {volume} {13}},\ \bibinfo {pages} {215} (\bibinfo {year} {2022})}\BibitemShut {NoStop}%
\bibitem [{\citenamefont {Dabelow}\ \emph {et~al.}(2019)\citenamefont {Dabelow}, \citenamefont {Bo},\ and\ \citenamefont {Eichhorn}}]{dabelow2019irreversibility}%
  \BibitemOpen
  \bibfield  {author} {\bibinfo {author} {\bibfnamefont {L.}~\bibnamefont {Dabelow}}, \bibinfo {author} {\bibfnamefont {S.}~\bibnamefont {Bo}},\ and\ \bibinfo {author} {\bibfnamefont {R.}~\bibnamefont {Eichhorn}},\ }\href@noop {} {\bibfield  {journal} {\bibinfo  {journal} {Physical Review X}\ }\textbf {\bibinfo {volume} {9}},\ \bibinfo {pages} {021009} (\bibinfo {year} {2019})}\BibitemShut {NoStop}%
\bibitem [{\citenamefont {Andrieux}\ and\ \citenamefont {Gaspard}(2008)}]{andrieux2008temporal}%
  \BibitemOpen
  \bibfield  {author} {\bibinfo {author} {\bibfnamefont {D.}~\bibnamefont {Andrieux}}\ and\ \bibinfo {author} {\bibfnamefont {P.}~\bibnamefont {Gaspard}},\ }\href@noop {} {\bibfield  {journal} {\bibinfo  {journal} {Physical Review E—Statistical, Nonlinear, and Soft Matter Physics}\ }\textbf {\bibinfo {volume} {77}},\ \bibinfo {pages} {031137} (\bibinfo {year} {2008})}\BibitemShut {NoStop}%
\bibitem [{\citenamefont {Polettini}\ and\ \citenamefont {Esposito}(2014)}]{polettini2014transient}%
  \BibitemOpen
  \bibfield  {author} {\bibinfo {author} {\bibfnamefont {M.}~\bibnamefont {Polettini}}\ and\ \bibinfo {author} {\bibfnamefont {M.}~\bibnamefont {Esposito}},\ }\href@noop {} {\bibfield  {journal} {\bibinfo  {journal} {Journal of Statistical Mechanics: Theory and Experiment}\ }\textbf {\bibinfo {volume} {2014}},\ \bibinfo {pages} {P10033} (\bibinfo {year} {2014})}\BibitemShut {NoStop}%
\bibitem [{\citenamefont {Kanazawa}\ \emph {et~al.}(2013)\citenamefont {Kanazawa}, \citenamefont {Sagawa},\ and\ \citenamefont {Hayakawa}}]{kanazawa2013heat}%
  \BibitemOpen
  \bibfield  {author} {\bibinfo {author} {\bibfnamefont {K.}~\bibnamefont {Kanazawa}}, \bibinfo {author} {\bibfnamefont {T.}~\bibnamefont {Sagawa}},\ and\ \bibinfo {author} {\bibfnamefont {H.}~\bibnamefont {Hayakawa}},\ }\href@noop {} {\bibfield  {journal} {\bibinfo  {journal} {Phys. Rev. E}\ }\textbf {\bibinfo {volume} {87}},\ \bibinfo {pages} {052124} (\bibinfo {year} {2013})}\BibitemShut {NoStop}%
\bibitem [{\citenamefont {Kanazawa}\ \emph {et~al.}(2015{\natexlab{a}})\citenamefont {Kanazawa}, \citenamefont {Sano}, \citenamefont {Sagawa},\ and\ \citenamefont {Hayakawa}}]{kanazawa2015minimal}%
  \BibitemOpen
  \bibfield  {author} {\bibinfo {author} {\bibfnamefont {K.}~\bibnamefont {Kanazawa}}, \bibinfo {author} {\bibfnamefont {T.~G.}\ \bibnamefont {Sano}}, \bibinfo {author} {\bibfnamefont {T.}~\bibnamefont {Sagawa}},\ and\ \bibinfo {author} {\bibfnamefont {H.}~\bibnamefont {Hayakawa}},\ }\href@noop {} {\bibfield  {journal} {\bibinfo  {journal} {Phys. Rev. Lett.}\ }\textbf {\bibinfo {volume} {114}},\ \bibinfo {pages} {090601} (\bibinfo {year} {2015}{\natexlab{a}})}\BibitemShut {NoStop}%
\bibitem [{\citenamefont {Kanazawa}\ \emph {et~al.}(2015{\natexlab{b}})\citenamefont {Kanazawa}, \citenamefont {Sano}, \citenamefont {Sagawa},\ and\ \citenamefont {Hayakawa}}]{kanazawa2015asymptotic}%
  \BibitemOpen
  \bibfield  {author} {\bibinfo {author} {\bibfnamefont {K.}~\bibnamefont {Kanazawa}}, \bibinfo {author} {\bibfnamefont {T.~G.}\ \bibnamefont {Sano}}, \bibinfo {author} {\bibfnamefont {T.}~\bibnamefont {Sagawa}},\ and\ \bibinfo {author} {\bibfnamefont {H.}~\bibnamefont {Hayakawa}},\ }\href@noop {} {\bibfield  {journal} {\bibinfo  {journal} {J. Stat. Phys.}\ }\textbf {\bibinfo {volume} {160}},\ \bibinfo {pages} {1294} (\bibinfo {year} {2015}{\natexlab{b}})}\BibitemShut {NoStop}%
\bibitem [{\citenamefont {Kanazawa}(2017)}]{kanazawa2017statistical}%
  \BibitemOpen
  \bibfield  {author} {\bibinfo {author} {\bibfnamefont {K.}~\bibnamefont {Kanazawa}},\ }\href@noop {} {\emph {\bibinfo {title} {Statistical mechanics for athermal fluctuation: Non-Gaussian noise in physics}}}\ (\bibinfo  {publisher} {Springer},\ \bibinfo {year} {2017})\BibitemShut {NoStop}%
\bibitem [{\citenamefont {Touchette}\ and\ \citenamefont {Cohen}(2007)}]{touchette2007fluctuation}%
  \BibitemOpen
  \bibfield  {author} {\bibinfo {author} {\bibfnamefont {H.}~\bibnamefont {Touchette}}\ and\ \bibinfo {author} {\bibfnamefont {E.}~\bibnamefont {Cohen}},\ }\href@noop {} {\bibfield  {journal} {\bibinfo  {journal} {Phys. Rev. E}\ }\textbf {\bibinfo {volume} {76}},\ \bibinfo {pages} {020101} (\bibinfo {year} {2007})}\BibitemShut {NoStop}%
\bibitem [{\citenamefont {Touchette}\ and\ \citenamefont {Cohen}(2009)}]{touchette2009anomalous}%
  \BibitemOpen
  \bibfield  {author} {\bibinfo {author} {\bibfnamefont {H.}~\bibnamefont {Touchette}}\ and\ \bibinfo {author} {\bibfnamefont {E.}~\bibnamefont {Cohen}},\ }\href@noop {} {\bibfield  {journal} {\bibinfo  {journal} {Phys. Rev. E}\ }\textbf {\bibinfo {volume} {80}},\ \bibinfo {pages} {011114} (\bibinfo {year} {2009})}\BibitemShut {NoStop}%
\bibitem [{\citenamefont {Baule}\ and\ \citenamefont {Cohen}(2009)}]{baule2009fluctuation}%
  \BibitemOpen
  \bibfield  {author} {\bibinfo {author} {\bibfnamefont {A.}~\bibnamefont {Baule}}\ and\ \bibinfo {author} {\bibfnamefont {E.}~\bibnamefont {Cohen}},\ }\href@noop {} {\bibfield  {journal} {\bibinfo  {journal} {Phys. Rev. E}\ }\textbf {\bibinfo {volume} {79}},\ \bibinfo {pages} {030103} (\bibinfo {year} {2009})}\BibitemShut {NoStop}%
\bibitem [{\citenamefont {Budini}(2012)}]{budini2012generalized}%
  \BibitemOpen
  \bibfield  {author} {\bibinfo {author} {\bibfnamefont {A.~A.}\ \bibnamefont {Budini}},\ }\href@noop {} {\bibfield  {journal} {\bibinfo  {journal} {Phys. Rev. E}\ }\textbf {\bibinfo {volume} {86}},\ \bibinfo {pages} {011109} (\bibinfo {year} {2012})}\BibitemShut {NoStop}%
\bibitem [{\citenamefont {Lucente}\ \emph {et~al.}(2023)\citenamefont {Lucente}, \citenamefont {Puglisi}, \citenamefont {Viale},\ and\ \citenamefont {Vulpiani}}]{lucente2023statistical}%
  \BibitemOpen
  \bibfield  {author} {\bibinfo {author} {\bibfnamefont {D.}~\bibnamefont {Lucente}}, \bibinfo {author} {\bibfnamefont {A.}~\bibnamefont {Puglisi}}, \bibinfo {author} {\bibfnamefont {M.}~\bibnamefont {Viale}},\ and\ \bibinfo {author} {\bibfnamefont {A.}~\bibnamefont {Vulpiani}},\ }\href@noop {} {\bibfield  {journal} {\bibinfo  {journal} {J. Stat. Mech.-Theory Exp.}\ }\textbf {\bibinfo {volume} {2023}},\ \bibinfo {pages} {113202} (\bibinfo {year} {2023})}\BibitemShut {NoStop}%
\bibitem [{\citenamefont {Faria}\ \emph {et~al.}(2025)\citenamefont {Faria}, \citenamefont {Bonanca},\ and\ \citenamefont {Lutz}}]{faria2025nonequilibrium}%
  \BibitemOpen
  \bibfield  {author} {\bibinfo {author} {\bibfnamefont {A.~M.}\ \bibnamefont {Faria}}, \bibinfo {author} {\bibfnamefont {M.~V.}\ \bibnamefont {Bonanca}},\ and\ \bibinfo {author} {\bibfnamefont {E.}~\bibnamefont {Lutz}},\ }\href@noop {} {\bibfield  {journal} {\bibinfo  {journal} {arXiv preprint arXiv:2502.20179}\ } (\bibinfo {year} {2025})}\BibitemShut {NoStop}%
\bibitem [{\citenamefont {Huang}\ \emph {et~al.}(2025)\citenamefont {Huang}, \citenamefont {Zhou},\ and\ \citenamefont {Duan}}]{huang2024probability}%
  \BibitemOpen
  \bibfield  {author} {\bibinfo {author} {\bibfnamefont {Y.}~\bibnamefont {Huang}}, \bibinfo {author} {\bibfnamefont {X.}~\bibnamefont {Zhou}},\ and\ \bibinfo {author} {\bibfnamefont {J.}~\bibnamefont {Duan}},\ }\href@noop {} {\bibfield  {journal} {\bibinfo  {journal} {SIAM J. Appl. Math.}\ }\textbf {\bibinfo {volume} {85}},\ \bibinfo {pages} {524} (\bibinfo {year} {2025})}\BibitemShut {NoStop}%
\bibitem [{\citenamefont {Huang}\ \emph {et~al.}(2024)\citenamefont {Huang}, \citenamefont {Liu},\ and\ \citenamefont {Zhou}}]{huang2024vy}%
  \BibitemOpen
  \bibfield  {author} {\bibinfo {author} {\bibfnamefont {Y.}~\bibnamefont {Huang}}, \bibinfo {author} {\bibfnamefont {C.}~\bibnamefont {Liu}},\ and\ \bibinfo {author} {\bibfnamefont {X.}~\bibnamefont {Zhou}},\ }\href@noop {} {\bibfield  {journal} {\bibinfo  {journal} {arXiv preprint arXiv:2412.19520}\ } (\bibinfo {year} {2024})}\BibitemShut {NoStop}%
\bibitem [{\citenamefont {Sagawa}\ and\ \citenamefont {Ueda}(2012)}]{sagawa2012fluctuation}%
  \BibitemOpen
  \bibfield  {author} {\bibinfo {author} {\bibfnamefont {T.}~\bibnamefont {Sagawa}}\ and\ \bibinfo {author} {\bibfnamefont {M.}~\bibnamefont {Ueda}},\ }\href@noop {} {\bibfield  {journal} {\bibinfo  {journal} {Phys. Rev. Lett.}\ }\textbf {\bibinfo {volume} {109}},\ \bibinfo {pages} {180602} (\bibinfo {year} {2012})}\BibitemShut {NoStop}%
\bibitem [{\citenamefont {Seifert}(2005)}]{seifert2005entropy}%
  \BibitemOpen
  \bibfield  {author} {\bibinfo {author} {\bibfnamefont {U.}~\bibnamefont {Seifert}},\ }\href@noop {} {\bibfield  {journal} {\bibinfo  {journal} {Phys. Rev. Lett.}\ }\textbf {\bibinfo {volume} {95}},\ \bibinfo {pages} {040602} (\bibinfo {year} {2005})}\BibitemShut {NoStop}%
\bibitem [{\citenamefont {Boffi}\ and\ \citenamefont {Vanden-Eijnden}(2024)}]{boffi2024deep}%
  \BibitemOpen
  \bibfield  {author} {\bibinfo {author} {\bibfnamefont {N.~M.}\ \bibnamefont {Boffi}}\ and\ \bibinfo {author} {\bibfnamefont {E.}~\bibnamefont {Vanden-Eijnden}},\ }\href@noop {} {\bibfield  {journal} {\bibinfo  {journal} {Proc. Natl. Acad. Sci.}\ }\textbf {\bibinfo {volume} {121}},\ \bibinfo {pages} {e2318106121} (\bibinfo {year} {2024})}\BibitemShut {NoStop}%
\bibitem [{\citenamefont {Kim}\ \emph {et~al.}(2020)\citenamefont {Kim}, \citenamefont {Bae}, \citenamefont {Lee},\ and\ \citenamefont {Jeong}}]{kim2020learning}%
  \BibitemOpen
  \bibfield  {author} {\bibinfo {author} {\bibfnamefont {D.-K.}\ \bibnamefont {Kim}}, \bibinfo {author} {\bibfnamefont {Y.}~\bibnamefont {Bae}}, \bibinfo {author} {\bibfnamefont {S.}~\bibnamefont {Lee}},\ and\ \bibinfo {author} {\bibfnamefont {H.}~\bibnamefont {Jeong}},\ }\href@noop {} {\bibfield  {journal} {\bibinfo  {journal} {Physical Review Letters}\ }\textbf {\bibinfo {volume} {125}},\ \bibinfo {pages} {140604} (\bibinfo {year} {2020})}\BibitemShut {NoStop}%
\bibitem [{\citenamefont {Otsubo}\ \emph {et~al.}(2020)\citenamefont {Otsubo}, \citenamefont {Ito}, \citenamefont {Dechant},\ and\ \citenamefont {Sagawa}}]{otsubo2020estimating}%
  \BibitemOpen
  \bibfield  {author} {\bibinfo {author} {\bibfnamefont {S.}~\bibnamefont {Otsubo}}, \bibinfo {author} {\bibfnamefont {S.}~\bibnamefont {Ito}}, \bibinfo {author} {\bibfnamefont {A.}~\bibnamefont {Dechant}},\ and\ \bibinfo {author} {\bibfnamefont {T.}~\bibnamefont {Sagawa}},\ }\href@noop {} {\bibfield  {journal} {\bibinfo  {journal} {Physical Review E}\ }\textbf {\bibinfo {volume} {101}},\ \bibinfo {pages} {062106} (\bibinfo {year} {2020})}\BibitemShut {NoStop}%
\bibitem [{\citenamefont {Ro}\ \emph {et~al.}(2022)\citenamefont {Ro}, \citenamefont {Guo}, \citenamefont {Shih}, \citenamefont {Phan}, \citenamefont {Austin}, \citenamefont {Levine}, \citenamefont {Chaikin},\ and\ \citenamefont {Martiniani}}]{ro2022model}%
  \BibitemOpen
  \bibfield  {author} {\bibinfo {author} {\bibfnamefont {S.}~\bibnamefont {Ro}}, \bibinfo {author} {\bibfnamefont {B.}~\bibnamefont {Guo}}, \bibinfo {author} {\bibfnamefont {A.}~\bibnamefont {Shih}}, \bibinfo {author} {\bibfnamefont {T.~V.}\ \bibnamefont {Phan}}, \bibinfo {author} {\bibfnamefont {R.~H.}\ \bibnamefont {Austin}}, \bibinfo {author} {\bibfnamefont {D.}~\bibnamefont {Levine}}, \bibinfo {author} {\bibfnamefont {P.~M.}\ \bibnamefont {Chaikin}},\ and\ \bibinfo {author} {\bibfnamefont {S.}~\bibnamefont {Martiniani}},\ }\href@noop {} {\bibfield  {journal} {\bibinfo  {journal} {Physical Review Letters}\ }\textbf {\bibinfo {volume} {129}},\ \bibinfo {pages} {220601} (\bibinfo {year} {2022})}\BibitemShut {NoStop}%
\bibitem [{\citenamefont {Crooks}(1999)}]{crooks1999entropy}%
  \BibitemOpen
  \bibfield  {author} {\bibinfo {author} {\bibfnamefont {G.~E.}\ \bibnamefont {Crooks}},\ }\href@noop {} {\bibfield  {journal} {\bibinfo  {journal} {Phys. Rev. E}\ }\textbf {\bibinfo {volume} {60}},\ \bibinfo {pages} {2721} (\bibinfo {year} {1999})}\BibitemShut {NoStop}%
\bibitem [{\citenamefont {Qian}(2001)}]{qian2001mesoscopic}%
  \BibitemOpen
  \bibfield  {author} {\bibinfo {author} {\bibfnamefont {H.}~\bibnamefont {Qian}},\ }\href@noop {} {\bibfield  {journal} {\bibinfo  {journal} {Phys. Rev. E}\ }\textbf {\bibinfo {volume} {65}},\ \bibinfo {pages} {016102} (\bibinfo {year} {2001})}\BibitemShut {NoStop}%
\bibitem [{\citenamefont {Applebaum}(2009)}]{applebaum2009levy}%
  \BibitemOpen
  \bibfield  {author} {\bibinfo {author} {\bibfnamefont {D.}~\bibnamefont {Applebaum}},\ }\href@noop {} {\emph {\bibinfo {title} {L\'{e}vy Processes and Stochastic Calculus}}},\ \bibinfo {edition} {2nd}\ ed.,\ \bibinfo {series} {Cambridge Studies in Advanced Mathematics}, Vol.~\bibinfo {volume} {93}\ (\bibinfo  {publisher} {Cambridge University Press},\ \bibinfo {year} {2009})\BibitemShut {NoStop}%
\bibitem [{Sto()}]{StoIntNote}%
  \BibitemOpen
  \href@noop {} {}\bibinfo {note} {The Di Paola–Falsone calculus \cite{di1993stochastic,di1993ito} and the $(*)$-calculus \cite{kanazawa2012stochastic,fodor2018non} represent two additional stochastic calculi that preserve the chain rule for systems with jump noise. Although their definitions take different forms, both have been shown to be mathematically equivalent to the Marcus integral \cite{li2013marcus,falsone2018stochastic}.}\BibitemShut {Stop}%
\bibitem [{sup()}]{supplemental_material}%
  \BibitemOpen
  \href@noop {} {}\bibinfo {note} {See Supplemental Material for full derivations of the main results, details on the numerical case studies.}\BibitemShut {Stop}%
\bibitem [{\citenamefont {Privault}\ and\ \citenamefont {Zambrini}(2004)}]{privault2004markovian}%
  \BibitemOpen
  \bibfield  {author} {\bibinfo {author} {\bibfnamefont {N.}~\bibnamefont {Privault}}\ and\ \bibinfo {author} {\bibfnamefont {J.-C.}\ \bibnamefont {Zambrini}},\ }\href@noop {} {\bibfield  {journal} {\bibinfo  {journal} {Annales de l'IHP Probabilit{\'e}s et statistiques}\ }\textbf {\bibinfo {volume} {40}},\ \bibinfo {pages} {599} (\bibinfo {year} {2004})}\BibitemShut {NoStop}%
\bibitem [{\citenamefont {Conforti}\ and\ \citenamefont {L{\'e}onard}(2022)}]{conforti2022time}%
  \BibitemOpen
  \bibfield  {author} {\bibinfo {author} {\bibfnamefont {G.}~\bibnamefont {Conforti}}\ and\ \bibinfo {author} {\bibfnamefont {C.}~\bibnamefont {L{\'e}onard}},\ }\href@noop {} {\bibfield  {journal} {\bibinfo  {journal} {Stoch. Process. Their Appl.}\ }\textbf {\bibinfo {volume} {144}},\ \bibinfo {pages} {85} (\bibinfo {year} {2022})}\BibitemShut {NoStop}%
\bibitem [{\citenamefont {Baule}\ and\ \citenamefont {Sollich}(2023)}]{baule2023exponential}%
  \BibitemOpen
  \bibfield  {author} {\bibinfo {author} {\bibfnamefont {A.}~\bibnamefont {Baule}}\ and\ \bibinfo {author} {\bibfnamefont {P.}~\bibnamefont {Sollich}},\ }\href@noop {} {\bibfield  {journal} {\bibinfo  {journal} {Sci Rep}\ }\textbf {\bibinfo {volume} {13}},\ \bibinfo {pages} {3853} (\bibinfo {year} {2023})}\BibitemShut {NoStop}%
\bibitem [{cod()}]{code}%
  \BibitemOpen
  \href@noop {} {\bibinfo {title} {\url{https://github.com/cliu687/sbtm-levy}}}\BibitemShut {NoStop}%
\bibitem [{\citenamefont {Fodor}\ \emph {et~al.}(2016)\citenamefont {Fodor}, \citenamefont {Nardini}, \citenamefont {Cates}, \citenamefont {Tailleur}, \citenamefont {Visco},\ and\ \citenamefont {Van~Wijland}}]{fodor2016far}%
  \BibitemOpen
  \bibfield  {author} {\bibinfo {author} {\bibfnamefont {{\'E}.}~\bibnamefont {Fodor}}, \bibinfo {author} {\bibfnamefont {C.}~\bibnamefont {Nardini}}, \bibinfo {author} {\bibfnamefont {M.~E.}\ \bibnamefont {Cates}}, \bibinfo {author} {\bibfnamefont {J.}~\bibnamefont {Tailleur}}, \bibinfo {author} {\bibfnamefont {P.}~\bibnamefont {Visco}},\ and\ \bibinfo {author} {\bibfnamefont {F.}~\bibnamefont {Van~Wijland}},\ }\href@noop {} {\bibfield  {journal} {\bibinfo  {journal} {Physical Review Letters}\ }\textbf {\bibinfo {volume} {117}},\ \bibinfo {pages} {038103} (\bibinfo {year} {2016})}\BibitemShut {NoStop}%
\bibitem [{\citenamefont {Di~Paola}\ and\ \citenamefont {Falsone}(1993{\natexlab{a}})}]{di1993stochastic}%
  \BibitemOpen
  \bibfield  {author} {\bibinfo {author} {\bibfnamefont {M.}~\bibnamefont {Di~Paola}}\ and\ \bibinfo {author} {\bibfnamefont {G.}~\bibnamefont {Falsone}},\ }\href@noop {} {\bibfield  {journal} {\bibinfo  {journal} {J. Appl. Mech.}\ }\textbf {\bibinfo {volume} {60}},\ \bibinfo {pages} {141} (\bibinfo {year} {1993}{\natexlab{a}})}\BibitemShut {NoStop}%
\bibitem [{\citenamefont {Di~Paola}\ and\ \citenamefont {Falsone}(1993{\natexlab{b}})}]{di1993ito}%
  \BibitemOpen
  \bibfield  {author} {\bibinfo {author} {\bibfnamefont {M.}~\bibnamefont {Di~Paola}}\ and\ \bibinfo {author} {\bibfnamefont {G.}~\bibnamefont {Falsone}},\ }\href@noop {} {\bibfield  {journal} {\bibinfo  {journal} {Probabilistic Engineering Mechanics}\ }\textbf {\bibinfo {volume} {8}},\ \bibinfo {pages} {197} (\bibinfo {year} {1993}{\natexlab{b}})}\BibitemShut {NoStop}%
\bibitem [{\citenamefont {Kanazawa}\ \emph {et~al.}(2012)\citenamefont {Kanazawa}, \citenamefont {Sagawa},\ and\ \citenamefont {Hayakawa}}]{kanazawa2012stochastic}%
  \BibitemOpen
  \bibfield  {author} {\bibinfo {author} {\bibfnamefont {K.}~\bibnamefont {Kanazawa}}, \bibinfo {author} {\bibfnamefont {T.}~\bibnamefont {Sagawa}},\ and\ \bibinfo {author} {\bibfnamefont {H.}~\bibnamefont {Hayakawa}},\ }\href@noop {} {\bibfield  {journal} {\bibinfo  {journal} {Physical Review Letters}\ }\textbf {\bibinfo {volume} {108}},\ \bibinfo {pages} {210601} (\bibinfo {year} {2012})}\BibitemShut {NoStop}%
\bibitem [{\citenamefont {Fodor}\ \emph {et~al.}(2018)\citenamefont {Fodor}, \citenamefont {Hayakawa}, \citenamefont {Tailleur},\ and\ \citenamefont {van Wijland}}]{fodor2018non}%
  \BibitemOpen
  \bibfield  {author} {\bibinfo {author} {\bibfnamefont {{\'E}.}~\bibnamefont {Fodor}}, \bibinfo {author} {\bibfnamefont {H.}~\bibnamefont {Hayakawa}}, \bibinfo {author} {\bibfnamefont {J.}~\bibnamefont {Tailleur}},\ and\ \bibinfo {author} {\bibfnamefont {F.}~\bibnamefont {van Wijland}},\ }\href@noop {} {\bibfield  {journal} {\bibinfo  {journal} {Physical Review E}\ }\textbf {\bibinfo {volume} {98}},\ \bibinfo {pages} {062610} (\bibinfo {year} {2018})}\BibitemShut {NoStop}%
\bibitem [{\citenamefont {Li}\ \emph {et~al.}(2013)\citenamefont {Li}, \citenamefont {Min},\ and\ \citenamefont {Wang}}]{li2013marcus}%
  \BibitemOpen
  \bibfield  {author} {\bibinfo {author} {\bibfnamefont {T.}~\bibnamefont {Li}}, \bibinfo {author} {\bibfnamefont {B.}~\bibnamefont {Min}},\ and\ \bibinfo {author} {\bibfnamefont {Z.}~\bibnamefont {Wang}},\ }\href@noop {} {\bibfield  {journal} {\bibinfo  {journal} {The Journal of Chemical Physics}\ }\textbf {\bibinfo {volume} {138}},\ \bibinfo {pages} {104118} (\bibinfo {year} {2013})}\BibitemShut {NoStop}%
\bibitem [{\citenamefont {Falsone}(2018)}]{falsone2018stochastic}%
  \BibitemOpen
  \bibfield  {author} {\bibinfo {author} {\bibfnamefont {G.}~\bibnamefont {Falsone}},\ }\href@noop {} {\bibfield  {journal} {\bibinfo  {journal} {Communications in Nonlinear Science and Numerical Simulation}\ }\textbf {\bibinfo {volume} {56}},\ \bibinfo {pages} {198} (\bibinfo {year} {2018})}\BibitemShut {NoStop}%
\end{thebibliography}%


\begin{thebibliography}{1}

  \bibitem[S1]{app:applebaum2009levy}
  Applebaum, D. L\'{e}vy Processes and Stochastic Calculus. Cambridge Studies in Advanced Mathematics. 2009

  \bibitem[S2]{app:huang2024probability}
  Huang Y., Zhou X., Duan J.. Probability flow approach to the Onsager–Machlup functional for jump-diffusion processes. SIAM Journal on Applied Mathematics 85 (2), 524-547, 2025.

  \bibitem[S3]{app:baule2025ge}
  Adrian Baule. Generative modelling with jump-diffusions. arXiv preprint 	arXiv:2503.06558, 2025.

  \bibitem[S4]{app:privault2004markovian}
  Privault, Nicolas and Zambrini, Jean-Claude. Markovian bridges and reversible diffusion processes with jumps. Annales de l'IHP Probabilit{\'e}s et statistiques. 40(5), 599--633, 2004.

  \bibitem[S5]{app:conforti2022time}
  Conforti, Giovanni and L{\'e}onard, Christian. Time reversal of {M}arkov processes with jumps under a finite entropy condition. Stochastic Processes and their Applications. 144, 85--124, 2022.

\bibitem[S6]{app:huang2025transition}
  Huang Y, Zhou X. Transition Path Theory For L\'{e}vy-Type Processes: SDE Representation and Statistics[J]. arXiv preprint arXiv:2506.09462, 2025.

  

  

  \bibitem[S7]
  {app:figueroa2018small}
  Figueroa-L{\'o}pez, Jos{\'e} E and Luo, Yankeng. Small-time expansions for state-dependent local jump--diffusion models with infinite jump activity. Stochastic Processes and their Applications. 128(12), 4207--4245, 2018.

  \bibitem[S8]{app:Hunt1981path}
  Hunt K L C, Ross J. Path integral solutions of stochastic equations for nonlinear irreversible processes: The uniqueness of the thermodynamic Lagrangian. The Journal of Chemical Physics,  75(2): 976-984, 1981.


  \bibitem[S9]{app:Ishikawa2023stochastic}
  Ishikawa Y. Stochastic calculus of variations: For jump processes. Walter de Gruyter GmbH \& Co KG, 2023.

\bibitem[S10]{app:Fujisaki2021generalized}
  Fujisaki M, Komatsu T. Generalized Girsanov transform of processes and Zakai equation with jumps. Journal of Stochastic Analysis, 2021, 2(3): 17.

\bibitem[S11]{app:zhang2025entropy}
  Zhang, Qi and Lu, Yubin. Entropy production rate and time-reversibility for general jump diffusions on $R^n$. Chaos: An Interdisciplinary Journal of Nonlinear Science, 35, 10, 2025.

\bibitem[S12]{app:seifert2005entropy}
  Seifert, Udo. Entropy production along a stochastic trajectory and an integral fluctuation theorem. Phys. Rev. Lett. 95, 040602 (2005)
\end{thebibliography}

\clearpage
\newpage

\begin{center}
   END MATTER
\end{center}
This End Matter serves as the appendices to the main text. Appendix~A provides a detailed discussion of the physical interpretation of the quantity $B_\mathrm{act}$ in Eq.~\eqref{FT}. Appendix~B presents the derivation of the DFT and includes the numerical verification for the first example discussed in the main text. Appendix~C provides supplementary description for the second example.

\setcounter{equation}{0}  
\renewcommand{\theequation}{A.\arabic{equation}}

{\itshape Appendix A: Physical interpretation of $B_\mathrm{act}$---}Let $\{\bm{x}(t)\}_{0\leq t\leq T}$ denote a trajectory of \eqref{eqn:SDE}.
By using Girsanov theorem, we find that (\cite[Eq.~\eqref{app:Bact=Q+P}]{supplemental_material}):
{\small
\begin{equation}\label{Endmatter:Bact}
    \begin{aligned}
        &B_{\mathrm{act}}(\{\bm{x}(t)\}_{0\leq t\leq T})\\
        =& \int_0^T\d t \frac{\bm{V}(\bm{x}(t), T-t)}{D_\mathrm{th}} \diamond \bigg( \bm{\eta}_\mathrm{act}(t)  + \bm{S}_\mathrm{L}(\bm{x}(t), T-t) \bigg) \\
        & + \log \frac{\tilde{P}\left[ \{\bm{x}(t)\}_{0\leq t\leq T} \mid \bm{x}_0\right]}{\tilde{P}^{\dagger}\left[ \{\bm{x}(t)\}_{0\leq t\leq T} \mid \bm{x}_0\right]} + \log\frac{ P_0(\bm{x}_0)}{ P_T(\bm{x}_0)}\\
        \equiv& \mathcal{Q}(\{\bm{x}(t)\}_{0\leq t\leq T}) + \mathcal{P}(\{\bm{x}(t)\}_{0\leq t\leq T})+ \log\frac{ P_0(\bm{x}_0)}{ P_T(\bm{x}_0)},
\end{aligned}
\end{equation}
}where $\bm{\eta}_\mathrm{act}(t)$ is the jump noise used to construct the trajectory $\{\bm{x}(t)\}_{0\leq t\leq T}$ using Eq. \eqref{eqn:SDE}. The path measure $\tilde{P}$ is the path law of the reversed process $\{\bm{\tilde{r}}(t)\}_{0\leq t\leq T}$ of \eqref{eqn:SDEreversed} in the main text, and the definition of the path measure $\tilde{P}^\dagger$ will be given in Appendix A.2. We will give more discussion on the quantities $\mathcal{Q}$ and $\mathcal{P}$ in Appendix A.1 and A.2, respectively. 

Generally speaking, the quantity $\mathcal{Q}$ represents the accumulation of the differences between the trajectory- and ensemble-level active heat dissipations along $\{\bm{x}(t)\}_{0\leq t\leq T}$. 
And the quantity 
$\mathcal{P}$ captures the non-Gaussian contribution to the 
irreversibility of $\bm{r}(t)$ in Eq.~\eqref{eqn:SDE}, namely the discrepancy 
between the reversed process $\bm{\tilde{r}}(t)$ in Eq.~\eqref{eqn:SDEreversed} 
and the auxiliary process $\bm{\tilde{r}}^{\dagger}(t)$ in 
Eq.~\eqref{appeqn:SDEreverse} which will be given later.  The final term, $\log\!\left[\frac{P_0(\bm{x}(0))}{P_T(\bm{x}(0))}\right]$, captures the contribution arising from the system not yet having relaxed to its stationary state. In particular, when $\bm{\eta}_\mathrm{act}=0$ in \eqref{eqn:SDE}, both $\mathcal{Q}$ and $\mathcal{P}$ vanish. 

In addition, a detailed but less intuitive expression of $B_\mathrm{act}$ is
(\cite[Eq.~\eqref{app:p/tildep2}]{supplemental_material}):
{\footnotesize\begin{equation}
    \begin{aligned}
        &B_\mathrm{act}(\{\bm{x}(t)\}_{0\leq t\leq T}) 
        =  \bm{S}_\mathrm{L}(\bm{x}(t),T-t)\diamond\int_0^T  \bigg(   \nabla\log P(\bm{x}(t),T-t)\d t  \\
        & \left. +  \frac{\int\bm{z}\mathcal{N}(\d t,\d\bm{z})}{ D_\mathrm{th}}  \right)  + \int_0^T\frac{\bm{V}(\bm{x}(t),T-t)}{D_\mathrm{th}} \diamond\int\bm{z}\mathcal{N}(\d t,\d\bm{z})\\
        & - \int_0^T\nabla\cdot\bm{S}_\mathrm{L}(\bm{x}(t),T-t) \d t   +   \int_0^T \frac{\bm{S}_\mathrm{L}(\bm{x}(t),T-t)}{ D_\mathrm{th}} \diamond  \d\bm{x}(t)\\
        & +  \int_0^T\int \left[\log\left( \frac{P(\bm{x}(t)+\bm{z},T-t)\nu(-\bm{z})}{P(\bm{x}(t),T-t)\nu(\bm{z})}\right)\mathcal{N}(\d t,\d\bm{z}) \right. \\
        & \left.  -\left(  \frac{P(\bm{x}(t)+\bm{z},T-t)\nu(-\bm{z})}{P(\bm{x}(t),T-t)\nu(\bm{z})}-1\right)\nu(\d \bm{z})\d t\right]+ \log\frac{ P_0(\bm{r}(0))}{ P_T(\bm{r}(0))},
    \end{aligned}
\end{equation}}where $\mathcal{N}(\d t,\d\bm{z})$ is a Poisson random measure with L\'{e}vy measure $\lambda_0\nu_A(\d\bm{z})\d t$. 


\setcounter{equation}{0}  
\renewcommand{\theequation}{A.1.\arabic{equation}}
{\itshape Appendix A.1: Active heat dissipation---}The heat exchange rate between $\bm{r}(t)$ of \eqref{eqn:SDE} and the medium at position $\bm{r}$ and time $t$ is:
{\footnotesize\begin{equation}
    \begin{aligned}
        \dot{Q}(\bm{r},t) = \bm{V}(\bm{r},t)\cdot \frac{\bm{F}(\bm{r})}{k_B\mathcal{T}}.
    \end{aligned}
\end{equation}
}Based on the definition of $\bm{V}$ at different times $t$ and $T-t$, we know that 
{\footnotesize\begin{equation}
    \begin{aligned}
        \frac{\bm{F}(\bm{r})}{k_B\mathcal{T}}=& \frac{\bm{V}(\bm{r},t)}{D_\mathrm{th}}+\nabla\log P(\bm{r},t) + \frac{\bm{S}_\mathrm{L}(\bm{r},t)}{D_\mathrm{th}}\\
        =& \frac{\bm{V}(\bm{r},T-t)}{D_\mathrm{th}}+\nabla\log P(\bm{r},T-t) + \frac{\bm{S}_\mathrm{L}(\bm{r},T-t)}{D_\mathrm{th}}.
    \end{aligned}
\end{equation}
}Thus, the heat dissipation rate in medium is 
{\footnotesize\begin{align}
        \dot{Q}(\bm{r},t)
        =&  \left(\frac{\bm{V}(\bm{r},t)}{D_\mathrm{th}} +  \nabla\log P(\bm{r}, t) \right)\cdot\bm{V}(\bm{r}, t) \label{app:activeheat1V}\\
        & +  \frac{ \bm{S}_\mathrm{L}(\bm{r},t) \cdot \bm{V}(\bm{r},t) }{D_\mathrm{th}}  \label{app:activeheat1}\\
        =&  \left(\frac{\bm{V}(\bm{r},T-t)}{D_\mathrm{th}} +\nabla\log P(\bm{r},T-t)\right)\cdot\bm{V}(\bm{r}, t) \label{app:activeheatV}\\
        &+  \frac{\bm{S}_\mathrm{L}(\bm{r},T-t) \cdot \bm{V}(\bm{r},t)}{D_\mathrm{th}}   \label{app:activeheat}.
\end{align}
}The terms in \eqref{app:activeheat1} and~\eqref{app:activeheat} describe the heat dissipation originating from active fluctuations. As noted in Eq.~\eqref{eqn:LFP},  $-\bm{S}_\mathrm{L}(\bm{r}, t)$ acts as an effective force induced by the active fluctuation $\bm{\eta}_\mathrm{act}$ and is derived from the entire ensemble of trajectories $\bm{r}(t)$ of Eq.~\eqref{eqn:SDE}. Consequently,
{\small\begin{align}
    \frac{ -\bm{S}_\mathrm{L}(\bm{r},t) \cdot \bm{V}(\bm{r},t) }{D_\mathrm{th}}\quad\mbox{and}\quad\frac{-\bm{S}_\mathrm{L}(\bm{r},t) \cdot \bm{V}(\bm{r},T-t)}{D_\mathrm{th}}
\end{align}
}are called the \emph{ensemble-level active heat dissipation}. In contrast, we define the \emph{trajectory-level active heat dissipation} as
{\small\begin{equation}
    \bm{\eta}_\mathrm{act}(t)\diamond\frac{\bm{V}(\bm{r},t)}{D_\mathrm{th}}\quad \mbox{and}\quad \bm{\eta}_\mathrm{act}(t)\diamond\frac{\bm{V}(\bm{r},T-t)}{D_\mathrm{th}}.
\end{equation}}It can be shown~\cite[Appendix I]{supplemental_material} that the ensemble average of the trajectory-level active heat dissipation coincides with the ensemble-level active heat dissipation under $P$. Thus the quantity $\mathcal{Q}(\{\bm{x}(t)\}_{0\leq t\leq T})$ in Eq. \eqref{Endmatter:Bact} represents the accumulation of the differences between the trajectory- and ensemble-level active heat dissipations along $\{\bm{x}(t)\}_{0\leq t\leq T}$. Its ensemble average is given by
{\footnotesize\begin{equation}\label{EM:f}
    \begin{aligned}
        \left\langle \mathcal{Q} \right\rangle 
        = &\int_0^T \d t\int \d\bm{r}\frac{\bm{V}(\bm{r}, T-t)}{D_\mathrm{th}} \cdot\bigg(  \bm{S}_\mathrm{L}(\bm{r}, T-t)   -\bm{S}_\mathrm{L}(\bm{r},t) \bigg)P(\bm{r},t).
    \end{aligned}
\end{equation}
}In steady state, the ensemble average of $\mathcal{Q}$ vanishes,
{\footnotesize\begin{align}
    \langle\mathcal{Q}\rangle^\mathrm{s}= & T\int \d\bm{r}\frac{\bm{V}^\mathrm{s}(\bm{r} )}{D_\mathrm{th}} \cdot\bigg(  \bm{S}^\mathrm{s}_\mathrm{L}(\bm{r} )  -\bm{S}^\mathrm{s}_\mathrm{L}(\bm{r}) \bigg)P^\mathrm{s}(\bm{r})=0.
\end{align}
}

\setcounter{equation}{0}  
\renewcommand{\theequation}{A.2.\arabic{equation}}

{\itshape Appendix A.2: Symmetry-breaking when reversing non-Gaussian process---}In the time-reversed process \eqref{eqn:SDEreversed}, the L\'{e}vy noise $\tilde{\bm{\eta}}_\mathrm{act}(t)$
has the different statistical property from the original non-Gaussian noise  
${\bm{\eta}}_\mathrm{act}(t)$, which is a signature of the non-Gaussian system.  

If we use the same non-Gaussian noise  
${\bm{\eta}}_\mathrm{act}(t)$ to drive the time-reversed process, we need to  define the following stochastic process $\bm{\tilde{r}}^\dagger(t)$, which is used in our derivation of $B_\mathrm{act}$:
{\small\begin{equation}\label{appeqn:SDEreverse}
\begin{aligned}
    \frac{\d\bm{\tilde{r}}^{\dagger}(t)}{\d t} =&  \frac{\bm{F}(\bm{\tilde{r}}^{\dagger}(t)) }{\Gamma} -2\bm{V}(\bm{\tilde{r}}^{\dagger}(t),T-t) + \bm{\eta}_\mathrm{th}(t)+ \bm{\eta}_\mathrm{act}(t),\\
         \bm{\tilde{r}}^{\dagger}(0) \sim& P(\sbullet,T).
         \end{aligned}
\end{equation}
}Here, $\bm{\eta}_\mathrm{th}$ and $\bm{\eta}_\mathrm{act}$ represent the thermal and active fluctuations, respectively, and they share the same statistical properties as those in Eq.~\eqref{eqn:SDE}.
The distribution of $\bm{\tilde{r}}^{\dagger}$ is denoted by $\tilde{P}^\dagger$.
For any $t \in [0,T]$, $\tilde{P}^\dagger(\bm{r},t) = \tilde{P}(\bm{r},t) = P(\bm{r},T-t)$, indicating that $\bm{\tilde{r}}^\dagger(t)$ shares the same probability flow as $\bm{\tilde{r}}(t)$ of Eq.~\eqref{eqn:SDEreversed}, but their underlying dynamics differ. For the derivation of the process $\bm{\tilde{r}}^\dagger(t)$, see \cite[Appendix II]{supplemental_material}.

When $\bm{\eta}_\mathrm{act}=0$ in Eq.~\eqref{eqn:SDE}, the  process $\bm{\tilde{r}}^\dagger(t)$ coincides exactly with $\bm{\tilde{r}}(t)$, implying $\tilde{P}^\dagger=\tilde{P}$. In this case, thermal noise drives stochastic trajectories through local diffusion motions, which  makes $\tilde{\bm{\eta}}_\mathrm{th}$ in the reversed dynamics ~\eqref{eqn:SDEreversed} retain the same statistical properties as $\bm{\eta}_\mathrm{th}$ in the original system Eq.~\eqref{eqn:SDE}.

In contrast, L\'evy-type active fluctuations produce  nonlocal  jumps, thereby  breaking  the time-reversal symmetry presented in purely diffusive dynamics. This symmetry breaking is manifested through the fact that the active noise $\tilde{\bm{\eta}}_\mathrm{act}$ driving the reversed dynamics in Eq.~\eqref{eqn:SDEreversed} is statistically different from $\bm{\eta}_\mathrm{act}$ in Eq.~\eqref{eqn:SDE}. Consequently, the non-Gaussian character of $\bm{\eta}_\mathrm{act}$ leads to $\tilde{P}^\dagger$ and $\tilde{P}$ being two \emph{distinct} path measures, even though they share identical marginals.

\setcounter{equation}{0}  
\renewcommand{\theequation}{B.\arabic{equation}}
{\itshape Appendix B: Derivation of detailed fluctuation theorem and numerical verification---}Recall the process $\bm{r}(t)$ of Eq. \eqref{eqn:SDE}, its time reversal process $\bm{\tilde{r}}(t)=\bm{r}(T-t)$, and the definition of $\mathcal{R}$ in Eq. \eqref{eqn:Rn}. For a trajectory $\{\bm{x}(t)\}_{0\leq t\leq T}$,
{\small\begin{equation}
    \begin{aligned}
        \mathcal{R}(\{\bm{x}(t)\}_{0\leq t\leq T})
        =&  \log\frac{\Pr\left[\bm{r}(t)=\bm{x}(t),\forall t\in[0,T]\right]}{\Pr\left[\bm{\tilde{r}}(t)=\bm{x}(t),\forall t\in[0,T]\right]}\\
        =&  \log\frac{\Pr\left[\bm{r}(t)=\bm{x}(t),\forall t\in[0,T]\right]}{\Pr\left[ \bm{r}(T-t)=\bm{x}(t),\forall t\in[0,T]\right]}\\
        =&  \log\frac{\Pr\left[\bm{r}(t)=\bm{x}(t),\forall t\in[0,T]\right]}{\Pr\left[ \bm{r}(t)=\bm{x}(T-t),\forall t\in[0,T]\right]}\\
        =& \log\frac{P[\{\bm{x}(t)\}_{0\leq t\leq T}]}{P[\{\tilde{\bm{x}}(t)\}_{0\leq t\leq T}]},
    \end{aligned}
\end{equation}
}where $\tilde{\bm{x}}(t)=\bm{x}(T-t)$ and $\Pr[A]$ denotes the probability of event $A$. Therefore, we know that $\mathcal{R}(\{\tilde{\bm{x}}(t)\}_{0\leq t\leq T})=-\mathcal{R}(\{\bm{x}(t)\}_{0\leq t\leq T})$, and recall $\rho_{\mathcal{R}}$ the distribution of the random variable $\mathcal{R}$ in Eq. \eqref{eqn:Rn}, then we have
{\small\begin{equation}
    \begin{aligned}
        \rho_{\mathcal{R}}(\Sigma)=&\int\delta(\mathcal{R}(\{\bm{x}(t)\}_{0\leq t\leq T}) -\Sigma)P(\{\bm{x}(t)\}_{0\leq t\leq T})\mathcal{D}\bm{x}\\
        =&\int\delta(\mathcal{R}(\{\bm{x}(t)\}_{0\leq t\leq T}) -\Sigma)P(\{\tilde{\bm{x}}(t)\}_{0\leq t\leq T})e^{\mathcal{R}}\mathcal{D}\bm{x}\\
        =& e^{\Sigma}\int\delta(-\mathcal{R}(\{\tilde{\bm{x}}(t)\}_{0\leq t\leq T}) -\Sigma)P(\{\tilde{\bm{x}}(t)\}_{0\leq t\leq T}) \mathcal{D}\tilde{\bm{x}}\\
        =& e^{\Sigma}\int\delta( \mathcal{R}(\{\tilde{\bm{x}}(t)\}_{0\leq t\leq T}) + \Sigma)P(\{\tilde{\bm{x}}(t)\}_{0\leq t\leq T}) \mathcal{D}\tilde{\bm{x}}\\
        =& e^{\Sigma}\rho_{\mathcal{R}}(-\Sigma).
    \end{aligned}
\end{equation}
}Thus the detailed fluctuation theorem \eqref{DFT1} is proved.

\begin{figure}[htbp]
  \centering
  \begin{minipage}[t]{0.47\linewidth}
    \raggedright (a)\\[-0.5ex]
    \includegraphics[width=\linewidth]{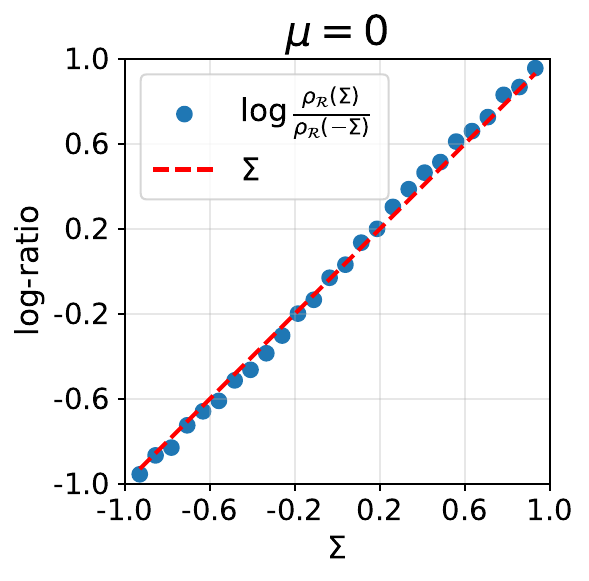}
  \end{minipage}
  \begin{minipage}[t]{0.47\linewidth}
    \raggedright (b)\\[-0.5ex]
    \includegraphics[width=\linewidth]{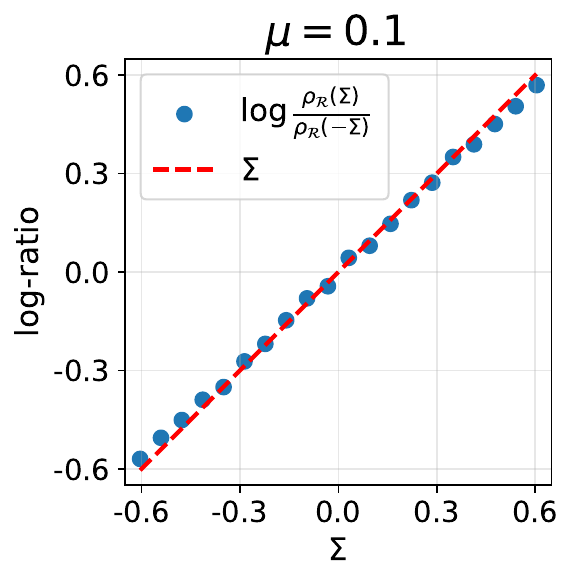}
  \end{minipage}
  \caption{The verification of DFT for the first example in the main text is shown in (a) for $\mu = 0$, and in (b) for $\mu = 0.1$. The blue dots represent the numerical results, i.e., the empirical probability ratios $\log\!\left[\rho_{\mathcal{R}}(\Sigma) / \rho_{\mathcal{R}}(-\Sigma)\right]$ evaluated at the corresponding values of $\Sigma$. The red dashed line indicates the theoretical prediction $y = \Sigma$. The fitted slopes of the data points show excellent agreement with the theoretical value.
}
  \label{fig:verificationDFT}
\end{figure}

We verify the detailed fluctuation theorem \eqref{DFT1} for the first example of the main text in Fig.~\ref{fig:verificationDFT}(a) (for $\mu = 0$) and Fig.~\ref{fig:verificationDFT}(b) (for $\mu = 0.1$). The fitted slopes of the data points exhibit excellent agreement with the theoretical prediction, see \cite[Appendix VII]{supplemental_material} for simulation details. 

\setcounter{equation}{0}  
\renewcommand{\theequation}{C.\arabic{equation}}

{\itshape Appendix C:  Active polyme model.---}The dynamics of the active polymer is
{\footnotesize\begin{equation}\label{eqn:example2SDE}
    \begin{aligned}
        \Gamma \frac{\d\bm{r}_\mathrm{A}}{\d t}=& - k\sum_{l=1}^m (\bm{r}_\mathrm{A} - \bm{r}^{(l)}_1 ) + \bm{\eta}_\mathrm{th}(t) + \bm{\eta}_\mathrm{act}(t),\\
        \Gamma\frac{\d\bm{r}^{(l)}_s}{\d t}=& - k\left(2 \bm{r}^{(l)}_s  -\bm{r}^{(l)}_{s+1} -\bm{r}^{(l)}_{s-1} \right)+ \bm{\eta}^{(l)}_\mathrm{th}(t).
    \end{aligned}
\end{equation}
}The first equation describes the motion of the ABP cross-linker
$\bm{r}_\mathrm{A}(t)$ ($\equiv \bm{r}_0^{(\cdot)}(t)$ with the index 0 denoting the
center bead); $\bm{r}^{(l)}_s$ denotes the position of $s$-th monomer in the $l$-th linear chain ($s\in\{1,\cdots,n\}$ and $l\in\{1,\cdots,m\}$). $k$ is the spring constant for the harmonic potential between neighboring beads. The active fluctuation $\bm{\eta}_\mathrm{act}$ is modeled as the compound Poisson process $\bm{\eta}_\mathrm{act}(t)= \sum_{j=1}^{N_t}v_{0}\bm{\sigma}_{\mathrm{D}}(t)\delta(t-t_j)$ where  $N_t$ is a Poisson process with
parameter $\lambda_0$, $v_{0}$ is the constant speed of self-propulsion and the random vector  $\bm{\sigma}_{\mathrm{D}}(t)$ takes four possible   directions corresponding to the positive and negative \( x \)-axis and \( y \)-axis, respectively, with probability weight \( r_i \), \( i = 1, 2, 3, 4 \). 
The equal weight $r_\mathrm{i} = 1/4$ ($i=1,\cdots,4$) gives the  the active fluctuations   of the unbiased type.  The second equation describes the dynamics
of the Rouse chains in the polymer network. $\bm{\eta}_\mathrm{th}$ and $\bm{\eta}_\mathrm{th}^{(l)}$ are independent
thermal white noises and both have a variance of $2 k_B\mathcal{T}/\Gamma$ for each Cartesian component. The boundary bead (the last $(n+1)$-th bead) in each  arm  is fixed in space. We set the initial state of the system to follow a Gaussian distribution, where the mean configuration ensures a distance of 0.5 between adjacent particles, and the covariance matrix is specified as the identity matrix. 
More details can be found in \cite[Appendix IX]{supplemental_material}.

\newpage
\clearpage

\onecolumngrid

\appendix

\setcounter{secnumdepth}{2}
\renewcommand{\thesection}{\Roman{section}}

\setcounter{equation}{0} 
\renewcommand{\theequation}{S\Roman{section}.\arabic{equation}}

\begin{center}
  \Large \bf Supplemental Material
\end{center}

In this Supplemental Material accompanying the Letter, we provide a rigorous derivation of the entropy decomposition formula and the fluctuation theorems, utilizing the probability flow equivalence technique and tools from stochastic analysis. Additionally, we include detailed algorithms for computing EPRs, as well as comprehensive explanations of the example discussed in the main text.

\section{Proof of Entropy production decomposition formula}
This Appendix is devoted to providing the proof of the entropy production decomposition formula. The main objective is to show that the ensemble–averaged velocity of $\boldsymbol{r}(t)$ is given by $\boldsymbol{V}(\boldsymbol{r},t)$, as defined by the LFPE~\eqref{eqn:LFP}.

Given  $\bm{F}(\bm{r})$, $\bm{\eta}_\mathrm{th}(t)$, and $\bm{\eta}_\mathrm{act}(t)$   as in the Langevin equation   of the Letter,  the process $\bm{r}(t)$ of 
\begin{eqnarray}\label{app:SDE}
 \d \bm{r}(t)/\d t = \bm{F}(\bm{r}(t))/\Gamma + \bm{\eta}_{\mathrm{th}}(t) + \bm{\eta}_{\mathrm{act}}(t),
\end{eqnarray}
is not differentiable, which necessitates justification for the relationship
\begin{eqnarray}
    \langle \dot{\bm{r}}\mid \bm{r},t\rangle = \bm{J}(\bm{r},t)/P(\bm{r},t)=\bm{V}(\bm{r},t),
\end{eqnarray}
where $\bm{V}(\bm{r}(t),t) = \bm{F}(\bm{r}(t))/\Gamma - D_{\mathrm{th}} \nabla \log P(\bm{r}(t),t) - \mathcal{S}_\mathrm{L}(\bm{r}(t),t)$. To this end, consider any vector-valued test function $\bm{\phi}$ and any finite interval $T > 0$. The velocity $\bm{V}(\bm{r}(t),t)$ is shown to satisfy the following calculation (here $\mathcal{S}_\mathrm{L}(\bm{r}(t),t)$ is the L\'{e}vy score given in Eq. \eqref{levyscore} in the main text):
\begin{equation}\label{app:<V>}
\begin{aligned}
    & \left\langle \int_0^T \bm{\phi}(\bm{r}(t))\diamond\d\bm{r}(t) \right\rangle\\ 
    =& 
     \left\langle  \int_0^T \d t \left(\bm{\phi}(\bm{r}(t))\cdot \bm{F}(\bm{r}(t))/\Gamma  +  \bm{\phi}(\bm{r}(t))\diamond\bm{\eta}_{\mathrm{th}}(t)  + \bm{\phi}(\bm{r}(t))\diamond\bm{\eta}_{\mathrm{act}}(t) \right) \right\rangle \quad\quad\mbox{(using SDE Eqn. )}\\
    =& 
     \left\langle  \int_0^T \d t \Big\{ \bm{\phi}(\bm{r}(t))\cdot \bm{F}(\bm{r}(t))/\Gamma  +  \bm{\phi}(\bm{r}(t))\cdot\bm{\eta}_{\mathrm{th}}(t) +  \mathrm{tr}[D_{\mathrm{th}}\nabla\bm{\phi}(\bm{r}(t))] \right.\quad\quad\mbox{(Marcus to It\^{o})} \\
     & \left.  + \bm{\phi}(\bm{r}(t))\cdot\bm{\eta}_{\mathrm{act}}(t) \Big\}   +    \sum_{0\leq t\leq T}\left[\int_0^1\d\theta \Delta\bm{r}(t)\cdot\bm{\phi}(\bm{r}(t) + \theta\Delta\bm{r}(t)) -\Delta\bm{r}(t)\cdot\bm{\phi}(\bm{r}(t)) \right]\right\rangle \\ & \quad\quad\mbox{(Marcus to It\^{o} and Taylor's expansion for the second term)}\\
     =& 
     \left\langle  \int_0^T \d t \left\{ \bm{\phi}(\bm{r}(t))\cdot \bm{F}(\bm{r}(t))/\Gamma  +  \mathrm{tr}[D_{\mathrm{th}}\nabla\bm{\phi}(\bm{r}(t))]  \right\} + \sum_{0\leq t\leq T}  \int_0^1\d\theta \Delta\bm{r}(t)\cdot\bm{\phi}(\bm{r}(t) + \theta\Delta\bm{r}(t))\right\rangle\\
& \left(\mbox{zero expectation for  integral of Brownian white noise and}\ \int_0^T dt\ \bm{\phi}(\bm{r}(t))\cdot\bm{\eta}_{\mathrm{act}}(t)-\sum_{0\leq t\leq T}\Delta\bm{r}(t)\cdot\bm{\phi}(\bm{r}(t))=0\right)\\
     =&  
     \int\d\bm{r} \int_0^T \d t \left\{ \bm{\phi}(\bm{r} )\cdot \bm{F}(\bm{r} ) /\Gamma +  \mathrm{tr}[D_{\mathrm{th}}\nabla\bm{\phi}(\bm{r} )]   \right\}P(\bm{r} ,t)  \\
     & +  \int\d\bm{r} \int_0^T\d t\int_0^1\d\theta\int \nu(\d\bm{z})\bm{z}\bm{\phi}(\bm{r} + \theta\bm{z})P(\bm{r} ,t) \quad\quad\mbox{(the probability density of $\bm{r} $ is $P(\bm{r} ,t)$)}\\
     =& 
     \int\d\bm{r} \int_0^T \d t \bm{\phi}(\bm{r} )\cdot\left\{ \bm{F}(\bm{r} )/\Gamma  - D_{\mathrm{th}}\nabla\log P(\bm{r} ,t)   \right\}P(\bm{r} ,t) \quad\quad\mbox{(integrating by parts)} \\
     & +  \int\d\bm{r}  \int_0^T\d t \int_0^1\d\theta\int \nu(\d\bm{z})\bm{z}\bm{\phi}(\bm{r} )\frac{P(\bm{r} -\theta\bm{z},t) }{P(\bm{r} ,t)}P(\bm{r} ,t)\quad\quad\mbox{(change of variables)}\\
    \equiv& \int\d\bm{r} \int_0^T\d t ~\bm{\phi}(\bm{r} )\cdot\bm{V}(\bm{r} ,t)P(\bm{r} ,t)\qquad\mbox{(by the definition of $\bm{V}(\bm{r},t)$)},
    \end{aligned}
\end{equation}
where $\Delta\bm{r}(t) = \bm{r}(t) - \bm{r}(t^-)$, and the Marcus integrals are converted to It\^{o} integrals using standard techniques  \citeappendix[Theorem 4.4.28]{app:applebaum2009levy}. The remaining derivation for the EP decomposition formula will follow the same way in the Letter. As a direct application, Eq. \eqref{EM:f} in the End Matter can be proved.

\section{Probability flow equivalence technique}
In this Appendix, we provide a brief introduction to the probability–flow technique developed in Ref.~\cite{app:huang2024probability}, and we outline its connection to several classes of stochastic dynamics. These results form a technical foundation for the derivations presented in the main text and will also be used in the subsequent Appendices.

We rewrite the Langevin equation \eqref{app:SDE} as the form of stochastic differential equation
{\footnotesize\begin{equation}\label{eqn:SDEforward}
\begin{aligned}
    \d\bm{r}(t) =&\frac{\bm{F}(\bm{r}(t))}{\Gamma}\d t  + \d \bm{W}_t + \int\bm{z}\mathcal{N}(\d t,\d\bm{z}),
\end{aligned}
\end{equation}
}with initial probability distribution $\bm{r}(0)\sim P_0(\bm{r})$ and 
$\bm{W}_t$ is a Brownian motion in $\mathbb{R}^d$ with covariance matrix $\mathbb{E}|\bm{W}_t-\bm{W}_s|^2=2D_\mathrm{th}|t-s|I_d$, and $\mathcal{N}(\d t,\d\bm{z})$ is an independent Poisson random measure. Its L\'{e}vy measure is given by $\nu(\d\bm{z})\d t$ (state-independent for $\mathcal{N}$). 

The probability density $P(\bm{r},t)$ describes the distribution of $\bm{r}(t)$ at time $t$, and it satisfies the following L\'{e}vy-Fokker--Planck equation:
{\footnotesize\begin{equation}
    \begin{aligned}
        \partial_t P(\bm{r},t)=& -\nabla\cdot \left( \frac{\bm{F}(\bm{r} )}{\Gamma}    P(\bm{r},t ) \right)  + D_\mathrm{th}\triangle P(\bm{r},t)  + \int\nu(\d\bm{z})\left(P(\bm{r}-\bm{z},t)-P(\bm{r},t)\right).
    \end{aligned}
\end{equation}
}Now we consider the probability flow in backward time, as $\tilde{P}(\bm{r},t)\equiv P(\bm{r},T-t)$, which satisfies
{\footnotesize\begin{equation}
    \begin{aligned}
        \partial_t \left( P(\bm{r},T-t)\right)=&  \nabla\cdot \left( \frac{\bm{F}(\bm{r} )}{\Gamma}   P(\bm{r},T-t ) \right)   - D_\mathrm{th}\triangle P(\bm{r},T-t)  - \int\nu(\d\bm{z})\left(P(\bm{r}-\bm{z},T-t)-P(\bm{r},T-t)\right)\\
        =& -\nabla\cdot \left[\left( \frac{\bm{-F}(\bm{r} )}{\Gamma}   + 2D_\mathrm{th}\nabla\log P(\bm{r},T-t)\right)P(\bm{r},T-t ) \right]   + D_\mathrm{th}\triangle P(\bm{r},T-t) \\
        & - 2\int\nu(\d\bm{z})\left(P(\bm{r}-\bm{z},T-t)-P(\bm{r},T-t)\right)  + \int\nu(\d\bm{z})\left(P(\bm{r}-\bm{z},T-t)-P(\bm{r},T-t)\right)\\
         =& -\nabla\cdot \left[\left( \frac{\bm{-F}(\bm{r} )}{\Gamma}   + 2D_\mathrm{th}\nabla\log P(\bm{r},T-t) - 2\int\nu(\d\bm{z})\int_0^1\d\theta \bm{z}\frac{P(\bm{r}-\theta\bm{z},T-t)}{P(\bm{r},T-t)} \right)P(\bm{r},T-t ) \right]  \\
         &  + D_\mathrm{th}\triangle P(\bm{r},T-t)  + \int\nu(\d\bm{z})\left(P(\bm{r}-\bm{z},T-t)-P(\bm{r},T-t)\right)\\
        =& -\nabla\cdot \left[\left( \frac{\bm{-F}(\bm{r} )}{\Gamma}   + 2D_\mathrm{th}\nabla\log P(\bm{r},T-t) - 2\int\nu(\d\bm{z})\int_0^1\d\theta \bm{z}\frac{P(\bm{r}-\theta\bm{z},T-t)}{P(\bm{r},T-t)} \right)\tilde{P}(\bm{r}, t ) \right]  \\
        &  + D_\mathrm{th}\triangle \tilde{P}(\bm{r},t )  + \int\nu(\d\bm{z})\left(\tilde{P}(\bm{r}-\bm{z},t )-\tilde{P}(\bm{r},t )\right),
    \end{aligned}
\end{equation}
}which is also a L\'{e}vy--Fokker-Planck equation that governs the evolutions of a stochastic process $\bm{\tilde{r}}^{\dagger}(t)$ as
{\footnotesize\begin{equation}\label{eqn:SDEreversalflow}
    \begin{aligned}
        \d\bm{\tilde{r}}^{\dagger}(t) =& \left( \frac{\bm{-F}(\bm{\tilde{r}}^{\dagger} )}{\Gamma}   + 2D_\mathrm{th}\nabla\log P(\bm{\tilde{r}}^{\dagger},T-t)   + 2\bm{S}_\mathrm{L}(\bm{\tilde{r}}^\dagger,T-t) \right)\d t +  \d\bm{W}_t + \int\bm{z}\mathcal{N}(\d t,\d\bm{z})\\
        =&  \frac{\bm{F}(\bm{\tilde{r}}^{\dagger} )}{\Gamma} - 2 \bm{V}(\bm{\tilde{r}}^{\dagger} ,T-t) \d t +  \d\bm{W}_t + \int\bm{z}\mathcal{N}(\d t,\d\bm{z}),\quad t:0\to T,\quad
        \bm{\tilde{r}}^{\dagger}(0)\sim \tilde{P}(\bm{r},0)= P(\bm{r},T),
    \end{aligned}
\end{equation}
}where $\bm{W}_t$ and $\mathcal{N}(\d t,\d\bm{z})$ are identical to those of \eqref{eqn:SDEforward} in distribution, see also \citeappendix{app:baule2025ge}. Here, $\bm{S}_\mathrm{L}(\bm{r},T-t)\equiv -\int\nu(\d\bm{z})\int_0^1\d\theta \bm{z}\frac{P(\bm{r}-\theta\bm{z},T-t)}{P(\bm{r},T-t)} $ is the L\'{e}vy score function at time $T-t$ and $\bm{V}$ is the velocity field defined in \eqref{eqn:LFP} of the main text. This is the auxiliary process $\bm{\tilde{r}}^\dagger(t)$ in the main text. It is important to see that, the process $\bm{\tilde{r}}^{\dagger}(t)$ only shares the same probability flow with the one of $\bm{r}(t)$ of \eqref{eqn:SDEforward} backward in time. When the active noise vanishes, this process $\bm{\tilde{r}}^{\dagger}(t)$ is indeed the time reversal process $\bm{\tilde{r}}(t)\equiv\bm{r}(T-t)$ of $\bm{r}(t)$, however for general active fluctuation systems, they are not the same in path space while only share the same probability flows. 


The generator of the time reversal process $\bm{\tilde{r}}(t)\equiv\bm{r}(T-t)$ for $t=[0,T]$ is given by (see, e.g., \citeappendix{app:privault2004markovian,app:conforti2022time} and a simple derivation can be found in \cite[Appendix C]{app:huang2025transition}):
{\footnotesize\begin{equation}
    \begin{aligned}
        \tilde{\mathcal{L}}f(\bm{r})=& \left[-\frac{\bm{F}(\bm{r} )}{\Gamma} + 2D_\mathrm{th}\nabla\log P( \bm{r} ,T-t ) \right]\cdot\nabla f(\bm{r})  + D_\mathrm{th}\triangle f(\bm{r}) + \int\tilde{\nu}_{T-t}(\bm{r},\d\bm{z})\left[ f(\bm{r} + \bm{z}) - f(\bm{r})\right],
    \end{aligned}
\end{equation}
}where $\tilde{\nu}_{T-t}(\bm{r},\d\bm{z})\d t \equiv \frac{P(\bm{r}+\bm{z} ,T-t )}{P(\bm{r},T-t )}\nu(-\d\bm{z})\d t$ is a state-dependent L\'{e}vy measure and $P(\bm{r},t)$ is the probability density of $\bm{r}(t)$ to SDE \eqref{eqn:SDEforward}. The time reversal process $\bm{\tilde{r}}(t)$ satisfies the following SDE:
{\small\begin{align}
    \d\bm{\tilde{r}}(t) =&-\frac{\bm{F}(\bm{\tilde{r}}(t))}{\Gamma}\d t + 2D_\mathrm{th}\nabla\log P(\bm{\tilde{r}}(t),T-t )\d t +  \d \bm{W}_t + \int\bm{z}\tilde{\mathcal{N}}(\d t,\d\bm{z}),\
    \bm{\tilde{r}}(0) \sim  P(\bm{r},T), \label{eqn:SDEreversal2}
\end{align}
}where $\bm{W}_t$ is a Brownian motion that has the same statistical properties with the one in \eqref{eqn:SDEforward}, and $\tilde{\mathcal{N}}(\d t,\d\bm{z})$ is an independent Poisson random measure. The L\'{e}vy measure of $\tilde{\mathcal{N}}(\d t,\d\bm{z})$ is given by $\tilde{\nu}_{T-t}(\bm{r},\d\bm{z})\d t$.

Next, we consider the discrete time by setting  $0=t_0\leq t_1\leq\cdots\leq t_n=T$. 
Let the transition densities of $\bm{r}(t)$, $\bm{\tilde{r}}^{\dagger}(t)$ and $\bm{\tilde{r}}(t)$ from position $\bm{r}_j$ at time $t_j$ to $\bm{r}$ at time $t$ be denoted by $P(\bm{r},t|\bm{r}_j,t_j)$, $\tilde{P}^{\dagger}(\bm{r},t|\bm{r}_j,t_j)$ and $\tilde{P}(\bm{r},t|\bm{r}_j,t_j)$ respectively ($0 \leq t_j \leq t \leq T$).
{\small\begin{equation}\label{eqn:LFP3types}
\begin{aligned}
\partial_t P(\bm{r},t|\bm{r}_j,t_j)=&   -\nabla\cdot \left(\frac{\bm{F}(\bm{r} )}{\Gamma} P(\bm{r},t |\bm{r}_j,t_j) \right)+ D_\mathrm{th}\triangle P(\bm{r},t|\bm{r}_j,t_j) - \nabla\cdot\left(\int_0^1\d\theta\int\bm{z} \nu(\d\bm{z})P(\bm{r} - \theta\bm{z},t \mid \bm{r}_j,t_j) \right),\\
    \partial_t \tilde{P}^{\dagger}(\bm{r},t|\bm{r}_j,t_j)=&   -\nabla\cdot\left[\left(-\frac{\bm{F}(\bm{r} )}{\Gamma}  + 2D_\mathrm{th}\nabla\log P( \bm{r} ,T-t )  + 2\bm{S}_\mathrm{L}(\bm{r},T-t) \right)\tilde{P}^{\dagger}(\bm{r},t |\bm{r}_j,t_j)  \right] \\
    &+ D_\mathrm{th}\triangle \tilde{P}^{\dagger}(\bm{r},t|\bm{r}_j,t_j) - \nabla\cdot\left(\int_0^1\d\theta\int \nu(\d\bm{z})\bm{z}\tilde{P}^{\dagger}(\bm{r} - \theta\bm{z},t \mid \bm{r}_j,t_j) \right),\\
    \partial_t \tilde{P}(\bm{r},t|\bm{r}_j,t_j)=&   -\nabla\cdot\left[\left(-\frac{\bm{F}(\bm{r} )}{\Gamma}  + 2D_\mathrm{th}\nabla\log P( \bm{r} ,T-t )\right)\tilde{P}(\bm{r},t |\bm{r}_j,t_j)  \right] \\
    &+ D_\mathrm{th}\triangle \tilde{P}(\bm{r},t|\bm{r}_j,t_j) - \nabla\cdot\left(\int_0^1\d\theta\int \tilde{\nu}_{T-t}(\bm{r}-\theta\bm{z},\d\bm{z})\bm{z}\tilde{P}(\bm{r} - \theta\bm{z},t \mid \bm{r}_j,t_j) \right).
    \end{aligned}
\end{equation}
}Next, we  construct three It\^{o} SDEs starting from the time $t_j$ at any   $\bm{r}_j$ so that   their (transition) densities,
$P_\mathrm{d}(\bm{r},t|\bm{r}_j,t_j)$, $\tilde{P}^{\dagger}_\mathrm{d}(\bm{r},t|\bm{r}_j,t_j)$ and $\tilde{P}_\mathrm{d}(\bm{r},t|\bm{r}_j,t_j)$,
exactly match $P(\bm{r},t|\bm{r}_j,t_j)$, $\tilde{P}^{\dagger}(\bm{r},t|\bm{r}_j,t_j)$ and $\tilde{P}(\bm{r},t|\bm{r}_j,t_j)$ respectively:
{\begin{equation}\label{app:rd1}
\begin{aligned}
    \d\bm{r}_\mathrm{d}(t)=&\left[ \bm{F}(\bm{r}_\mathrm{d}(t))/\Gamma + \int_0^1\d\theta\int\nu(\d\bm{z})\bm{z}\frac{P(\bm{r}_\mathrm{d}(t)-\theta\bm{z},t\mid \bm{r}_j,t_j)}{P(\bm{r}_\mathrm{d}(t),t\mid \bm{r}_j,t_j)}   \right]\d t  +  \d \bm{W}_t,\quad 
    \bm{r}_\mathrm{d}(t_j)= \bm{r}_j;\\
    \d\bm{\tilde{r}}^\dagger_\mathrm{d}(t)=&\left[-\bm{F}(\bm{\tilde{r}}^\dagger_\mathrm{d}(t))/\Gamma  + 2D_\mathrm{th}\nabla\log P(\bm{\tilde{r}}^\dagger_\mathrm{d}(t),T-t )  + 2\bm{S}_\mathrm{L}(\bm{\tilde{r}}^\dagger_\mathrm{d}(t),T-t)   -\bm{S}_{\mathrm{L},\bm{r}_j}(\bm{\tilde{r}}^\dagger_\mathrm{d}(t),t_j) \right]\d t +  \d \bm{W}_t, \quad  \bm{\tilde{r}}^\dagger_\mathrm{d}(t_j)=  \bm{r}_j,\\
    \d\bm{\tilde{r}}_\mathrm{d}(t)=&\left[-\bm{F}(\bm{\tilde{r}}_\mathrm{d}(t))/\Gamma  + 2D_\mathrm{th}\nabla\log P(\bm{\tilde{r}}_\mathrm{d}(t),T-t ) - \tilde{\bm{S}}_{\mathrm{L},\bm{r}_j}(\bm{\tilde{r}}_\mathrm{d}(t),t) \right]\d t +  \d \bm{W}_t, \quad
     \bm{\tilde{r}}_\mathrm{d}(t_j)= \bm{r}_j 
 \end{aligned}
\end{equation}}
where  we have defined the following L\'{e}vy score functions associated with the conditional transition densities:
{ \begin{align}
        \bm{S}_{\mathrm{L},\bm{r}(t_i)}(\bm{r},t)\equiv& -\int_0^1\d\theta\int\nu(\d\bm{z})\bm{z}\frac{\tilde{P}^\dagger(\bm{r} -\theta\bm{z},t \mid \bm{r} (t_i),t_i)}{\tilde{P}^\dagger (\bm{r},t \mid \bm{r} (t_i),t_i)}, \label{app:defS_L}\\
        \tilde{\bm{S}}_{\mathrm{L}, \bm{r}(t_i)}(\bm{r},t )\equiv& -\int_0^1\d\theta\int \tilde{\nu}_{T-t}(\bm{r}-\theta\bm{z},\d\bm{z})\bm{z} \frac{\tilde{P}(\bm{r} -\theta\bm{z},t \mid \bm{r} (t_i),t_i)}{\tilde{P}(\bm{r},t \mid \bm{r} (t_i),t_i)}.
\end{align}
}

We justify the notion   $\bm{S}_{\mathrm{L},\bm{r}(t_i )} (\bm{r},t)$ in \eqref{app:defS_L}  by showing  the approximation in short time  
\begin{equation}
  \label{eq:shortimePasym}
\bm{S}_{\mathrm{L},\bm{r}(t_i )}(\bm{r},t_{i+1})   =  -\int_0^1\d\theta\int\nu(\d\bm{z})\bm{z}\frac{P(\bm{r} -\theta\bm{z},t \mid \bm{r} (t_i),t_i)}{P(\bm{r},t \mid \bm{r} (t_i),t_i)} + \mathrm{O}(\Delta t).
\end{equation}
Eq. \eqref{eq:shortimePasym} 
holds because the jump noises in the L\'{e}vy SDE \eqref{eqn:SDEforward} of $\bm{r}(t)$    and the L\'{e}vy SDE \eqref{eqn:SDEreversalflow} of $\bm{\tilde{r}}^\dagger(t)$ share the same L\'{e}vy measure. 
The rigorous proof of \eqref{eq:shortimePasym}  is based on the short time asymptotic expansion  of the transition densities for   L\'{e}vy SDE \citeappendix{app:figueroa2018small}, which is quoted below:
{\footnotesize\begin{equation}\label{expansion:p}
        P(x,t\mid y,0)=\ t^{-d/2}\exp\left(-\frac{C^{(-1)}(y,x)}{t}\right)\sum_{k=0}^\infty C^{(k)}(y,x)t^k + \sum_{k=1}^\infty D^{(k)}(y,x)t^k,
    \end{equation}
}for some functions $\{C^{(k)}\}_{k=-1}^\infty$ and $\{D^{(k)}\}_{k=1}^\infty$, and in particular,{\footnotesize\begin{equation*}
        \begin{aligned}
            C^{(-1)}(y,x)\geq&\ 0,\quad \forall x,y\in\mathbb{R}^d;\
            C^{(-1)}(y,x)=\ 0,\quad \mbox{if and only if}\quad x=y,\\
            C^{(0)}(x,x)=&\ \frac{1}{((2\pi)^{d}\det \Sigma(x))^{\frac{1}{2}}},\ 
            D^{(1)}(y,x)=\  \lambda_0(y)\nu_J(x-y),\\
\lim_{t\downarrow0}\frac{P(x,t\mid y,0)}{t}=&\ \lambda_0(y)\nu_A(x-y),\quad x\neq y;\quad
            \lim_{t\downarrow0}\frac{P(x,t\mid y,0)}{t^{-\frac{d}{2}}}=\ \frac{1}{((2\pi)^{d}\det \Sigma(x))^{\frac{1}{2}}}.
        \end{aligned}
    \end{equation*}}

Eq. \eqref{eq:shortimePasym} is used to approximate the second  term in the drift of the first Ito SDE in \eqref{app:rd1}.
As the final result,
the three transition probability densities from $t_i$ to $t_{i+1}$ for  \eqref{app:rd1}  are given by:   
{\footnotesize\begin{equation}\label{eqn:shortestimate}
\begin{aligned}
    P_\mathrm{d}[\bm{r}(t_{i+1}),t_{i+1}&\mid\bm{r}(t_i),t_i] 
    = \frac{1}{\sqrt{(2\pi)^d\Delta t}}\exp\left\{-\left|\frac{ \bm{r}(t_{i+1}) -\bm{r}(t_i) }{\Delta t}  - \frac{\bm{F}\left(\bm{r}(t_{i+1}) \right)}{\Gamma} \right.  + \bm{S}_{\mathrm{L},\bm{r}(t_i) }(\bm{r}(t_{i+1}),t_{i+1})    \right|^2\frac{\Delta t}{4D_\mathrm{th}} \\
    & \left. + \nabla\cdot\left[ \frac{\bm{F}\left( \bm{r}(t_{i+1}) \right)}{\Gamma}  - \bm{S}_{\mathrm{L},\bm{r}(t_i) }(\bm{r}(t_{i+1}),t_{i+1}) \right] + \mathrm{o}(\Delta t)\right\},\\
     \tilde{P}^\dagger_\mathrm{d}[ \bm{r}(t_{i+1}),t_{i+1}&\mid \bm{r}(t_{i}),t_i] 
     = \frac{1}{\sqrt{(2\pi)^d\Delta t}} \exp\left\{-\left|\frac{ \bm{r}(t_{i+1})  - \bm{r}(t_{i}) }{\Delta t} + \frac{\bm{F}\left(  \bm{r}(t_{i+1}) \right)}{\Gamma}   - 2D_\mathrm{th}\nabla\log P( \bm{r}(t_{i+1}),T-t_{i+1} ) \right.\right.\\
     &   + \bm{S}_{\mathrm{L},\bm{r}(t_i)}(\bm{r}(t_{i+1},t_{i+1})  - 2\bm{S}_\mathrm{L}(\bm{r}(t_{i+1}),T-t_{i+1}) \Bigg|^2\frac{\Delta t}{4D_\mathrm{th}} \\
     &  \left. +  \nabla\cdot\left[ -\frac{\bm{F}\left( \bm{r} (t_{i+1}) \right)}{\Gamma}  + 2D_\mathrm{th}\nabla\log P(\bm{r} (t_{i+1}),T-t_{i+1} ) + 2\bm{S}_\mathrm{L}(\bm{r}(t_{i+1}),T-t_{i+1}) - \bm{S}_{\mathrm{L},\bm{r}(t_i)}(\bm{r}(t_{i+1},t_{i+1}) \right]  + \mathrm{o}(\Delta t)\right\}\\
     \tilde{P}_\mathrm{d} [\bm{r}(t_{i+1}),t_{i+1}&\mid \bm{r}(t_{i}),t_i] 
     =  \frac{1}{\sqrt{(2\pi)^d\Delta t}} \exp\left\{-\left|\frac{  \bm{r}(t_{i+1})  - \bm{r}(t_{i}) }{\Delta t} + \frac{ \bm{F}\left(  \bm{r}(t_{i+1}) \right)}{\Gamma}   - 2D_\mathrm{th}\nabla\log P( \bm{r}(t_{i+1}),T-t_{i+1} ) \right.\right.\\
     &  + \tilde{\bm{S}}_{\mathrm{L},\bm{r}(t_i)}(\bm{r}(t_{i+1}),T-t_{i+1}) \bigg|^2\frac{\Delta t}{4D_\mathrm{th}}  + \nabla\cdot \left[ -\frac{\bm{F}\left( \bm{r} (t_{i+1}) \right)}{\Gamma}  + 2D_\mathrm{th}\nabla\log P(\bm{r} (t_{i+1}),T-t_{i+1} ) \right. \\
     &  -  \tilde{\bm{S}}_{\mathrm{L},\bm{r}(t_i)} (\bm{r} (t_{i+1}),T-t_{i+1} ) \bigg] + \mathrm{o}(\Delta t)\bigg\}. 
\end{aligned}
\end{equation}}

\section{Proof of the fluctuation theorem: Path integral viewpoint}\label{app:FT}
In this appendix, we prove the integral fluctuation theorem using the path integral approach and the probability flow technique introduced before.

By Bayes' theorem, we find that,
{\footnotesize\begin{equation}\label{app:totaldecom}
    \begin{aligned}
          \log \frac{P[\bm{r}(t_1) \cdots \bm{r}(t_n) \mid \bm{r}_0]}{\tilde{P} [ \bm{r}(t_1) \cdots \bm{r}(t_n) \mid \bm{r}_0] }
         =& \log \frac{P[\bm{r}(t_1) \cdots \bm{r}(t_n) \mid \bm{r}_0]}{\tilde{P}^{\dagger} [ \bm{r}(t_1) \cdots \bm{r}(t_n) \mid \bm{r}_0] } + \log \frac{\tilde{P}^{\dagger}[\bm{r}(t_1) \cdots \bm{r}(t_n) \mid \bm{r}_0]}{\tilde{P} [ \bm{r}(t_1) \cdots \bm{r}(t_n) \mid \bm{r}_0] } \\
    =& \log \frac{P[\bm{r}(t_1) \mid \bm{r}_0]\cdots P [\bm{r}(t_n) \mid \bm{r}(t_{n-1})]}{\tilde{P}^{\dagger} [ \bm{r}(t_1) \mid  \bm{r}_0]\cdots \tilde{P}^{\dagger} [\bm{r}(t_n) \mid  \bm{r}(t_{n-1})]}  + \log \frac{\tilde{P}^{\dagger}[\bm{r}(t_1) \mid \bm{r}_0]\cdots \tilde{P}^{\dagger} [\bm{r}(t_n) \mid \bm{r}(t_{n-1})]}{\tilde{P} [ \bm{r}(t_1) \mid  \bm{r}_0]\cdots \tilde{P} [\tilde{ \bm{r}}(t_n) \mid  \bm{r}(t_{n-1})]}\\
    =& \log \frac{ P_\mathrm{d}[\bm{r}(t_1) \mid \bm{r}_0]\cdots P_\mathrm{d}[\bm{r}(t_n) \mid \bm{r}(t_{n-1})]}{ \tilde{P}^\dagger_\mathrm{d}[ \bm{r}(t_1) \mid  \bm{r}_0]\cdots \tilde{P}^\dagger_\mathrm{d}[ \bm{r}(t_n) \mid \bm{r}(t_{n-1})]} + \log \frac{ \tilde{P}^\dagger_\mathrm{d}[\bm{r}(t_1) \mid \bm{r}_0]\cdots \tilde{P}^\dagger_\mathrm{d}[\bm{r}(t_n) \mid \bm{r}(t_{n-1})]}{ \tilde{P}_\mathrm{d}[ \bm{r}(t_1) \mid  \bm{r}_0]\cdots \tilde{P}_\mathrm{d}[ \bm{r}(t_n) \mid \bm{r}(t_{n-1})]}.
    \end{aligned}
\end{equation} }Here we introduce a new stochastic process, $\tilde{P}^\dagger$, to provide a clearer and more intuitive interpretation of the quantity $B_{\mathrm{act}}$ defined in the main text.

Substituting the estimates \eqref{eqn:shortestimate} into the above equalities, we obtain that,
{\footnotesize\begin{equation}\label{app:PPpf}
    \begin{aligned}
    &\log \frac{ P_\mathrm{d}[\bm{r}(t_1) \mid \bm{r}_0]\cdots P_\mathrm{d}[\bm{r}(t_n) \mid \bm{r}(t_{n-1})]}{ \tilde{P}^\dagger_\mathrm{d}[ \bm{r}(t_1) \mid  \bm{r}_0]\cdots \tilde{P}^\dagger_\mathrm{d}[ \bm{r}(t_n) \mid \bm{r}(t_{n-1})]}\\
    =&   -\sum_{i=0}^{n-1}\left|\frac{ \bm{r}(t_{i+1}) -\bm{r}(t_i) }{\Delta t}  - \frac{\bm{F} \left(\bm{r}(t_{i+1}) \right)}{\Gamma} +\bm{S}_{\mathrm{L},\bm{r}(t_i)}(\bm{r}(t_{i+1}),t_{i+1}) \right|^2\frac{\Delta t}{4D_\mathrm{th}}   + \sum_{i=0}^{n-1}\left|\frac{ \bm{r}(t_{i+1})  - \bm{r}(t_{i}) }{\Delta t} + \frac{\bm{F}\left(  \bm{r}(t_{i+1}) \right)}{\Gamma} \right. \\
    &  + \bm{S}_{\mathrm{L},\bm{r}(t_i)}(\bm{r}(t_{i+1}),t_{i+1})   - 2D_\mathrm{th}\nabla\log P( \bm{r}(t_{i+1}),T-t_{i+1} )   - 2\bm{S}_\mathrm{L}(\bm{r}(t_{i+1}),T-t_{i+1})   \bigg|^2\frac{\Delta t}{4D_\mathrm{th}} \\
    & - \sum_{i=0}^{n-1}\nabla\cdot\left( 2\frac{\bm{F}\left(  \bm{r}(t_{i+1}) \right)}{\Gamma}   - 2D_\mathrm{th}\nabla\log P( \bm{r}(t_{i+1}),T-t_{i+1} ) -2\bm{S}_\mathrm{L}(\bm{r}(t_{i+1}),T-t_{i+1})\right)\Delta t+ o(\Delta t)  \\
    & (\mbox{Path integral representations, the last summation is due to the variable change:}\ \{\bm{W}(t_i)\}\to\{\bm{r}(t_i)\})\\
     =&  \sum_{i=0}^{n-1}\left(\frac{ \bm{r}(t_{i+1}) -\bm{r}(t_i) }{\Delta t}  + \bm{S}_{\mathrm{L},\bm{r}(t_i)}(\bm{r}(t_{i+1}),t_{i+1}) - D_\mathrm{th}\nabla\log P( \bm{r}(t_{i+1}),T-t_{i+1} )   - \bm{S}_\mathrm{L}(\bm{r}(t_{i+1}),T-t_{i+1}) \right)   \\
     &   \cdot\left( \frac{\bm{F}\left(  \bm{r}(t_{i+1}) \right)}{\Gamma}   - D_\mathrm{th}\nabla\log P( \bm{r}(t_{i+1}),T-t_{i+1} )   - \bm{S}_\mathrm{L}(\bm{r}(t_{i+1}),T-t_{i+1})   \right)\frac{\Delta t}{D_\mathrm{th}} \\
     & - 2\sum_{i=0}^{n-1}\nabla\cdot\bm{V}(\bm{r}(t_{i+1}),T-t_{i+1})\Delta t+ o(\Delta t)  \quad \quad (\mbox{direct calculation})\\
    \overset{(1)}{=}&  \sum_{i=0}^{n-1} \frac{ \bm{r}(t_{i+1}) -\bm{r}(t_i) }{\Delta t}\cdot\left( \frac{\bm{F}\left(  \bm{r}(t_{i+1}) \right)}{\Gamma}   - D_\mathrm{th}\nabla\log P( \bm{r}(t_{i+1}),T-t_{i+1} )   - \bm{S}_\mathrm{L}(\bm{r}(t_{i+1}),T-t_{i+1})   \right)\frac{\Delta t}{D_\mathrm{th}}  \\
    & + \sum_{i=0}^{n-1}\left(  \bm{S}_{\mathrm{L},\bm{r}(t_i)}(\bm{r}(t_{i+1}),t_{i+1}) - D_\mathrm{th}\nabla\log P( \bm{r}(t_{i+1}),T-t_{i+1} )   - \bm{S}_\mathrm{L}(\bm{r}(t_{i+1}),T-t_{i+1}) \right)    \\
    &  \cdot\left( \frac{\bm{F}\left(  \bm{r}(t_{i+1}) \right)}{\Gamma}   - D_\mathrm{th}\nabla\log P( \bm{r}(t_{i+1}),T-t_{i+1} )   - \bm{S}_\mathrm{L}(\bm{r}(t_{i+1}),T-t_{i+1})   \right)\frac{\Delta t}{D_\mathrm{th}} \\
    & - 2\sum_{i=0}^{n-1}\nabla\cdot\bm{V}(\bm{r}(t_{i+1}),T-t_{i+1})\Delta t+ o(\Delta t)  \quad \quad (\mbox{direct calculation}) \\
    \overset{(2)}{=}&  \sum_{i=0}^{n-1}  (\bm{r}(t_{i+1}) -\bm{r}(t_i) )\cdot\left( \frac{\bm{F}\left( \bm{r} \right)}{\Gamma}   - D_\mathrm{th}\nabla\log P\left( \bm{r},T-t_{i+1} \right)  \right.  - \bm{S}_\mathrm{L}\left(\bm{r},T-t_{i+1}\right)   \Bigg)\bigg|_{\bm{r}=\frac{\bm{r}(t_i)+\bm{r}(t_{i+1})}{2}}\frac{1}{D_\mathrm{th}}  \\
    & + \sum_{i=0}^{n-1}\nabla\cdot\bm{V}\left(\frac{\bm{r}(t_{i+1})+\bm{r}(t_i)}{2},T-t_{i+1}\right) \frac{|\bm{r}(t_{i+1})-\bm{r}(t_i)|^2}{2D_\mathrm{th}} - 2\sum_{i=0}^{n-1}\nabla\cdot\bm{V}(\bm{r}(t_{i+1}),T-t_{i+1})\Delta t \\
    & + \sum_{i=0}^{n-1}\left(  \bm{S}_{\mathrm{L},\bm{r}(t_i)}(\bm{r}(t_{i+1}),t_{i+1}) - D_\mathrm{th}\nabla\log P( \bm{r}(t_{i+1}),T-t_{i+1} )   - \bm{S}_\mathrm{L}(\bm{r}(t_{i+1}),T-t_{i+1}) \right)  \\
    &  \cdot\left( \frac{\bm{F}\left(  \bm{r}(t_{i+1}) \right)}{\Gamma}   - D_\mathrm{th}\nabla\log P( \bm{r}(t_{i+1}),T-t_{i+1} )   - \bm{S}_\mathrm{L}(\bm{r}(t_{i+1}),T-t_{i+1})   \right)\frac{\Delta t}{D_\mathrm{th}} + o(\Delta t) .\\ & \quad(\mbox{Anti-It\^{o} sheme to Stratonovich scheme to obtain stochastic entropy})
    \end{aligned}
\end{equation} 
}The change of variable from $\{\bm{W}(t_i)\}$ to $\{\bm{r}(t_i)\}$ was conducted using the classical path integral method \cite{app:Hunt1981path}. Note that the exponential terms in the above equality $\overset{(1)}{=}$ correspond to the anti-It\^{o} integrals when $n\to\infty$. Since the definition of EP relies on a stochastic interpretation of that preserves the chain rule, which for the diffusion case corresponds to the Stratonovich scheme. Thus, we converted the above anti-It\^{o}  integrals to Stratonovich integrals in equality $\overset{(2)}{=}$. 

Recall that the probability density $P(\bm{r},T-t)$ satisfies
{\small \begin{equation}\label{app:FPET-t}
        \begin{aligned}
            &\frac{\partial_t\left( P(\bm{r}(t),T-t)\right)}{P(\bm{r}(t),T-t)}= \nabla\cdot\left[\left( \frac{\bm{F}(\bm{r}(t))}{\Gamma}   - D_\mathrm{th}\nabla\log P(\bm{r}(t),T-t)  + \bm{S}_\mathrm{L}(\bm{r}(t),T-t) \right) P(\bm{r},T-t)\right]/P(\bm{r}(t),T-t).
        \end{aligned}
    \end{equation}
}By the discrete scheme of $\bm{r}(t)$, we have the following approximate,
{\small\begin{equation}\label{app"V-2V}
    \begin{aligned}
        & \sum_{i=0}^{n-1}\nabla\cdot\bm{V}\left(\frac{\bm{r}(t_{i+1})+\bm{r}(t_i)}{2},T-t_{i+1}\right) \frac{|\bm{r}(t_{i+1})-\bm{r}(t_i)|^2}{2D_\mathrm{th}} - 2\sum_{i=0}^{n-1}\nabla\cdot\bm{V}(\bm{r}(t_{i+1}),T-t_{i+1})\Delta t \\
        =& - \sum_{i=0}^{n-1}\nabla\cdot\bm{V}\left(\frac{\bm{r}(t_{i+1})+\bm{r}(t_i)}{2},T-t_{i+1}\right)\Delta t + o(\Delta t),
    \end{aligned}
\end{equation}
}
where we have used the relation $|\Delta \bm{W}_t|^2\simeq 2D_\mathrm{th}\Delta t$. Thus, from \eqref{app:PPpf} and \eqref{app"V-2V}, we have
{\small\begin{equation}\label{app:p/psimple}
    \begin{aligned}
       & \log \frac{P[\bm{r}(t_1) \cdots \bm{r}(t_n) \mid \bm{r}_0]}{\tilde{P}^{\dagger} [ \bm{r}(t_1) \cdots \bm{r}(t_n) \mid \bm{r}_0] } \\
        =& \sum_{i=0}^{n-1} \left[ (\bm{r}(t_{i+1}) -\bm{r}(t_i) )\cdot\left( \frac{\bm{F}\left( \bm{r} \right)}{\Gamma}   - D_\mathrm{th}\nabla\log P\left( \bm{r},T-t_{i+1} \right)  \right.  - \bm{S}_\mathrm{L}\left(\bm{r},T-t_{i+1}\right)   \Bigg)\frac{1}{D_\mathrm{th}} -  \frac{\partial_t P(\bm{r},T-t)}{P(\bm{r},T-t) }\right]\bigg|_{\bm{r}=\frac{\bm{r}(t_{i+1})+\bm{r}(t_i)}{2}}\\
        & + \sum_{i=0}^{n-1}\left(  \bm{S}_{\mathrm{L},\bm{r}(t_i)}(\bm{r}(t_{i+1}),t_{i+1})  - \bm{S}_\mathrm{L}(\bm{r}(t_{i+1}),T-t_{i+1}) \right)  \\
    &  \cdot\left( \frac{\bm{F}\left(  \bm{r}(t_{i+1}) \right)}{\Gamma}   - D_\mathrm{th}\nabla\log P( \bm{r}(t_{i+1}),T-t_{i+1} )   - \bm{S}_\mathrm{L}(\bm{r}(t_{i+1}),T-t_{i+1})   \right)\frac{\Delta t}{D_\mathrm{th}} + o(\Delta t) 
    \end{aligned}
\end{equation}}

Recall that when considering the system in backward time, for any trajectory $\{\bm{r}(t)\}_{0\leq t\leq T}$, we define the ``new'' system entropy as 
\begin{equation}
    \tilde{s}_\mathrm{sys}(t)\equiv -\log \tilde{P}(\bm{r}(t),t)=-\log P(\bm{r}(t),T-t),
\end{equation}
the time derivative of the system EP is given as
{\small\begin{equation}\label{app:newentropy}
     \begin{aligned}
         \dot{\tilde{s}}_{\mathrm{sys}}(t)
         =& -\frac{\partial_t \left(P(\bm{r},T-t)\right)}{P(\bm{r},T-t) }\bigg|_{\bm{r}(t)} - \frac{\nabla P(\bm{r},T-t)}{P(\bm{r},T-t) }\bigg|_{\bm{r}(t)}\diamond \dot{\bm{r}}\\
        =& \underbrace{-\frac{\partial_t \left(P(\bm{r},T-t)\right)}{P(\bm{r},T-t) }\bigg|_{\bm{r}(t)} + \frac{\bm{J}(\bm{r},T-t)}{D_{\mathrm{th}} P(\bm{r},T-t)}\bigg|_{\bm{r}(t)} \diamond\dot{\bm{r}}}_{\dot{\tilde{s}}_\mathrm{tot}(t)} - \underbrace{\frac{\bm{F}(\bm{r})}{\Gamma D_{\mathrm{th}} }\bigg|_{\bm{r}(t)}\diamond \dot{\bm{r}}}_{\dot{\tilde{s}}_\mathrm{m}}  - \underbrace{\frac{ \int_0^1\d\theta\int \nu(\d \bm{z}) \bm{z} P(\bm{r} -\theta\bm{z},T-t)  }{D_{\mathrm{th}} P(\bm{r},T-t)}\bigg|_{\bm{r}(t)} \diamond\dot{\bm{r}}}_{\dot{\tilde{s}}_\mathrm{act}}.
     \end{aligned}
 \end{equation}
}Combining the results of \eqref{app:p/psimple} and \eqref{app:newentropy},
we obtain that:
{\small\begin{equation}\label{app:entropy1}
    \begin{aligned}
         \log \frac{P[\bm{r}(t_1) \cdots \bm{r}(t_n) \mid \bm{r}_0]}{\tilde{P}^{\dagger} [ \bm{r}(t_1) \cdots \bm{r}(t_n) \mid \bm{r}_0] }   = &   \sum_{i=0}^{n-1}\left( \dot{\tilde{s}}_\mathrm{m}(t_i) +  \dot{\tilde{s}}_\mathrm{act}(t_i) +  \dot{\tilde{s}}_\mathrm{sys}(t_i) \right)\Delta t + \sum_{i=0}^{n-1}\left( \bm{S}_{\mathrm{L},\bm{r}(t_i)}(\bm{r}(t_{i+1}),t_{i+1}) - \bm{S}_\mathrm{L}(\bm{r}(t_{i+1}),T-t_{i+1}) \right) \\
     & \cdot\left( \frac{\bm{F}\left(  \bm{r}(t_{i+1}) \right)}{\Gamma}  - D_\mathrm{th}\nabla\log P( \bm{r}(t_{i+1}),T-t_{i+1} )  - \bm{S}_\mathrm{L}(\bm{r}(t_{i+1}),T-t_{i+1})   \right)\frac{\Delta t}{D_\mathrm{th}} + o(\Delta t)  \\
     =: & \Delta\tilde{s}_\mathrm{tot} - \mathcal{Q} + o(\Delta t).
    \end{aligned}
\end{equation}}


Next, we are ready for the second part of \eqref{app:totaldecom} involving  $\tilde{P}^{\dagger}$ and $\tilde{P}$, 
{\footnotesize\begin{equation}\label{app:entropy2}
        \begin{aligned}
            &\log \frac{\tilde{P}^{\dagger}[\bm{r}(t_1) \cdots \bm{r}(t_n) \mid \bm{r}_0]}{\tilde{P} [ \bm{r}(t_1) \cdots \bm{r}(t_n) \mid \bm{r}_0] } \\
            =&   -  \sum_{i=0}^{n-1}\left|\frac{ \bm{r}(t_{i+1})  - \bm{r}(t_{i}) }{\Delta t} + \frac{\bm{F}\left(  \bm{r}(t_{i+1}) \right)}{\Gamma} \right.  + \bm{S}_{\mathrm{L},\bm{r}(t_i)}(\bm{r}(t_{i+1}),t_{i+1})   - 2D_\mathrm{th}\nabla\log P( \bm{r}(t_{i+1}),T-t_{i+1} )  - 2\bm{S}_\mathrm{L}(\bm{r}(t_{i+1}),T-t_{i+1})   \bigg|^2\frac{\Delta t}{4D_\mathrm{th}}  \\
            & + \sum_{i=0}^{n-1}\left|\frac{ \bm{r}(t_{i+1}) -\bm{r}(t_i) }{\Delta t}  +  \frac{\bm{F} \left(\bm{r}(t_{i+1}) \right)}{\Gamma}  - 2D_\mathrm{th}\nabla\log P( \bm{r}(t_{i+1}),T-t_{i+1} ) \right.  +  \tilde{\bm{S}}_{\mathrm{L},\bm{r}(t_i)}(\bm{r}(t_{i+1}), t_{i+1}) \bigg|^2\frac{\Delta t}{4D_\mathrm{th}}  \\
            & - \sum_{i=0}^{n-1}\nabla\cdot\left(  \tilde{\bm{S}}_{\mathrm{L},\bm{r}(t_i)}(\bm{r}(t_{i+1}), t_{i+1}) +2\bm{S}_\mathrm{L}(\bm{r}(t_{i+1}),T-t_{i+1}) -\bm{S}_{\mathrm{L},\bm{r}(t_i)}(\bm{r}(t_{i+1}),t_{i+1}) \right)\Delta t+ o(\Delta t)  \\
    & (\mbox{Path integral representations, the last summation is due to the variable change:}\ \{\bm{W}(t_i)\}\to\{\bm{r}(t_i)\})\\
    =&    \sum_{i=0}^{n-1}\left(\frac{ \bm{r}(t_{i+1})  - \bm{r}(t_{i}) }{\Delta t} + \frac{\bm{F}\left(  \bm{r}(t_{i+1}) \right)}{\Gamma}  + \frac{\bm{S}_{\mathrm{L},\bm{r}(t_i)}(\bm{r}(t_{i+1}),t_{i+1}) }{2}  - 2D_\mathrm{th}\nabla\log P( \bm{r}(t_{i+1}),T-t_{i+1} ) - \bm{S}_\mathrm{L}(\bm{r}(t_{i+1}),T-t_{i+1})  \right. \\
    &\left.  +\frac{ \tilde{\bm{S}}_{\mathrm{L},\bm{r}(t_i)}(\bm{r}(t_{i+1}), t_{i+1}) }{2} \right)\cdot\left( \bm{S}_\mathrm{L}(\bm{r}(t_{i+1}),T-t_{i+1}) - \frac{\bm{S}_{\mathrm{L},\bm{r}(t_i)}(\bm{r}(t_{i+1}),t_{i+1}) }{2} + \frac{ \tilde{\bm{S}}_{\mathrm{L},\bm{r}(t_i)}(\bm{r}(t_{i+1}), t_{i+1}) }{2} \right)\frac{\Delta t}{D_\mathrm{th}}\\
    & - \sum_{i=0}^{n-1}\nabla\cdot\left(  \tilde{\bm{S}}_{\mathrm{L},\bm{r}(t_i)}(\bm{r}(t_{i+1}), t_{i+1}) +2\bm{S}_\mathrm{L}(\bm{r}(t_{i+1}),T-t_{i+1}) -\bm{S}_{\mathrm{L},\bm{r}(t_i)}(\bm{r}(t_{i+1}),t_{i+1}) \right)\Delta t+ o(\Delta t). \quad(\mbox{direct calculation})
        \end{aligned}
    \end{equation} }

Finally, combining \eqref{app:entropy1} and \eqref{app:entropy2} we immediately obtain that
    \begin{equation}\label{app:p/tildep}
        \begin{aligned}
     \log \frac{P[\bm{r}(t_1) \cdots \bm{r}(t_n) \mid \bm{r}_0]}{\tilde{P} [ \bm{r}(t_1) \cdots \bm{r}(t_n) \mid \bm{r}_0] }
     \simeq \sum_{i=0}^{n-1}\left( \dot{\tilde{s}}_\mathrm{m}(t_i) +  \dot{\tilde{s}}_\mathrm{act}(t_i) +  \dot{\tilde{s}}_\mathrm{sys}(t_i) \right)\Delta t - B_{\mathrm{act}}(\{\bm{r}(t_i)\}_{i=0}^n),
        \end{aligned}
    \end{equation}
where
{\small\begin{equation}\label{Bact1}
    \begin{aligned}
        &B_{\mathrm{act}}(\{\bm{r}(t_i)\}_{i=0}^n) 
        = -\sum_{i=0}^{n-1}\left( \frac{ \bm{r}(t_{i+1})  - \bm{r}(t_{i}) }{\Delta t} -D_\mathrm{th}\nabla\log P( \bm{r}(t_{i+1}),T-t_{i+1} ) + \frac{\bm{S}_{\mathrm{L},\bm{r}(t_i)}(\bm{r}(t_{i+1}),t_{i+1}) }{2}  \right.\\
     & \left. + \frac{ \tilde{\bm{S}}_{\mathrm{L},\bm{r}(t_i)}(\bm{r}(t_{i+1}), t_{i+1}) }{2} \right) \cdot\left( \bm{S}_\mathrm{L}(\bm{r}(t_{i+1}),T-t_{i+1}) - \frac{\bm{S}_{\mathrm{L},\bm{r}(t_i)}(\bm{r}(t_{i+1}),t_{i+1}) }{2} + \frac{ \tilde{\bm{S}}_{\mathrm{L},\bm{r}(t_i)}(\bm{r}(t_{i+1}), t_{i+1}) }{2} \right)\frac{\Delta t}{D_\mathrm{th}}\\
     & - \sum_{i=0}^{n-1}\bm{V}(\bm{r}(t_{i+1}),T-t_{i+1})\cdot\left( \frac{\bm{S}_{\mathrm{L},\bm{r}(t_i)}(\bm{r}(t_{i+1}),t_{i+1}) }{2} + \frac{ \tilde{\bm{S}}_{\mathrm{L},\bm{r}(t_i)}(\bm{r}(t_{i+1}), t_{i+1}) }{2} \right)\frac{\Delta t}{D_\mathrm{th}}\\
     & - \sum_{i=0}^{n-1}\nabla\cdot\left(  \tilde{\bm{S}}_{\mathrm{L},\bm{r}(t_i)}(\bm{r}(t_{i+1}), t_{i+1}) +2\bm{S}_\mathrm{L}(\bm{r}(t_{i+1}),T-t_{i+1}) -\bm{S}_{\mathrm{L},\bm{r}(t_i)}(\bm{r}(t_{i+1}),t_{i+1}) \right)\Delta t .
    \end{aligned}
\end{equation}}
Such representation is complicated. In the next section, we show another expression of it using Girsanov theorem.

\section{\label{app:FT2}Proof of the fluctuation theorem: continuous viewpoint}

The purpose of this appendix is to clarify the definition of $B_{\mathrm{act}}$ and to derive an alternative representation of it using the Girsanov theorem. We further demonstrate that the resulting expression is fully consistent with the path–integral formulation developed in the preceding Appendix \ref{app:FT}.

The ratio of path measures can be evaluated using the Girsanov transform. Now we present the approach here. Recall the original process $\bm{r}(t)$ \eqref{eqn:SDEforward}, its reversed flow process $\bm{\tilde{r}}^{\dagger}(t)$ \eqref{eqn:SDEreversalflow} and its time reversal process $\tilde{r}(t)$ \eqref{eqn:SDEreversal2}. For clearness, the equations they satisfy are listed again, respectively,
{\small\begin{equation}
    \begin{aligned}
         \d\bm{r}(t) =&\frac{\bm{F}(\bm{r}(t))}{\Gamma}\d t  +  \d \bm{W}_t + \int\bm{z}\mathcal{N}(\d t,\d\bm{z}),\quad r(0)\sim P_0(\bm{r}),\\
         \d\bm{\tilde{r}}^{\dagger}(t) =& \left( \frac{\bm{-F}(\bm{\tilde{r}}^{\dagger}(t) )}{\Gamma}   + 2D_\mathrm{th}\nabla\log P(\bm{\tilde{r}}^{\dagger}(t),T-t)   - 2\int_0^1\d\theta \int\nu(\d\bm{z})\bm{z}\frac{P(\bm{\tilde{r}}^{\dagger}(t) -\theta\bm{z},T-t)}{P(\bm{\tilde{r}}^{\dagger}(t),T-t)} \right)\d t\\
        & +  \d\bm{W}_t + \int\bm{z}\mathcal{N}(\d t,\d\bm{z}),\quad
        \bm{\tilde{r}}^{\dagger}(0)\sim  \tilde{P}^{\dagger}(\bm{r},0)= P(\bm{r},T),\\
        \d\bm{\tilde{r}}(t) =&-\frac{\bm{F}(\bm{\tilde{r}}(t))}{\Gamma}\d t + 2D_\mathrm{th}\nabla\log P(\bm{\tilde{r}}(t),T-t )\d t +  \d \bm{W}_t + \int\bm{z}\tilde{\mathcal{N}}(\d t,\d\bm{z}),\
    \bm{\tilde{r}}(0) \sim  \tilde{P}(\bm{r},0)=P(\bm{r},T).
    \end{aligned}
\end{equation}
}Consider the c\`{a}dl\`{a}g path space $D[0,T]$ (all paths on time interval $[0,T]$ that are right continuous with left limit). The process $\bm{r}$, $\bm{\tilde{r}}^{\dagger}$ and $\bm{\tilde{r}}$ induce path measure $P$, $\tilde{P}^{\dagger}$ and $\tilde{P}$ respectively on $D[0,T]$. Recall that, for a trajectory $\bm{r}$,
{\small\begin{equation}\label{app:totalP/P}
    \begin{aligned}
        \log\frac{ P(\bm{r})}{ \tilde{P}(\bm{r})}=\log\frac{  P(\bm{r}\mid\bm{r}(0))P_0(\bm{r}(0))}{ \tilde{P}(\bm{r}\mid\bm{r}(0))P_T(\bm{r}(0))}= \log\frac{\d P(\bm{r}\mid\bm{r}(0))}{\d \tilde{P}^{\dagger}(\bm{r}\mid\bm{r}(0))} + \log\frac{\d \tilde{P}^{\dagger}(\bm{r}\mid\bm{r}(0))}{\d \tilde{P}(\bm{r}\mid\bm{r}(0))} + \log\frac{P_0(\bm{r}(0))}{P_T(\bm{r}(0))}.
    \end{aligned}
\end{equation}
}According to the Girsanov theorem (see e.g. \cite[Chapter 1, Theorem 1.4]{app:Ishikawa2023stochastic} or \cite[Theorem 2.4]{app:Fujisaki2021generalized}), we have
{\small\begin{equation}\label{app:logp/pfgir}
    \begin{aligned}
        &-\left\langle\log\frac{\d \tilde{P}^{\dagger}(\bm{r}\mid\bm{r}(0))}{\d P(\bm{r}\mid\bm{r}(0))} \right\rangle  
        =  \left\langle \frac{1}{4D_\mathrm{th}}\int_0^T \left| \frac{2\bm{F}(\bm{r} )}{\Gamma}   - 2D_\mathrm{th}\nabla\log P(\bm{r} ,T-t)   + 2\int_0^1\d\theta \int\nu(\d\bm{z})\bm{z}\frac{P(\bm{r}-\theta\bm{z},T-t)}{P( \bm{r} ,T-t)} \right|^2\d t \right\rangle\\
        & + \left\langle\frac{1}{2D_\mathrm{th}}\int_0^T\left(\frac{2\bm{F}(\bm{r} )}{\Gamma}   - 2D_\mathrm{th}\nabla\log P(\bm{r} ,T-t)   + 2\int_0^1\d\theta \int\nu(\d\bm{z})\bm{z}\frac{P(\bm{r}-\theta\bm{z},T-t)}{P( \bm{r} ,T-t)} \right)\cdot\d\bm{W}_t\right\rangle\\
        &\quad\quad(\mbox{Girsanov theorem})\\
        =& \int_0^T\int\frac{1}{D_\mathrm{th}}|\bm{V}(\bm{r},T-t)|^2P(\bm{r},t)\d\bm{r}\d t +   \left\langle\int_0^T\frac{1}{D_\mathrm{th}}\bm{V}(\bm{r}(t),T-t) \cdot\d\bm{W}_t\right\rangle\quad\quad(\mbox{definition of } \bm{V}(\bm{r},T-t))\\
        =& \int_0^T\int\frac{1}{D_\mathrm{th}}|\bm{V}(\bm{r},T-t)|^2P(\bm{r},t)\d\bm{r}\d t,\quad\quad(\mbox{It\^{o} integral has zero mean})
    \end{aligned}
\end{equation}
}where we use the fact that, $\bm{W}_t$ is a standard Brownian motion under the measure $P$ induced by $\bm{r}(t)$, thus the expectation of the It\^{o} integral vanishes. This result coincides with \eqref{app:PPpf} obtained by path integral approach, which is verified as follows. By the second equality of {\eqref{app:PPpf},
{\footnotesize\begin{equation}\label{app:discreteP/Pf}
    \begin{aligned}
    & \left\langle\log \frac{P[\bm{r}(t_1) \cdots \bm{r}(t_n) \mid \bm{r}_0]}{\tilde{P}^{\dagger} [ \bm{r}(t_1) \cdots \bm{r}(t_n) \mid \bm{r}_0] }\right\rangle \\
    \simeq&  \left\langle\sum_{i=0}^{n-1}\left(\frac{ \bm{r}(t_{i+1}) -\bm{r}(t_i) }{\Delta t}  + \bm{S}_{\mathrm{L},\bm{r}(t_i)}(\bm{r}(t_{i+1}),t_{i+1}) - D_\mathrm{th}\nabla\log P( \bm{r}(t_{i+1}),T-t_{i+1} )   - \bm{S}_\mathrm{L}(\bm{r}(t_{i+1}),T-t_{i+1}) \right) \right.  \\
    &  \left. \cdot\left( \frac{\bm{F}\left(  \bm{r}(t_{i+1}) \right)}{\Gamma}   - D_\mathrm{th}\nabla\log P( \bm{r}(t_{i+1}),T-t_{i+1} )   - \bm{S}_\mathrm{L}(\bm{r}(t_{i+1}),T-t_{i+1})   \right)\frac{\Delta t}{D_\mathrm{th}}  - 2\sum_{i=0}^{n-1}\nabla\cdot\bm{V}(\bm{r}(t_{i+1}),T-t_{i+1})\Delta t\right\rangle\\
    &\quad (\mbox{by the second equality of \eqref{app:PPpf}})\\
    =& \left\langle\sum_{i=0}^{n-1}\left(\frac{ \bm{r}(t_{i+1}) -\bm{r}(t_i) }{\Delta t} -\frac{\bm{F}(\bm{r}(t_{i+1}))}{\Gamma} + \bm{S}_{\mathrm{L},\bm{r}(t_i)}(\bm{r}(t_{i+1}),t_{i+1}) + \bm{V}(\bm{r}(t_{i+1},T-t_{i+1})) \right)   \cdot \bm{V}(\bm{r}(t_{i+1},T-t_{i+1}))  \frac{\Delta t}{D_\mathrm{th}}  \right\rangle\\
    &  -\left\langle  2\sum_{i=0}^{n-1}\nabla\cdot\bm{V}(\bm{r}(t_{i+1}),T-t_{i+1})\Delta t\right\rangle\quad\quad(\mbox{direct calculation})\\
    \simeq& \left\langle\sum_{i=0}^{n-1} \frac{ \bm{W} (t_{i+1})-\bm{W}(t_i)}{\Delta t}\cdot \bm{V}(\bm{r}(t_{i},T-t_{i+1}))  \frac{\Delta t}{D_\mathrm{th}}  \right\rangle + \left\langle\sum_{i=0}^{n-1} |\bm{V}(\bm{r}(t_{i+1},T-t_{i+1}))|^2  \frac{\Delta t}{D_\mathrm{th}}  \right\rangle\\
    &\quad (\mbox{anti-It\^{o} to It\^{o}, and omit the higher order terms})\\
    \simeq & \left\langle\int_0^T\frac{1}{D_\mathrm{th}}\bm{V}(\bm{r}(t),T-t)\cdot\d\bm{W}_t\right\rangle + \left\langle\int_0^T\frac{1}{D_\mathrm{th}}|\bm{V}(\bm{r}(t),T-t)|^2\d t\right\rangle\quad\quad(\mbox{discrete version to continuous version})\\
    =& \int_0^T\int\frac{1}{D_\mathrm{th}}|\bm{V}(\bm{r},T-t)|^2P(\bm{r},t)\d\bm{r}\d t.\quad\quad(\mbox{It\^{o} integral has zero mean})
    \end{aligned}
\end{equation} 
}Here we have used the first discrete scheme of \eqref{app:rd1} to obtain the It\^{o} integral whose expectation is zero. This formulation for calculating the EP under averaging on all ensembles of $\bm{r}(t)$ is simpler than the way using stochastic entropy, i.e.,  \eqref{app:newentropy}, since the latter needs to calculate the expectation of the partial time of $P(\bm{r},t)$. On the other hand, the above calculation leads to another expression for $\mathcal{Q}$ as follows.
{\small\begin{equation}\label{app:continuouslogp/pf} 
    \begin{aligned}
    & \log \frac{P[\bm{r}(t_1) \cdots \bm{r}(t_n) \mid \bm{r}_0]}{\tilde{P}^{\dagger} [ \bm{r}(t_1) \cdots \bm{r}(t_n) \mid \bm{r}_0] }  
    \simeq \int_0^T\frac{1}{D_\mathrm{th}}\bm{V}(\bm{r}(t),T-t)\cdot\d\bm{W}_t +  \int_0^T\frac{1}{D_\mathrm{th}}|\bm{V}(\bm{r}(t),T-t)|^2\d t \quad\quad(\mbox{Eq.\eqref{app:logp/pfgir} or Eq. \eqref{app:discreteP/Pf}})\\
    =& \int_0^T\frac{1}{D_\mathrm{th}}\bm{V}(\bm{r}(t),T-t)\diamond\d\bm{W}_t - \int_0^T\nabla\bm{V}(\bm{r}(t),T-t)\d t +  \int_0^T\frac{1}{D_\mathrm{th}}|\bm{V}(\bm{r}(t),T-t)|^2\d t \quad\quad (\mbox{It\^{o} to Marcus})\\
    =&  \int_0^T\frac{1}{D_\mathrm{th}}\bm{V}(\bm{r}(t),T-t)\diamond\left(\d\bm{r}(t) - \frac{\bm{F}(\bm{r}(t))}{\Gamma}\d t - \int\bm{z}\mathcal{N}(\d t,\d\bm{z})\right) - \int_0^T\nabla\bm{V}(\bm{r}(t),T-t)\d t +  \int_0^T\frac{|\bm{V}(\bm{r}(t),T-t)|^2}{D_\mathrm{th}}\d t \\
    &\quad (\mbox{the SDE of $\bm{r}(t): \d\bm{W}_t=\d\bm{r}(t) - \frac{\bm{F}(\bm{r}(t))}{\Gamma}\d t - \int\bm{z}\mathcal{N}(\d t,\d\bm{z})$})\\
    =& \int_0^T\frac{1}{D_\mathrm{th}}\bm{V}(\bm{r}(t),T-t)\diamond\d\bm{r}(t) - \int_0^T\nabla\cdot\bm{V}(\bm{r}(t),T-t)\d t -\int_0^T\bm{V}(\bm{r}(t),T-t)\cdot\nabla\log P(\bm{r}(t),T-t)\d t\\
    & - \frac{1}{D_\mathrm{th}}\bm{V}(\bm{r}(t),T-t)\diamond\int_0^T\left( \int\bm{z}\mathcal{N}(\d t,\d\bm{z}) +  \bm{S}_\mathrm{L}(\bm{r}(t),T-t)\d t\right)\quad\quad(\mbox{direct calculation})\\
    =& \Delta \tilde{s}_\mathrm{tot} -  \int_0^T\frac{\bm{V}(\bm{r},T-t)}{D_\mathrm{th}}\diamond\left( \int\bm{z}\mathcal{N}(\d t,\d\bm{z}) +  \bm{S}_\mathrm{L}(\bm{r}(t),T-t)\d t\right). \quad(\mbox{the total entropy production in \eqref{app:newentropy}})
    \end{aligned}
\end{equation} 
}This coincides with \eqref{app:entropy1}, since the active fluctuation $\int\bm{z}\mathcal{N}(\d t,\d\bm{z})$ conditioned at $\bm{r}(t_i)$ has the equivalent local force $-\bm{S}_{\mathrm{L},\bm{r}(t_i)}(\bm{r},t)$ in the view of probability flow as shown in \eqref{app:rd1} and \eqref{eqn:shortestimate}.

By \eqref{app:totalP/P} and \eqref{app:continuouslogp/pf}, for a trajectory $\{\bm{x}(t)\}_{0\leq t\leq T}$, we derive
{\small\begin{align}\label{app:Bact=Q+P}
    \begin{aligned}
        B_{\mathrm{act}}(\{\bm{x}(t)\}_{0\leq t\leq T})
        =& \int_0^T\d t \frac{\bm{V}(\bm{x}(t), T-t)}{D_\mathrm{th}} \diamond \bigg( \bm{\eta}_\mathrm{act}(t)  + \bm{S}_\mathrm{L}(\bm{x}(t), T-t) \bigg) + \left( - \log \frac{\tilde{P}^{\dagger}\left[ \{\bm{x}(t)\}_{0\leq t\leq T} \mid \bm{x}_0\right]}{\tilde{P}\left[ \{\bm{x}(t)\}_{0\leq t\leq T} \mid \bm{x}_0\right]}\right)+ \log\frac{ P_0(\bm{x}(0))}{ P_T(\bm{x}(0))}\\
        \equiv& \mathcal{Q}(\{\bm{x}(t)\}_{0\leq t\leq T}) + \mathcal{P}(\{\bm{x}(t)\}_{0\leq t\leq T}) + \log\frac{ P_0(\bm{x}(0))}{ P_T(\bm{x}(0))}.
    \end{aligned}
\end{align}}

Furthermore, according to the Girsanov theorem \cite[Theorem 2.4]{app:Fujisaki2021generalized}, see also \cite{app:zhang2025entropy}, we have
{\footnotesize\begin{equation}\label{app:p/tildep2}
    \begin{aligned}
        &- \log\frac{\d \tilde{P} (\bm{r}\mid \bm{r}(0))}{\d P(\bm{r}\mid \bm{r}(0))}  
        = \frac{1}{4D_\mathrm{th}}\int_0^T \left| \frac{2\bm{F}(\bm{r}(t) )}{\Gamma}   - 2D_\mathrm{th}\nabla\log P(\bm{r}(t) ,T-t)     \right|^2\d t  \\
        & +  \frac{1}{ 2D_\mathrm{th}}\int_0^T\left(\frac{2\bm{F}(\bm{r}(t) )}{\Gamma}   - 2D_\mathrm{th}\nabla\log P(\bm{r}(t) ,T-t)    \right)\cdot\d\bm{W}_t \\
        & -  \int_0^T\int \left[\log\left( \frac{P(\bm{r}(t)+\bm{z},T-t)\nu(-\bm{z})}{P(\bm{r}(t),T-t)\nu(\bm{z})}\right)\mathcal{N}(\d t,\d\bm{z}) -\left(  \frac{P(\bm{r}(t)+\bm{z},T-t)\nu(-\bm{z})}{P(\bm{r}(t),T-t)\nu(\bm{z})}-1\right)\nu(\d \bm{z})\d t\right] \quad(\mbox{Girsanov theorem})\\
        =&   \frac{1}{D_\mathrm{th}}\int_0^T \left| \frac{\bm{F}(\bm{r}(t) )}{\Gamma}   - D_\mathrm{th}\nabla\log P(\bm{r} ,T-t)     \right|^2\d t -\int_0^T\nabla\cdot\left(\frac{\bm{F}(\bm{r}(t) ) }{\Gamma}   - D_\mathrm{th}\nabla\log P(\bm{r} ,T-t)\right)\d t \\
        & +  \frac{1}{ D_\mathrm{th}}\int_0^T\left(\frac{\bm{F}(\bm{r}(t) )}{\Gamma}   - D_\mathrm{th}\nabla\log P(\bm{r} ,T-t)    \right)\diamond \left(\d\bm{r}(t) -\frac{\bm{F}(\bm{r}(t))}{\Gamma}\d t - \int\bm{z}\mathcal{N}(\d t,\d\bm{z})\right)\\
        & -  \int_0^T\int \left[\log\left( \frac{P(\bm{r}(t)+\bm{z},T-t)\nu(-\bm{z})}{P(\bm{r}(t),T-t)\nu(\bm{z})}\right)\mathcal{N}(\d t,\d\bm{z}) -\left(  \frac{P(\bm{r}(t)+\bm{z},T-t)\nu(-\bm{z})}{P(\bm{r}(t),T-t)\nu(\bm{z})}-1\right)\nu(\d \bm{z})\d t\right] \quad(\mbox{It\^{o} to Marcus})\\
        =&  \frac{1}{D_\mathrm{th}}\int_0^T \left(   -\Bigg( \bm{V}(\bm{r}(t),T-t) + \bm{S}_\mathrm{L}(\bm{r}(t),T-t)     \right)\cdot D_\mathrm{th}\nabla\log P(\bm{r}(t),T-t)\d t \\
        & - \left( \bm{V}(\bm{r}(t),T-t)  + \bm{S}_\mathrm{L}(\bm{r}(t),T-t)     \right)\diamond\int\bm{z}\mathcal{N}(\d t,\d\bm{z})\Bigg)\\
        & -\int_0^T\nabla\cdot\left(\bm{V}(\bm{r}(t),T-t) + \bm{S}_\mathrm{L}(\bm{r}(t),T-t) \right)\d t  +  \frac{1}{ D_\mathrm{th}}\int_0^T\left(\bm{V}(\bm{r}(t),T-t)  + \bm{S}_\mathrm{L}(\bm{r}(t),T-t)   \right)\diamond  \d\bm{r}(t) \\
        & -  \int_0^T\int \left[\log\left( \frac{P(\bm{r}(t)+\bm{z},T-t)\nu(-\bm{z})}{P(\bm{r}(t),T-t)\nu(\bm{z})}\right)\mathcal{N}(\d t,\d\bm{z}) -\left(  \frac{P(\bm{r}(t)+\bm{z},T-t)\nu(-\bm{z})}{P(\bm{r}(t),T-t)\nu(\bm{z})}-1\right)\nu(\d \bm{z})\d t\right] \\
        &\quad\quad(\mbox{using the definition of $\bm{V}(\bm{r}(t),T-t)$ and open the square term and direct calculation})\\
        =& \int_0^T\frac{1}{D_\mathrm{th}}\bm{V}(\bm{r}(t),T-t)\diamond\d\bm{r}(t) - \int_0^T\nabla\cdot\bm{V}(\bm{r}(t),T-t)\d t -\int_0^T\bm{V}(\bm{r}(t),T-t)\cdot\nabla\log P(\bm{r},T-t)\d t\\
        & + \frac{1}{D_\mathrm{th}}\int_0^T  \left( -\bm{S}_\mathrm{L}(\bm{r}(t),T-t) \cdot D_\mathrm{th}\nabla\log P(\bm{r}(t),T-t)\d t - \bm{S}_\mathrm{L}(\bm{r}(t),T-t) \diamond\int\bm{z}\mathcal{N}(\d t,\d\bm{z})\right) \\
        & -\frac{1}{D_\mathrm{th}}\int_0^T\bm{V}(\bm{r}(t),T-t)\diamond\int\bm{z}\mathcal{N}(\d t,\d\bm{z}) - \int_0^T\nabla\cdot\bm{S}_\mathrm{L}(\bm{r}(t),T-t) \d t +  \frac{1}{ D_\mathrm{th}}\int_0^T \bm{S}_\mathrm{L}(\bm{r}(t),T-t)\diamond  \d\bm{r}(t)\\
        & -  \int_0^T\int \left[\log\left( \frac{P(\bm{r}(t)+\bm{z},T-t)\nu(-\bm{z})}{P(\bm{r}(t),T-t)\nu(\bm{z})}\right)\mathcal{N}(\d t,\d\bm{z}) -\left(  \frac{P(\bm{r}(t)+\bm{z},T-t)\nu(-\bm{z})}{P(\bm{r}(t),T-t)\nu(\bm{z})}-1\right)\nu(\d \bm{z})\d t\right] \quad(\mbox{direct calculation})\\
        =& \Delta\tilde{s}_\mathrm{tot} + \frac{1}{D_\mathrm{th}}\int_0^T  \left( -\bm{S}_\mathrm{L}(\bm{r}(t),T-t) \cdot D_\mathrm{th}\nabla\log P(\bm{r}(t),T-t)\d t - \bm{S}_\mathrm{L}(\bm{r}(t),T-t) \diamond\int\bm{z}\mathcal{N}(\d t,\d\bm{z})\right)\d t \\
        & -\frac{1}{D_\mathrm{th}}\int_0^T\bm{V}(\bm{r}(t),T-t)\diamond\int\bm{z}\mathcal{N}(\d t,\d\bm{z}) - \int_0^T\nabla\cdot\bm{S}_\mathrm{L}(\bm{r}(t),T-t) \d t +  \frac{1}{ D_\mathrm{th}}\int_0^T \bm{S}_\mathrm{L}(\bm{r}(t),T-t)\diamond  \d\bm{r}(t)\\
        & -  \int_0^T\int \left[\log\left( \frac{P(\bm{r}(t)+\bm{z},T-t)\nu(-\bm{z})}{P(\bm{r}(t),T-t)\nu(\bm{z})}\right)\mathcal{N}(\d t,\d\bm{z}) -\left(  \frac{P(\bm{r}(t)+\bm{z},T-t)\nu(-\bm{z})}{P(\bm{r}(t),T-t)\nu(\bm{z})}-1\right)\nu(\d \bm{z})\d t\right]. \quad(\mbox{definition of $\Delta\tilde{s}_\mathrm{tot}$})
    \end{aligned}
\end{equation}
}

On the other hand, to obtain the expectation of $B_\mathrm{act}$, we know that,
{\footnotesize\begin{equation}
    \begin{aligned}
        &-\left\langle\log\frac{\d \tilde{P}(\bm{r}\mid\bm{r}(0))}{\d P(\bm{r}\mid\bm{r}(0))} \right\rangle  
        = \left\langle \frac{1}{4D_\mathrm{th}}\int_0^T \left| \frac{2\bm{F}(\bm{r} )}{\Gamma}   - 2D_\mathrm{th}\nabla\log P(\bm{r} ,T-t)     \right|^2\d t \right\rangle  + \left\langle\frac{1}{ 2D_\mathrm{th}}\int_0^T\left(\frac{2\bm{F}(\bm{r} )}{\Gamma}   - 2D_\mathrm{th}\nabla\log P(\bm{r} ,T-t)    \right)\cdot\d\bm{W}_t\right\rangle\\
        & - \left\langle\int_0^T\int \left[\log\left( \frac{P(\bm{r}+\bm{z},T-t)\nu(-\bm{z})}{P(\bm{r},T-t)\nu(\bm{z})}\right)\mathcal{N}(\d t,\d\bm{z}) -\left(  \frac{P(\bm{r}+\bm{z},T-t)\nu(-\bm{z})}{P(\bm{r},T-t)\nu(\bm{z})}-1\right)\nu(\d \bm{z})\d t\right]\right\rangle \quad\quad(\mbox{Girsanov theorem})\\
         =&   \frac{1}{D_\mathrm{th}}\int_0^T\int \left| \frac{\bm{F}(\bm{r} )}{\Gamma}   - D_\mathrm{th}\nabla\log P(\bm{r} ,T-t)     \right|^2P(\bm{r},t)\d\bm{r}\d t  \\
        &  - \left\langle\int_0^T\int \left[\log\left( \frac{P(\bm{r}+\bm{z},T-t)\nu(-\bm{z})}{P(\bm{r},T-t)\nu(\bm{z})}\right)\mathcal{N}(\d t,\d\bm{z}) -\left(  \frac{P(\bm{r}+\bm{z},T-t)\nu(-\bm{z})}{P(\bm{r},T-t)\nu(\bm{z})}-1\right)\nu(\d \bm{z})\d t\right]\right\rangle \quad(\mbox{It\^{o} integral has zero mean})\\
        =& \frac{1}{D_\mathrm{th}}\int_0^T\int|\bm{V}(\bm{r},T-t)|^2P(\bm{r},t)\d\bm{r}\d t - \frac{2}{D_\mathrm{th}}\int_0^T\int\bm{V}(\bm{r},T-t)\cdot\bm{S}_\mathrm{L}(\bm{r},T-t)P(\bm{r},t)\d\bm{r}\d t\\
        & + \frac{1}{D_\mathrm{th}}\int_0^T\int|\bm{S}(\bm{r},T-t)|^2P(\bm{r},t)\d\bm{r}\d t\\
        &  - \int_0^T\int \left[\log\left( \frac{P(\bm{r}+\bm{z},T-t)\nu(-\bm{z})}{P(\bm{r},T-t)\nu(\bm{z})}\right) -\left(  \frac{P(\bm{r}+\bm{z},T-t)\nu(-\bm{z})}{P(\bm{r},T-t)\nu(\bm{z})}-1\right)\right]\nu(\d \bm{z})\d t.\\
        &\quad\quad(\mbox{definition of the velocity field }\bm{V}(\bm{r},T-t),\ \mathcal{N}\mbox{ is a Poisson random measure with L\'{e}vy measure } \nu \mbox{ under } P)
    \end{aligned}
\end{equation}}

Recall \eqref{app:continuouslogp/pf} and \eqref{app:<V>}, we have
{\small\begin{equation}
    \langle \mathcal{Q}\rangle=  \frac{1}{D_\mathrm{th}}\int_0^T\int\bm{V}(\bm{r},T-t)\cdot\left(  \bm{S}_\mathrm{L}(\bm{r} ,T-t) - \bm{S}_\mathrm{L}(\bm{r} ,t)  \right)P(\bm{r},t)\d\bm{r}\d t.
\end{equation}
}Thus, the expectation $\left\langle B_\mathrm{act} \right\rangle$ is given by
{\small\begin{equation}\label{app:EBact}
    \begin{aligned}
        \left\langle B_\mathrm{act} \right\rangle =&  -\left(\left\langle\log\frac{\d P}{\d \tilde{P}}(\bm{r}) \right\rangle - \left\langle\log\frac{\d P}{\d \tilde{P}^{\dagger}}(\bm{r})  \right\rangle\right) + \langle \mathcal{Q}\rangle + \left\langle\frac{P_0(\bm{r}(0))}{P_T(\bm{r}(0))}\right\rangle\\
        =&  \frac{2}{D_\mathrm{th}}\int_0^T\int\bm{V}(\bm{r},T-t)\cdot\bm{S}_\mathrm{L}(\bm{r},T-t)P(\bm{r},t)\d\bm{r}\d t- \frac{1}{D_\mathrm{th}}\int_0^T\int|\bm{S}_\mathrm{L}(\bm{r},T-t)|^2P(\bm{r},t)\d\bm{r}\d t\\
       & + \frac{1}{D_\mathrm{th}}\int_0^T\int\bm{V}(\bm{r},T-t)\cdot\left(   \bm{S}_\mathrm{L}(\bm{r} ,T-t)  -\bm{S}_\mathrm{L}(\bm{r} ,t) \right)P(\bm{r},t)\d\bm{r}\d t + \left\langle\frac{P_0(\bm{r}(0))}{P_T(\bm{r}(0))}\right\rangle\\
        & +  \int_0^T\int\int\left[\log\left( \frac{P(\bm{r}+\bm{z},T-t)\nu(-\bm{z})}{P(\bm{r},T-t)\nu(\bm{z})}\right) -\left(  \frac{P(\bm{r}+\bm{z},T-t)\nu(-\bm{z})}{P(\bm{r},T-t)\nu(\bm{z})}-1\right)\right]P(\bm{r},t)\d\bm{r}\nu(\d \bm{z})\d t .
    \end{aligned}
\end{equation}}

We note that it is more practical to employ the first equality in \eqref{app:p/tildep2} to evaluate the quantity 
$\mathcal{R}\equiv \Delta \tilde{s}_{\mathrm{tot}}-B_{\mathrm{act}}$, or to obtain 
$B_{\mathrm{act}}= \Delta \tilde{s}_{\mathrm{tot}}-\mathcal{R} $.  
This approach involves only It\^{o} integrals, which are numerically more accessible than the corresponding, more intricate Marcus integrals.

\section{Recovering Gaussian stochastic thermodynamics and beyond}
This appendix demonstrates that  our framework   recovers the standard results in classical stochastic thermodynamics and, moreover, covers naturally   other classes of active–matter systems, including Active Ornstein–Uhlenbeck (Active OU) dynamics.

Let us now demonstrate how our results recover the classical stochastic thermodynamics framework. By setting $\bm{\eta}_\mathrm{act} \equiv 0$ in Eq.~\eqref{eqn:SDE}, i.e., the system is
\begin{equation}\label{app:Gaussiansde}
    \frac{\d \bm{r}(t)}{\d t} = \frac{\bm{F}(\bm{r}(t))}{\Gamma} + \bm{\eta}_{\mathrm{th}}(t)  ,\quad t>0.
\end{equation}
It is straightforward to see that the process $\bm{\tilde{r}}^{\dagger}(t)$ is identical to the reversed process $\bm{\tilde{r}}(t)$, as Eqs.~\eqref{eqn:SDEreversed} and \eqref{appeqn:SDEreverse} are equivalent in this case. This means that, $\bm{\tilde{r}}^{\dagger}(t) = \bm{\tilde{r}}(t) = \bm{r}(T-t)$, and all represent the solution process to the following Langevin equation:
 \begin{equation}
        \begin{aligned}
            \frac{\d\bm{\tilde{r}}(t)}{\d t} =& \frac{\bm{F}(\bm{\tilde{r}}(t)) }{\Gamma} -2\bm{V}(\bm{\tilde{r}}(t),T-t) + \bm{\eta}_\mathrm{th}(t), \\
            =& -\frac{\bm{F}(\bm{\tilde{r}}(t)) }{\Gamma} +  2D_\mathrm{th}\log P(\bm{\tilde{r}}(t),T-t) + \bm{\eta}_\mathrm{th}(t), \\
         \bm{\tilde{r}} (0) \sim& \tilde{P} (\bm{r},0)=P(\bm{r},T),\quad t:0\to T, 
        \end{aligned}
    \end{equation}
where the vector field is
\begin{equation*}
    \bm{V}(\bm{r},T-t)=\frac{\bm{F}(\bm{r})}{\Gamma} - D_{\mathrm{th}} \nabla\log P(\bm{r},T-t)=\bm{J} (\bm{r},T-t)/P(\bm{r},T-t). 
\end{equation*}
Now the ratio of the path density of process $\bm{r}(t)$ and $\bm{\tilde{r}}(t)$ is
\begin{equation}
    \begin{aligned}
        \frac{ P\left[\{\bm{x}(t)\}_{t\in[o,T]}\mid\bm{x}_0\right]}{\tilde{P}\left[\{\bm{x}(t)\}_{t\in[o,T]}\mid\bm{x}_0\right]}=\exp\left\{ -\int_0^TL(\bm{x}(t),\dot{\bm{x}}(t))\d t +\int_0^T\tilde{L}(\bm{x}(t),\dot{\bm{x}}(t))\d t\right\}
    \end{aligned}
\end{equation}
where the forward and reverse Lagrangians in It\^{o} sense are
\begin{equation}\label{lagrangian}
    \begin{aligned}
        L(\bm{x}(t),\dot{\bm{x}}(t))=&\frac{1}{4D_\mathrm{th}}\left|\dot{\bm{x}}(t)-\frac{\bm{F}(\bm{x}(t))}{\Gamma}\right|^2 ,\\
         \tilde{L}(\bm{x}(t),\dot{\bm{x}}(t))=&\frac{1}{4D_\mathrm{th}}\left|\dot{\bm{x}}(t)-\frac{\bm{F}(\bm{x}(t))}{\Gamma} + 2\bm{V}(\bm{x}(t),T-t) \right|^2 .
    \end{aligned}
\end{equation}
Recall that the ``new'' system entropy $\tilde{s}_\mathrm{sys}(t)\equiv-\log P(\bm{x}(t),T-t)$ satisfies
\begin{equation}\label{app:sysd}
     \begin{aligned}
         \d \tilde{s}_{\mathrm{sys}}(t)
         =& -\frac{\partial_t \left(P(\bm{x},T-t)\right)}{P(\bm{x},T-t) }\bigg|_{\bm{x}(t)}\d t - \frac{\nabla P(\bm{x},T-t)}{P(\bm{x},T-t) }\bigg|_{\bm{x}(t)}\circ \d \bm{x}\\
        =& \underbrace{-\frac{\partial_t \left(P(\bm{x},T-t)\right)}{P(\bm{x},T-t) }\bigg|_{\bm{x}(t)}\d t + \frac{\bm{J}(\bm{x},T-t)}{D_{\mathrm{th}} P(\bm{x},T-t)}\bigg|_{\bm{x}(t)} \circ\d \bm{x}}_{\d\tilde{s}_\mathrm{tot}} - \underbrace{\frac{\bm{F}(\bm{x})}{\Gamma D_{\mathrm{th}} }\bigg|_{\bm{x}(t)}\circ\d \bm{x} }_{\d\tilde{s}_\mathrm{m}}
     \end{aligned}
 \end{equation}
and we also know that $P(\bm{x}(t),T-t)$ satisfies
\begin{equation}\label{app:pt-t}
        \begin{aligned}
            \frac{\partial_t\left( P(\bm{x}(t),T-t)\right)}{P(\bm{x}(t),T-t)}=& \nabla\cdot\left[\left( \frac{\bm{F}(\bm{x}(t))}{\Gamma}   - D_\mathrm{th}\nabla\log P(\bm{x}(t),T-t)   \right) P(\bm{r},T-t)\right]/P(\bm{r}(t),T-t)\\
            =& \frac{\nabla\cdot \left(\bm{V}(\bm{x}(t),T-t) P(\bm{x}(t),T-t)\right)}{P(\bm{r}(t),T-t)}.
        \end{aligned}
    \end{equation}
Thus we have
\begin{equation}\label{app:Gaussianp/p}
    \begin{aligned}
        &\log\frac{ P\left[\{\bm{x}(t)\}_{t\in[o,T]}\mid\bm{x}_0\right]}{\tilde{P}\left[\{\bm{x}(t)\}_{t\in[o,T]}\mid\bm{x}_0\right]}\\
        =& -\int_0^TL(\bm{x}(t),\dot{\bm{x}}(t)) \d t+\int_0^T\tilde{L}(\bm{x}(t),\dot{\bm{x}}(t))\d t\quad(\mbox{substitute the Lagrangians}) \\
        =& \frac{1}{4D_\mathrm{th}}\int_0^T2\bm{V}(\bm{x}(t),T-t)\cdot\left(2\dot{\bm{x}}(t)-2\frac{\bm{F}(\bm{x}(t))}{\Gamma} + 2\bm{V}(\bm{x}(t),T-t)\right)\d t\quad(\mbox{direct calculation}) \\
        =& \frac{1}{D_\mathrm{th}}\int_0^T\bm{V}(\bm{x}(t),T-t)\cdot\d\bm{x}(t) - \int_0^T\bm{V}(\bm{x}(t),T-t)\cdot\nabla\log P(\bm{x}(t),T-t)\d t\quad(\mbox{direct calculation}) \\
        =& \frac{1}{D_\mathrm{th}}\int_0^T\bm{V}(\bm{x}(t),T-t)\circ\d\bm{x}(t) - \int_0^T\nabla\cdot\bm{V}(\bm{x}(t),T-t)\d t -\int_0^T\bm{V}(\bm{x}(t),T-t)\cdot\nabla\log P(\bm{x}(t),T-t)\d t\\
        &\quad\quad(\mbox{It\^{o} to Stratonovich}) \\
        =& \frac{1}{D_\mathrm{th}}\int_0^T\bm{V}(\bm{x}(t),T-t)\circ\d\bm{x}(t) -  \int_0^T\frac{\nabla\cdot \left(\bm{V}(\bm{x}(t),T-t) P(\bm{x}(t),T-t)\right)}{P(\bm{x}(t),T-t)}\d t\quad(\mbox{direct calculation})\\
        =& \int_0^T\d\tilde{s}_\mathrm{tot}(t)=\Delta \tilde{s}_\mathrm{tot}\quad(\mbox{Eqs. \eqref{app:sysd} and \eqref{app:pt-t}}).
    \end{aligned}
\end{equation}
Then the integral fluctuation theorem follows,
\begin{equation}
    \langle e^{-\Delta \tilde{s}_\mathrm{tot}}\rangle=1.
\end{equation}
It is easy to see, all results in this appendix follow directly from the discussions in the previous appendices. Specifically, the conventional stochastic thermodynamic system described by Eq.~\eqref{app:Gaussiansde} can be regarded as a special case of the generalized framework developed in our work. Furthermore, the detailed fluctuation relation for $\Delta \tilde{s}_\mathrm{tot}-\log\frac{P_0(\bm{x}(0))}{P_T(\bm{\bm{x}(0)})}$ can be proved using the same argument in End Matter Appendix B.
    

Other than the conventional stochastic thermodynamics system \eqref{app:Gaussiansde}, our framework also covers the Active Ornstein--Uhlenbeck (AOU) particle system.  The AOU process is described by the following Langevin equation
\begin{equation}
    \begin{aligned}
        \frac{\d \bm{r}(t)}{\d t} =&  \frac{\bm{F}(\bm{r}(t))}{\Gamma} + \bm{\eta}_{\mathrm{th}}^{(1)}(t) + \bm{\eta} (t),\\
        \frac{\d\bm{\eta} (t)}{\d t} =&  - \frac{\bm{\eta}(t)}{\tau} + \bm{\eta}_\mathrm{th}^{(2)}(t),
    \end{aligned}
\end{equation}
where $\bm{\eta}(t)$ is the active noise (for simplicity we omit the subscript $act$ in this special case), $\bm{\eta}_\mathrm{th}^{(1)}$ and $\bm{\eta}_\mathrm{th}^{(2)}$ are two independent Gaussian white noises with zero means and following variances respectively
\begin{equation}
    \begin{aligned}
        \langle \eta^{(1)}_{\mathrm{th,i}}(t) \eta^{(1)}_{\mathrm{th,j}}(t') \rangle = 2D^{(1)}_{\mathrm{th}} \delta_\mathrm{i,j} \delta(t-t'),\\
        \langle \eta^{(2)}_{\mathrm{th,i}}(t) \eta^{(2)}_{\mathrm{th,j}}(t') \rangle = 2D^{(2)}_{\mathrm{th}} \delta_\mathrm{i,j} \delta(t-t').
    \end{aligned}
\end{equation}
For simplicity, $\Gamma$, $\tau$, $D^{(1)}_\mathrm{th}$ and $D^{(2)}_\mathrm{th}$ are assumed to be constants. Then the AOU system can be rewritten as
\begin{equation}\label{app:AOUSDE}
     \d\left(\begin{matrix}
         \bm{r}(t)\\
         \bm{\eta} (t)
     \end{matrix}\right)= \left(\begin{matrix}
        \frac{\bm{F}(\bm{r}(t))}{\Gamma} + \bm{\eta} (t)\\
         -\frac{\bm{\eta} (t)}{\tau}
     \end{matrix}\right) \d t + \d \bm{W}_t,
\end{equation}
where $\bm{W}_t$ is a $2d$-dimensional Brownian motion with mean 0 and covariance matrix 
\begin{equation}
    \mathbb{E}|\bm{W}_t-\bm{W}_s|^2=\left(\begin{matrix}
        2D_\mathrm{th}^{(1)}(t-s)I_d &\ 0\\
         0 &\ 2D_\mathrm{th}^{(2)}(t-s)I_d
     \end{matrix}\right).
\end{equation}
where $I_d$ is the $d$-dimensional identity matrix. We now observe that the AOU system described by Eq.~\eqref{app:AOUSDE} is encompassed within the conventional stochastic thermodynamics framework given by Eq.~\eqref{app:Gaussiansde}, and is therefore also included in our generalized framework. Note that, in this special case, there is no non-Gaussian active entropy $s_\mathrm{act}$, the active forces $\bm{\eta}(t)$ contribute to the medium entropy $s_\mathrm{m}$ and the system entropy $s_\mathrm{sys}$. The stochastic Gibbs-Shannon entropy is 
\begin{equation}
    s_\mathrm{sys}=-\log P((\bm{r}(t),\bm{\eta}(t)),t),
\end{equation}
where $P((\bm{r}(t),\bm{\eta}(t)),t)$ is the joint probability density of $(\bm{r}(t),\bm{\eta}(t))$ at time $t$. Then the entropy decomposition formula reads
\begin{equation}
    \begin{aligned}
        \d s_\mathrm{sys}= -\frac{\partial_t P((\bm{r}(t),\bm{\eta} (t)),t)}{P((\bm{r}(t),\bm{\eta} (t)),t)} \d t - \frac{\nabla_{\bm{r}}P((\bm{r}(t),\bm{\eta} (t)),t)}{P((\bm{r}(t),\bm{\eta} (t)),t)}\circ \d \bm{r}(t) - \frac{\nabla_{\bm{\eta}}P((\bm{r}(t),\bm{\eta} (t)),t)}{P((\bm{r}(t),\bm{\eta} (t)),t)}\circ \d \bm{\eta} (t) .
    \end{aligned}
\end{equation}
The probability density 
$P((\bm{r}(t),\bm{\eta}(t)),t)$ satisfies the following Fokker--Planck equation
\begin{equation}
    \begin{aligned}
        \frac{\partial P((\bm{r} ,\bm{\eta}  ),t)}{\partial t} =&-\nabla_{\bm{r}}\cdot\left[\left(  
        \frac{\bm{F}(\bm{r} )}{\Gamma} + \bm{\eta}  - D^{(1)}_{\mathrm{th}} \nabla_{\bm{r}}\log P((\bm{r},\bm{\eta}),t)  \right)P((\bm{r},\bm{\eta}),t)\right] \\
     &-\nabla_{\bm{\eta}}\cdot\left[\left(  
         -\frac{\bm{\eta} }{\tau}
     - D^{(2)}_{\mathrm{th}} \nabla_{\bm{\eta}}\log P((\bm{r},\bm{\eta}),t)  \right)P((\bm{r},\bm{\eta}),t)\right] .
    \end{aligned}
\end{equation}
Then the entropy decomposition formula turns to
\begin{equation}
    \begin{aligned}
        \d s_\mathrm{sys}=& \underbrace{\frac{\nabla_{\bm{r}}\cdot\left[\left(  
        \frac{\bm{F}(\bm{r}(t) )}{\Gamma} + \bm{\eta}(t)  - D^{(1)}_{\mathrm{th}} \nabla_{\bm{r}}\log P((\bm{r}(t),\bm{\eta}(t)),t)  \right)P((\bm{r}(t),\bm{\eta}(t)),t)\right] \d t}{P((\bm{r}(t),\bm{\eta}(t)),t)}  - \frac{\nabla_{\bm{r}}P((\bm{r}(t),\bm{\eta} (t)),t)}{P((\bm{r}(t),\bm{\eta} (t)),t)}\circ \d \bm{r}(t)}_{\d s^{\bm{r}}_\mathrm{sys}}\\
     & + \underbrace{\frac{\nabla_{\bm{\eta}}\cdot\left[\left(  
         -\frac{\bm{\eta}(t)  }{\tau}
     - D^{(2)}_{\mathrm{th}} \nabla_{\bm{\eta}}\log P((\bm{r}(t),\bm{\eta}(t)),t)  \right)P((\bm{r}(t),\bm{\eta}(t)),t)\right]\d t}{P((\bm{r}(t),\bm{\eta}(t)),t)}  - \frac{\nabla_{\bm{\eta}}P((\bm{r}(t),\bm{\eta} (t)),t)}{P((\bm{r}(t),\bm{\eta} (t)),t)}\circ \d \bm{\eta} (t)}_{\d s^{\bm{\eta}}_\mathrm{sys}} \\
     =& \underbrace{\frac{\nabla_{\bm{r}}\cdot[\bm{V}^{\bm{r}}((\bm{r}(t),\bm{\eta}(t)),t)P((\bm{r}(t),\bm{\eta}(t)),t)]\d t}{P((\bm{r}(t),\bm{\eta}(t)),t)} + \frac{\bm{J}^{\bm{r}}((\bm{r}(t),\bm{\eta}(t)),t)}{D^{(1)}_\mathrm{th}P((\bm{r}(t),\bm{\eta} (t)),t)}\circ\d\bm{r}(t)}_{\d s^{\bm{r}}_\mathrm{tot}} - \underbrace{\frac{\bm{F}(\bm{r}(t))/\Gamma + \bm{\eta}(t)}{D^{(1)}_\mathrm{th}}\circ\d\bm{r}(t)}_{\d s^{\bm{r}}_\mathrm{m}}\\
     & + \underbrace{\frac{\nabla_{\bm{\eta}}\cdot[\bm{V}^{\bm{\eta}}((\bm{r}(t),\bm{\eta}(t)),t)P((\bm{r}(t),\bm{\eta}(t)),t)]\d t}{P((\bm{r}(t),\bm{\eta}(t)),t)} + \frac{\bm{J}^{\bm{\eta}}((\bm{r}(t),\bm{\eta}(t)),t)}{D^{(2)}_\mathrm{th}P((\bm{r}(t),\bm{\eta} (t)),t)}\circ\d\bm{\eta}(t)}_{\d s^{\bm{\eta}}_\mathrm{tot}} + \underbrace{\frac{  \bm{\eta}}{D^{(2)}_\mathrm{th}\tau}\circ\d\bm{\eta}(t)}_{\d s^{\bm{\eta}}_\mathrm{m}},
    \end{aligned}
\end{equation}
where 
\begin{equation}
\begin{aligned}
    \bm{V}^{\bm{r}}((\bm{r},\bm{\eta}),t)=\bm{J}^{\bm{r}}((\bm{r},\bm{\eta}),t)/P((\bm{r},\bm{\eta}),t)\equiv& \frac{\bm{F}(\bm{r} )}{\Gamma} + \bm{\eta}  - D^{(1)}_{\mathrm{th}} \nabla_{\bm{r}}\log P((\bm{r},\bm{\eta}),t) ,\\
    \bm{V}^{\bm{\eta}}((\bm{r},\bm{\eta}),t)=\bm{J}^{\bm{\eta}}((\bm{r},\bm{\eta}),t)/P((\bm{r},\bm{\eta}),t)\equiv &-\frac{\bm{\eta}  }{\tau}
     - D^{(2)}_{\mathrm{th}} \nabla_{\bm{\eta}}\log P((\bm{r},\bm{\eta}),t).
\end{aligned}
\end{equation}
The total entropy production can be separated into two distinct parts: one arising from the coordinate $\bm{r}(t)$ and the other from the coordinate $\bm{\eta}(t)$. Specifically, these two contributions consist of the system entropy production and the medium entropy production, respectively, as detailed below:
\begin{equation}
    \begin{aligned}
        \d s_\mathrm{tot}=& \d s^{\bm{r}}_\mathrm{tot} + \d s^{\bm{\eta}}_\mathrm{tot} \\
        =& \d s^{\bm{r}}_\mathrm{sys} + \d s^{\bm{r}}_\mathrm{m} + \d s^{\bm{\eta}}_\mathrm{sys}  + \d s^{\bm{\eta}}_\mathrm{m},
    \end{aligned}
\end{equation}
which can be considered as a special case of the entropy production decomposition formula \eqref{eqn:stotsmall} in the Letter. Then the fluctuation relation considered on a time interval $[0,T]$ follows directly,
\begin{equation}
    \langle e^{-\Delta \tilde{s}_\mathrm{tot}}\rangle=1,
\end{equation}
where $\Delta \tilde{s}_\mathrm{tot}$ can be derived from  $\tilde{s}_\mathrm{sys}\equiv-\log P((\bm{r}(t),\bm{\eta}(t)),T-t)$ as shown as before. Moreover, the detailed fluctuation relation for $\Delta \tilde{s}_\mathrm{tot}-\log\frac{P_0[(\bm{r}(0),\bm{\eta}(0))]}{P_T[(\bm{\bm{r}(0)},\bm{\eta}(0))]}$ can be proved using the same argument in End Matter Appendix B.

If we only focus on the dynamics of $\bm{r}(t)$, we have the expression of the system entropy for $\bm{r}(t)$ as follows,
\begin{align}
    s^*_\mathrm{sys}\equiv-\log \left(\int\d\bm{\eta}P\big((\bm{r}(t),\bm{\eta}),t\big)\right),
\end{align}
and its derivative reads
\begin{equation}
\begin{aligned}
   \dot{s}^*_\mathrm{sys}=& -\frac{\int\d\bm{\eta}\partial_tP\big((\bm{r}(t),\bm{\eta}),t\big)}{\int\d\bm{\eta} P\big((\bm{r}(t),\bm{\eta}),t\big)} - \frac{\int\d\bm{\eta}\nabla_{\bm{r}}P\big((\bm{r}(t),\bm{\eta}),t\big)}{\int\d\bm{\eta}P\big((\bm{r}(t),\bm{\eta}),t\big)}\diamond\dot{\bm{r}}\\
   =& -\frac{\int\d\bm{\eta}\partial_tP\big((\bm{r}(t),\bm{\eta}),t\big)}{\int\d\bm{\eta} P\big((\bm{r}(t),\bm{\eta}),t\big)}  -  \frac{1}{\int\d\bm{\eta} P\big((\bm{r}(t),\bm{\eta}),t\big)}\int\d\bm{\eta}\Bigg\{ \nabla_{\bm{r}}\cdot\left[\left(  
        \frac{\bm{F}(\bm{r}(t) )}{\Gamma} + \bm{\eta} \right.\right.\\
        & \left.\left. - D^{(1)}_{\mathrm{th}} \nabla_{\bm{r}}\log P((\bm{r}(t),\bm{\eta} ),t)  \right)P((\bm{r}(t),\bm{\eta} ),t)\right]    - \dot{s}^{\bm{r}}_\mathrm{sys}P((\bm{r}(t),\bm{\eta}),t)\Bigg\}\\
        =& -\underbrace{\left(\frac{\int\d\bm{\eta}\partial_tP\big((\bm{r}(t),\bm{\eta}),t\big)}{\int\d\bm{\eta} P\big((\bm{r}(t),\bm{\eta}),t\big)}  +  \frac{\int\d\bm{\eta}\Bigg\{ \nabla_{\bm{r}}\cdot\left[\left(  
        \frac{\bm{F}(\bm{r}(t) )}{\Gamma} + \bm{\eta} - D^{(1)}_{\mathrm{th}} \nabla_{\bm{r}}\log P((\bm{r}(t),\bm{\eta} ),t)  \right)P((\bm{r}(t),\bm{\eta} ),t)\right]     \Bigg\}}{\int\d\bm{\eta} P\big((\bm{r}(t),\bm{\eta}),t\big)} \right)}_{\equiv \dot{s}^*_\mathrm{act}}\\
        & + \underbrace{\frac{\int\d\bm{\eta}  \dot{s}^{\bm{r}}_\mathrm{tot}  P\big((\bm{r}(t),\bm{\eta}),t\big)}{\int\d\bm{\eta} P\big((\bm{r}(t),\bm{\eta}),t\big)}}_{\equiv \dot{s}^*_\mathrm{tot}} - \underbrace{\frac{\int\d\bm{\eta}   \dot{s}^{\bm{r}}_\mathrm{m} P\big((\bm{r}(t),\bm{\eta}),t\big)}{\int\d\bm{\eta} P\big((\bm{r}(t),\bm{\eta}),t\big)}}_{\equiv\dot{s}^*_\mathrm{m}}.
\end{aligned}
\end{equation}
This equation indicates that, when focusing only on the dynamics of $\bm{r}(t)$, the decomposition of its entropy production resembles that in the main text, although the definitions of the individual components differ. Nevertheless, they can be evaluated using our numerical method. Regarding the fluctuation theorem, however, the relation does not hold for $\Delta s^*_\mathrm{tot}$. Identifying the fluctuation relation satisfied by $\Delta s^*_\mathrm{tot}$ is more involved and will be addressed in future work.

\section{Deep learning algorithm for EPRs}
This Appendix provides additional details on the numerical deep–learning method introduced in the main text.

Recall that the LFPE of 
\eqref{app:SDE} can be written as a continuity equation, the general principle involved is to addresses the following fixed-point problem: For any given velocity field $\bm{V}^{in}(\bm{x},t)$, the flow dictated by the ODE
\begin{equation}\label{appeqn:interactingODE}
    \begin{aligned}
        \frac{\d \bm{X}_{s,t}(\bm{x})}{\d t}=\ \bm{V}(\bm{X}_{s,t}(\bm{x}),t),\quad \bm{X}_{s,s}(\bm{x})=\bm{x},\quad t\geq s\geq0.
    \end{aligned}
\end{equation}
will transport the initial density $P_0(\bm{x})$ to obtain $P(\bm{x},t)$, 
and this transported $P(\bm{x},t)$ furthermore induces  the new velocity field $\bm{V}^{out}$ defined via \eqref{eqn:LFP}.
It is evident that the true velocity field $\bm{V}$ is the fixed point of this map $\bm{V}^{in}\mapsto \bm{V}^{out}$. Thus, if we are provided with a set of vector fields $\{\bm{V}^\mathrm{NN}\}$, and obtain the corresponding probability flows $P^\mathrm{NN}$ via \eqref{appeqn:interactingODE}, the ideal choice of the optimal vector field that approximates the true vector field is the one that minimizes the following loss function with some samples from $ P^\mathrm{NN}(\bm{x},t)$:
\begin{align}
\int_{0}^T\d t\int \d\bm{x}|\bm{V}^\mathrm{NN}(\bm{x},t)-\bm{V} (\bm{x},t)|^2P^\mathrm{NN}(\bm{x},t).
\end{align} 
We use two neural networks, $\bm{S}^\mathrm{NN}_\mathrm{B}$ and $\bm{S}^\mathrm{NN}_\mathrm{L}$, in the same time during training. 
And write vector field $\bm{V}^\mathrm{NN}$ as 
\begin{equation}
    \bm{V}^\mathrm{NN}=\bm{F}/\Gamma  -D_\mathrm{th} \bm{S}^\mathrm{NN}_\mathrm{B} - \bm{S}^\mathrm{NN}_\mathrm{L} .
\end{equation}

Firstly, we note that
\begin{equation} 
    \begin{aligned}
       \left| \bm{V}^\mathrm{NN}(\bm{x},t) - \bm{V}(\bm{x},t)\right|^2 
        =&  \left| D_\mathrm{th} \bm{S}^\mathrm{NN}_\mathrm{B}(\bm{x},t) + \bm{S}^\mathrm{NN}_\mathrm{L}(\bm{x},t)   - D_\mathrm{th}\nabla\log P^\mathrm{NN}(\bm{x},t)  + \int\nu(\d\bm{z})\int_0^1\d\theta\bm{z}\frac{P^\mathrm{NN}(\bm{x} -\theta\bm{z},t)}{P^\mathrm{NN}(\bm{x} ,t)} \right|^2 \\
        \leq& 2D^2_\mathrm{th}\left| \bm{S}^\mathrm{NN}_\mathrm{B}(\bm{x},t) - \nabla\log P^\mathrm{NN}(\bm{x},t) \right|^2 + 2\left| \bm{S}^\mathrm{NN}_\mathrm{L}(\bm{x},t) +  \int\nu(\d\bm{z})\int_0^1\d\theta\bm{z}\frac{P^\mathrm{NN}(\bm{x} -\theta\bm{z},t)}{P^\mathrm{NN}(\bm{x},t )} \right|^2.
    \end{aligned}
\end{equation}
 
Define the loss functions:
    \begin{align*}
        Loss_\mathrm{B}(t)\equiv& \int\d\bm{x}\left|\bm{S}^\mathrm{NN}_\mathrm{B}(\bm{x},t) -\nabla\log P^\mathrm{NN}(\bm{x},t)\right|^2 P^\mathrm{NN}(\bm{x},t) ,\\
        Loss_\mathrm{L}(t)\equiv&\ \int\d\bm{x}P^\mathrm{NN}(\bm{x},t)\left|\bm{S}^\mathrm{NN}_\mathrm{L}(\bm{x},t) + \int\nu(\d\bm{z})\int_0^1\d\theta\bm{z}\frac{P^\mathrm{NN}(\bm{x} -\theta\bm{z},t)}{P^\mathrm{NN}(\bm{x},t )} \right|^2 .
\end{align*}

Expanding the squares in $Loss_\mathrm{B}(t)$ and $Loss_\mathrm{L}(t)$, we obtain
\begin{align}
    Loss_\mathrm{B}(t) =& \mathbb{E}_{\bm{x}\sim P^\mathrm{NN}(\bm{x},t)}\left( \left|\bm{S}^\mathrm{NN}_\mathrm{B}(\bm{x}\right|^2 \right) + \int\d\bm{x}\left|\nabla\log P^\mathrm{NN}(\bm{x},t)\right|^2 P^\mathrm{NN}(\bm{x},t) + 2D_\mathrm{th}\mathbb{E}_{\bm{x}\sim P^\mathrm{NN}(\bm{x},t)}\left[\nabla\cdot \bm{S}^\mathrm{NN}_\mathrm{B}(\bm{x}) \right],\\
    Loss_\mathrm{L}(t) =& \mathbb{E}_{\bm{x}\sim P^\mathrm{NN}(\bm{x},t)}\left(\left|\bm{S}^\mathrm{NN}_\mathrm{L}(\bm{x} \right|^2\right)  + \int\d\bm{x}\left| \int\nu(\d\bm{z})\int_0^1\d\theta\bm{z}\frac{P^\mathrm{NN}(\bm{x} -\theta\bm{z},t)}{P^\mathrm{NN}(\bm{x},t )} \right|^2 P^\mathrm{NN}(\bm{x},t) \\
    &  + 2 \mathbb{E}_{\bm{x}\sim P^\mathrm{NN}(\bm{x},t)}\left(\int \nu(\d\bm{z}) \int_0^1\d\theta \left(\bm{S}^\mathrm{NN}_\mathrm{L}(\bm{x}+\theta\bm{z} ,t)\cdot \bm{z}\right) \right).
\end{align}

We may neglect the square terms being independent of $\bm{S}^\mathrm{NN}_\mathrm{B}$ and $\bm{S}^\mathrm{NN}_\mathrm{L}$ during optimization and treat them as constant terms. Thus we have the total loss at time $t$ as
\eqref{opt:B} and \eqref{opt:L}. According to the above arguments, we design the Algorithm \ref{algorithm} to solve the nonlinear L\'{e}vy--Fokker--Planck equation \eqref{eqn:LFP}. At this point, we can use two neural networks to simultaneously approximate the Gaussian, and non-Gaussian scores. 

\begin{algorithm}\label{algorithm}
\caption{Sequential L\'{e}vy score-based transport modeling for EPR}
\SetKwInOut{Input}{Input}\SetKwInOut{Output}{Output}
\Input{An initial time $t_0=0$. A set of $N$ samples $\{\bm{x}^{(i)}\}_{i=1}^N$ from initial distribution $P(\cdot,t_0)$. A time step $\Delta t$ and the number of steps $N_T$.
Initialize sample locations $\bm{X}^{(i)}_{t_0}=\bm{x}^{(i)}$ for $i=1,\cdots,N$.}
\For{$k=0:N_T$}
{Optimize $(\bm{S}^\mathrm{NN}_\mathrm{B}(\cdot,t_k),\bm{S}^\mathrm{NN}_\mathrm{L}(\cdot,t_k))=\arg\min_{\bm{S}^\mathrm{NN}_\mathrm{B},\bm{S}^\mathrm{NN}_\mathrm{L}} \frac{1}{N} \sum_{i=1}^N\bigg[ D^2_\mathrm{th}  \left|\bm{S}^\mathrm{NN}_\mathrm{B}(\bm{X}_{t_k}^{(i)},t_k) \right|^2 + 2D^3_\mathrm{th}\nabla\cdot  \bm{S}^\mathrm{NN}_\mathrm{B}(\bm{X}_{t_k}^{(i)},t_k) +   \left|\bm{S}^\mathrm{NN}_\mathrm{L}(\bm{X}_{t_k}^{(i)},t_k) \right|^2  $\
$+ 2 \left( \int \nu(d\bm{z})\int_0^1 \d\theta\bm{S}^\mathrm{NN}_\mathrm{L} (\bm{X}_{t_k}^{(i)}+\theta\bm{z},t_k)\cdot \bm{z} \right)\bigg]$\;
Propagate the samples for $i=1,\cdots,N$:\ 
$\bm{X}^{(i)}_{t_{k+1}}=\bm{X}^{(i)}_{t_{k}} + \Delta t \big[ \bm{F}(\bm{X}^{(i)}_{t_{k}} )/\Gamma  -D_\mathrm{th}\bm{S}^\mathrm{NN}_\mathrm{B}(\bm{X}^{(i)}_{t_{k}},t_k) - \bm{S}^\mathrm{NN}_\mathrm{L}(\bm{X}^{(i)}_{t_{k}},t_k)  \big]$\;
Set $t_{k+1}=t_k+\Delta t$\; 
}
\Output{$N$ samples $\{\bm{X}^{(i)}_{t_k}\}_{i=1}^N$ from $p_{t_k}$ and the scores  $\{\bm{S}^\mathrm{NN}_\mathrm{B}(\cdot,t_k)(\bm{X}^{(i)}_{t_k},t_k)\}_{i=1}^N$ and $\{\bm{S}^\mathrm{NN}_\mathrm{L}(\cdot,t_k)(\bm{X}^{(i)}_{t_k},t_k)\}_{i=1}^N$ for all $\{t_k\}_{k=0}^{N_T}$.}
\end{algorithm}

When we use the Algorithm \ref{algorithm} to examine certain examples, the time interval \([0, T]\) is uniformly partitioned  into \( N_T \) sub-intervals \([t_k, t_{k+1}]\), where   \( t_k = k\frac{T}{N_T} \) for \( k = 0, 1, \dots, N_T \). 
On each sub-interval \([t_k, t_{k+1}]\), the transport map is approximated by the neural networks \( s_\mathrm{B}^{\theta_{k}}(\cdot,t_k),\ s_\mathrm{L}^{\theta_{k}}(\cdot,t_k): \mathbb{R}^d \to \mathbb{R}^d \), modeled as a multi-layer perceptron (MLP)  with $3$ hidden layers, $32$ neurons per layer, and the $Swish$ activation function. 
The algorithm is implemented with the following parameter settings: the time step size of \(\Delta t = T/N_T=t_{k+1}-t_{k}  = 10^{-3}\), and the sample size of \(N=4000\). 

The initial condition $P_0$ of the examples in the nest section is set as the   Gaussian distribution for its simplicity in generating initial samples \( \{ \bm{r}_0^{(i)} \}_{i=0}^{N} \) (unless otherwise specified, it is assumed to be the standard normal distribution). 
For each time step in training the scores $s_\mathrm{B}^{\theta_{k+1}}(\cdot, t_{k+1})$ and $s_\mathrm{L}^{\theta_{k+1}}(\cdot, t_{k+1})$, we use the warm start for the optimization by initializing the neural network parameter \( \theta_{{k+1}} \) by 
the obtained parameters \( \theta^*_{k} \) from the previous step, followed by  the standard   the Adam optimizer with a learning rate of \( 10^{-4} \) to optimize  \( \theta_{{k+1}} \). 

To evaluate our method, we 
use   the total variation (TV) distance of  the generated samples to compare out method  with the Monte Carlo simulation method.
At each time \(t_k\), we identify the smallest rectangular domain \(\Omega_{t_k}\) covering all sample points and discretize it into  uniform grid cells \(\{\Delta_i\}\). 
The distribution for each method is then approximated by a histogram:  
\begin{equation}
    P(\Delta_i) = \frac{\#(\text{samples} \in \Delta_i)}{\#\text{samples}}.
\end{equation} 
We denote these binned empirical distributions as \(P^{\mathrm{MC}}\) for Monte Carlo and \(P^{\mathrm{NN}}\) for our method, and the TV distance between these two distributions is then numerically computed by 
\begin{equation}
    d_{\mathrm{TV}}(P^{\mathrm{MC}}, P^{\mathrm{NN}}) = \sum_{i} \left|P^{\mathrm{MC}}(\Delta_i) - P^{\mathrm{NN}}(\Delta_i)\right|.
\end{equation}

\section{Example 1--A Brownian particle immersed in a periodic active bath}
This Appendix provides additional details on the first example discussed in the main text.

The first example considers a Brownian particle immersed in a periodic (active) bath, and the dynamics follows the SDE:
\begin{equation}\label{eqn:example1}
    \begin{aligned}
    \d r  =&  -\frac{V_0}{\Gamma}\left[\frac{2\pi}{L}\cos\left(\frac{2\pi r}{L}\right) + \frac{\pi}{L}\cos\left(\frac{4\pi r}{L}\right)  \right] \d t + \sqrt{2D_\mathrm{th}}\d W_t +   \int z\mathcal{N}(\d t,\d z),
\end{aligned}
\end{equation}
where $\mathcal{N}(\d t,\d z)$ is a Poisson random measure with L\'{e}vy measure $\lambda_0\nu_A(\d z)\d t$, here $\nu_A$ is the density of a Gaussian distribution with mean 0 and variance $\sigma^2$. The values of the parameters selected are listed in Table \ref{tab:table1}.


\begin{table*} 
\caption{\label{tab:table1}List of model parameters used in simulations for Example 1.}
\begin{ruledtabular}
\begin{tabular}{cccc}
 Parameter  & Notation & Value & Dimension 
 \\ \hline
 Thermal energy & $k_B\mathcal{T}$ & 4.114 & pN nm\\
 Viscous drag  & $\Gamma$  
 & 3.25 & pNs/nm \\
 Barrier height & $V_0$ & 5$\times$ 4.114 & nm
  \\
 Potential period  & $L$ & 40 & nm\\
 Poisson parameter & $\lambda_0$ &  30 & 1\\
Mean of jump amplitude & $\mu$ &  0.1  & nm\\
  Standard deviation of jump amplitude & $\sigma$ &   1/24   & nm\\
 Simulation time step & $\Delta t$ & $10^{-3}$ & $s$ 
\end{tabular}
\end{ruledtabular}
\end{table*}

\begin{figure*} 
    \begin{minipage}[t]{0.49\linewidth}
    \raggedright (a)\\[-0.5ex]
        \includegraphics[width=1\linewidth]{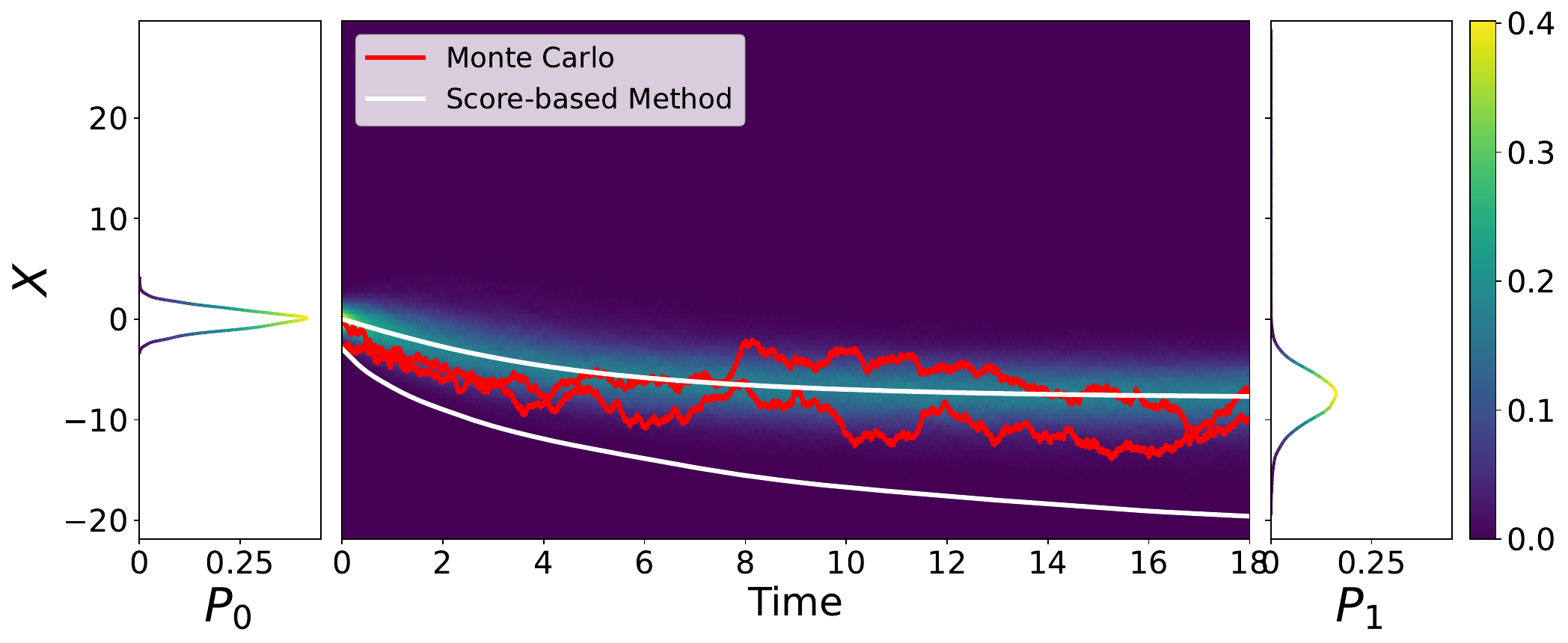} 
    \end{minipage}\hfill
    \begin{minipage}[t]{0.49\linewidth}
    \raggedright (b)\\[-0.5ex]
        \includegraphics[width=1\linewidth]{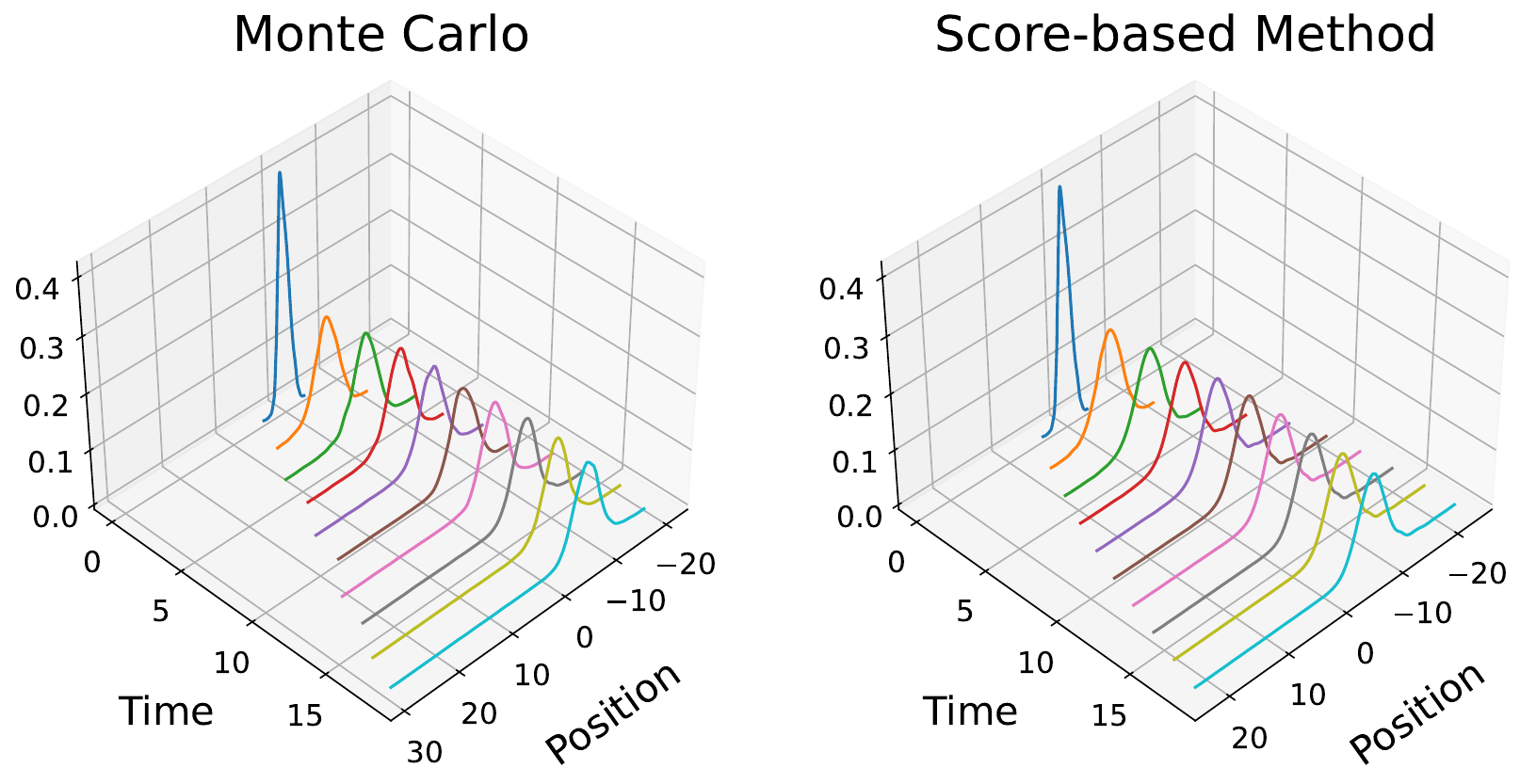} 
    \end{minipage}
    \begin{minipage}[t]{0.49\linewidth}
    \raggedright (c)\\[-0.5ex]
        \includegraphics[width=1\linewidth]{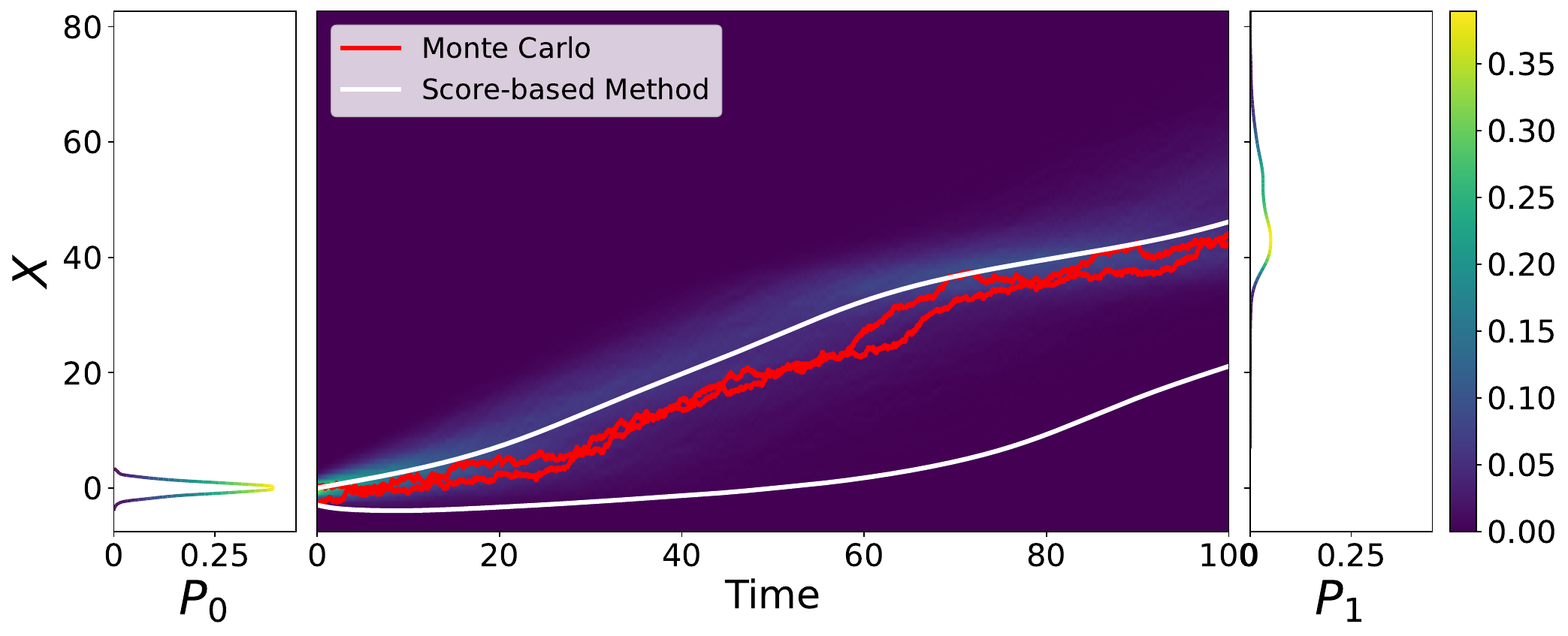} 
    \end{minipage}\hfill
    \begin{minipage}[t]{0.49\linewidth}
    \raggedright (d)\\[-0.5ex]
        \includegraphics[width=1\linewidth]{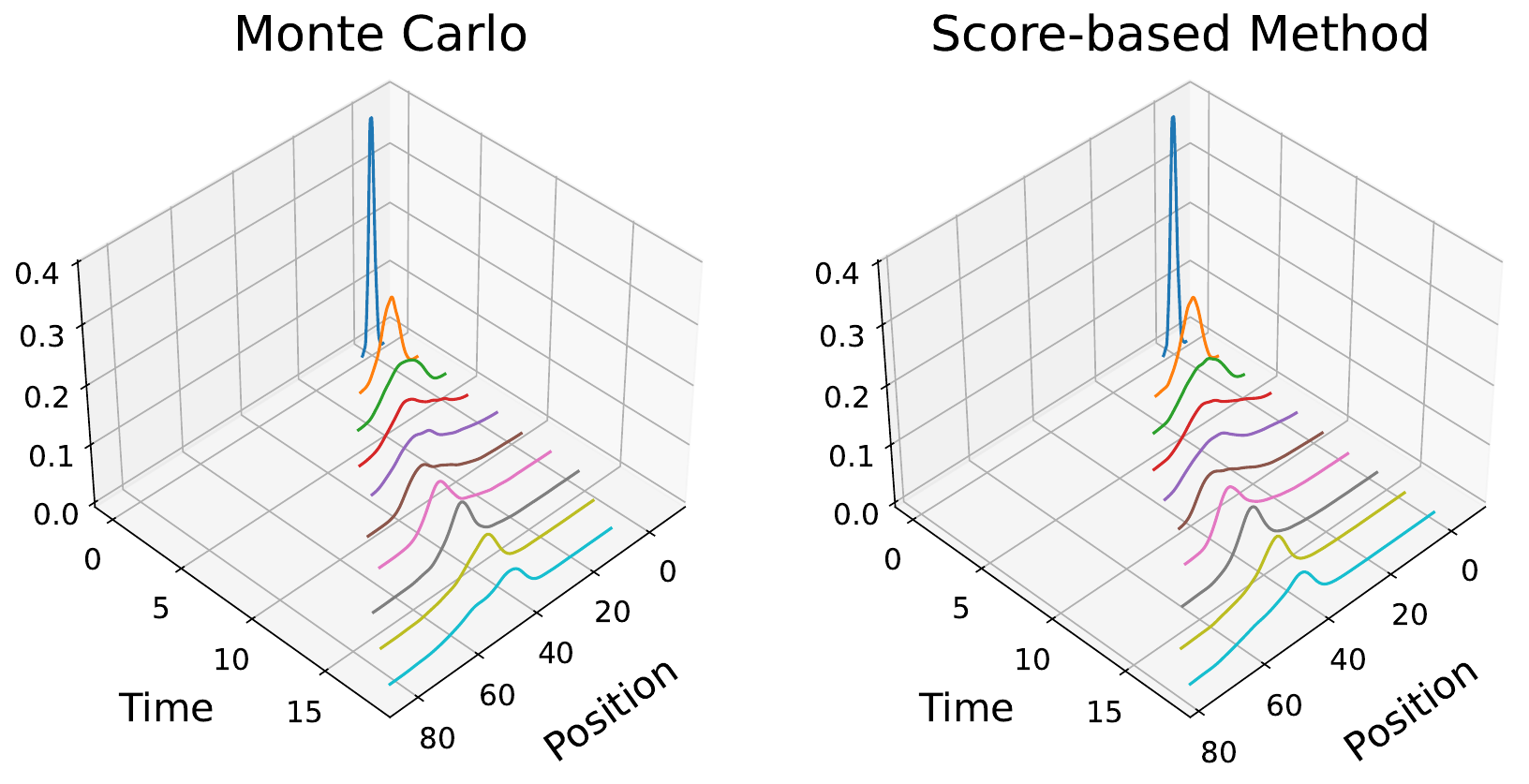} 
    \end{minipage} 
    \caption{Probability flows of the Brownian particle immersed in a periodic active bath. The top panel illustrates the temporal evolution of the probability distribution as a heat map, overlaid with two selective stochastic trajectories based on the Monte Carlo simulation (red) and the two deterministic trajectories based on the transport map \eqref{appeqn:interactingODE} (white). The bottom panels compare the probability distributions $P(\bm{r},t)$ from the Monte Carlo simulation and the proposed method in the time-state space.}\label{fig:example1-probabilityflow}
\end{figure*}

Figure \ref{fig:example1-probabilityflow}  illustrates the temporal evolution of probability flows and probability density functions for Equation \eqref{eqn:example1}, obtained via both the Monte Carlo simulation and the proposed numerical method. Specifically, the Monte Carlo simulation employs the following Euler–Maruyama discretization scheme:  
\begin{equation}\label{app:EMEXAMPLE1}
    \begin{aligned}
        r_{t+\Delta t}^{(i)} =& r_t^{(i)} -\frac{V_0}{\Gamma}\left[\frac{2\pi}{L}\cos\left(\frac{2\pi r_t^{(i)}}{L}\right) + \frac{\pi}{L}\cos\left(\frac{4\pi r_t^{(i)}}{L}\right)  \right]\Delta t + \sqrt{2D_\mathrm{th}} \xi_t +   \sum_{k=1}^{N_{\Delta t}} A_k, \quad i = 1, \dots, N,
    \end{aligned}
\end{equation}
where \( \xi_t \sim \mathrm{N}_{0, \Delta t} \) where $\mathrm{N}_{0, \Delta t} $ is a Gaussian distribution with mean 0 and variance $\Delta t$, \( N_{\Delta t} \sim \text{Po}(\lambda_0 \Delta t) \) is a Poisson random variable with rate \( \lambda_0 \Delta t \), and \( A_k \)'s are i.i.d.\ random variables for jump sizes distributed as $\mathrm{N}_{0,\sigma^2}$.

\begin{figure*} 
    \begin{minipage}[t]{1\linewidth}
    \raggedright (a)\\[-0.5ex]
        \includegraphics[width=1\linewidth]{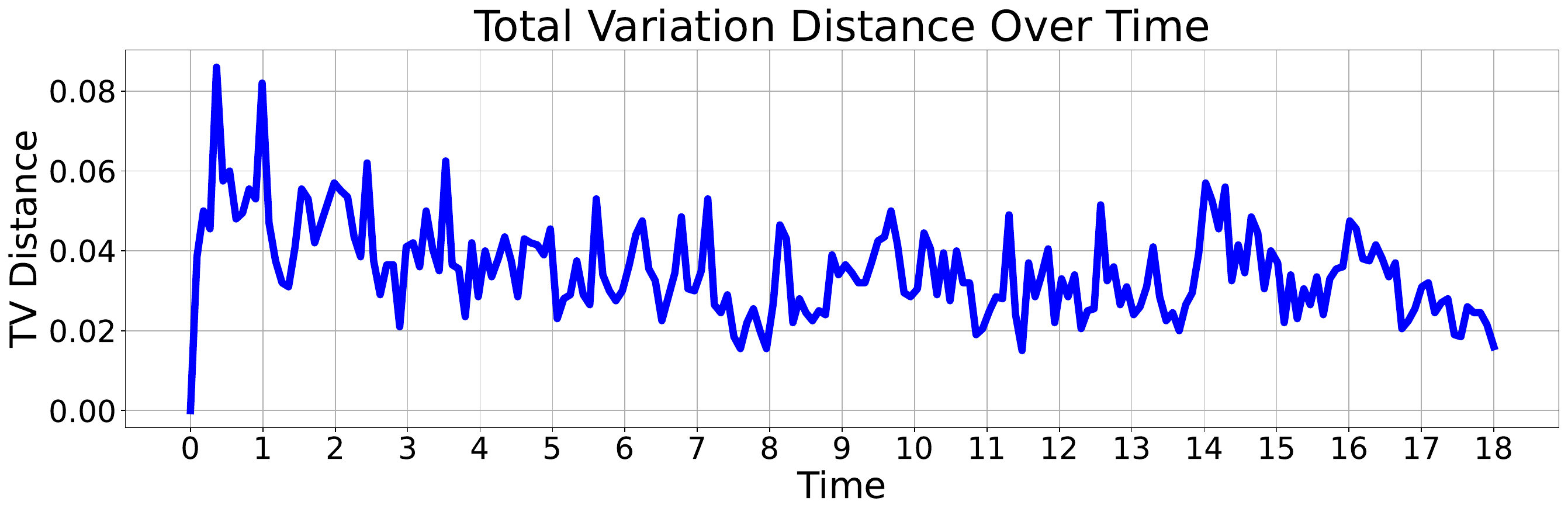} 
    \end{minipage}\hfill
    \begin{minipage}[t]{1\linewidth}
    \raggedright (b)\\[-0.5ex]
        \includegraphics[width=1\linewidth]{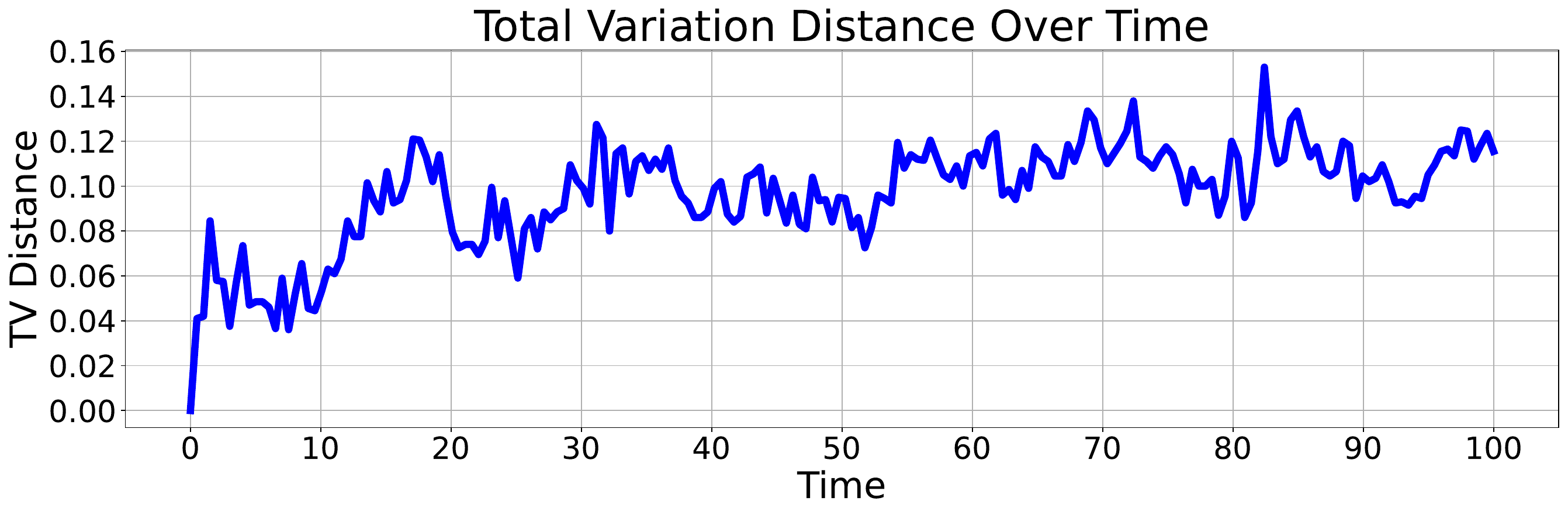}
    \end{minipage}
    \caption{Total variation distances between $P^\mathrm{MC}$ and $P^\mathrm{NN}$ for the Brownian particle immersed in an active bath. Top panel: $\mu=0$; bottom panel: $\mu=0.1$.}\label{fig:example1-TV}
\end{figure*}

FIG. \ref{fig:example1-TV} illustrates the total variation distances between \( P^{\mathrm{MC}} \) and \( P^{\mathrm{NN}} \) for both cases on time interval [0,18] ($\mu=0$) and [0,100] ($\mu=0.1$) respectively. The total variation distances remain consistently within the order of \( 10^{-2} \) to \( 10^{-1} \), demonstrating the robustness and relative accuracy of our method over time. 

FIG. \ref{fig:example1img1} shows the EPRs under different conditions on time interval [0,18]. When there is no active fluctuation, i.e., $\eta_\mathrm{act}=0$, all the EPRs decay to 0 meaning the system approaches the equilibrium state. When the active fluctuation exists and the mean jump height of the active noise is zero, the system reaches an equilibrium steady state relatively quickly; this behavior is very similar to the case when $\eta_\mathrm{act}=0$. However, when the mean jump height increases to 0.1, non-equilibrium behavior becomes evident.

\begin{figure}[htbp]
  \centering
  \begin{minipage}[t]{\linewidth}
    \raggedright (a)\\[-0.5ex]
    \includegraphics[width=.95\linewidth]{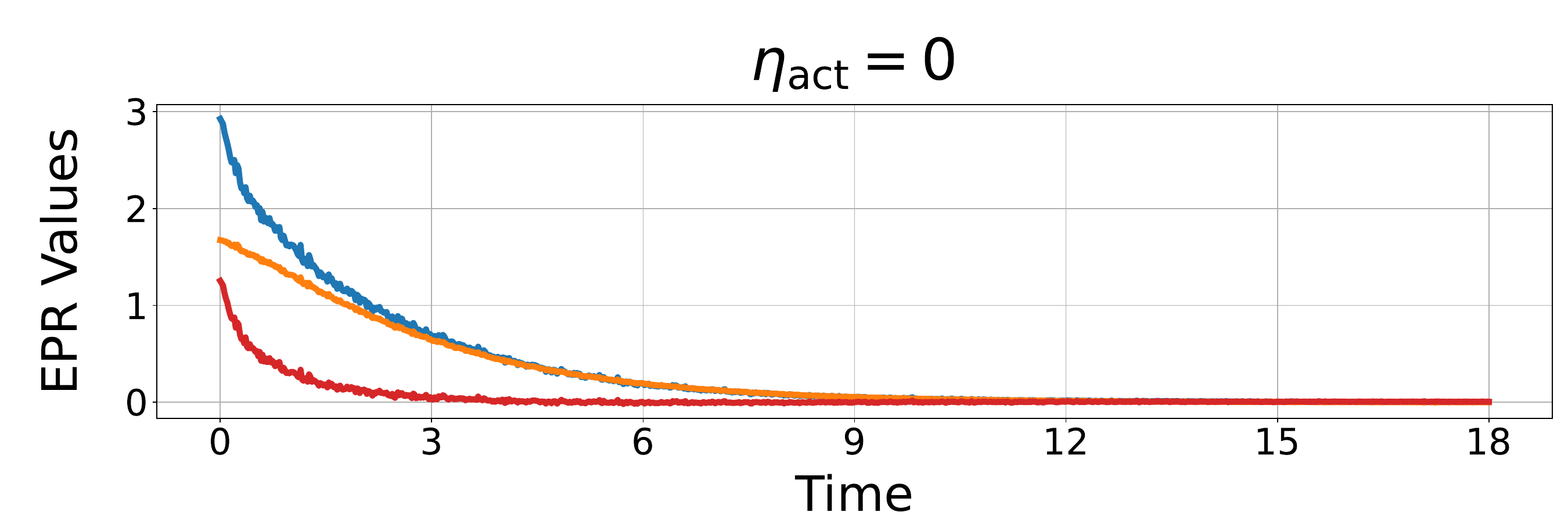}
  \end{minipage}\hfill
  \begin{minipage}[t]{\linewidth}
    \raggedright (b)\\[-0.5ex]
    \includegraphics[width=.95\linewidth]{Example1_1-epr.pdf}
  \end{minipage}\hfill
  \begin{minipage}[t]{\linewidth}
    \raggedright (c)\\[-0.5ex]
    \includegraphics[width=.95\linewidth]{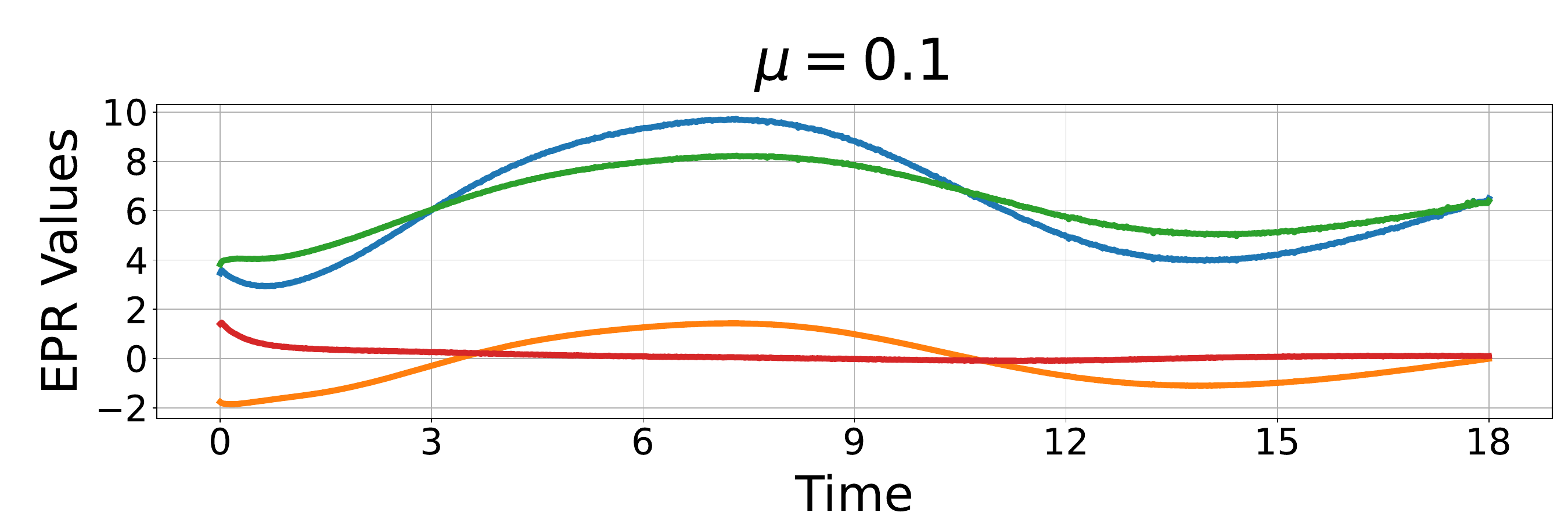}
  \end{minipage}
  \caption{EPRs of a Brownian particle in an (active) bath. Shown are $\dot{S}_\mathrm{tot}$ (blue) and its three components: $\dot{S}_\mathrm{m}$ (orange), $\dot{S}_\mathrm{act}$ (green), and $\dot{S}_\mathrm{sys}$ (red) in both plots. (a) $\eta_\mathrm{act}=0$ (there is no active fluctuation). All EPRs decay to zero, showing equilibrium-like behavior. (b)  $\mu=0$. All EPRs decay to zero, showing equilibrium-like behavior. (c)  $\mu=0.1$. $\dot{S}_\mathrm{tot}$ saturates at a positive value due to active contributions, showing a nonequilibrium steady state.}
  \label{fig:example1img1}
\end{figure}

\subsection{Validation via Gaussian Mixture Method with PINNs}

To further validate the accuracy and effectiveness of our proposed algorithm, we directly approximate the probability density function. In particular, for the one-dimensional example discussed in the main context, we represent the time-dependent density $P(x,t)$ using a Gaussian mixture model (GMM), where all parameters are modeled as functions of time and learned via a neural network.

Specifically, the density is approximated as
\begin{equation}
    P^{\mathrm{GMM}}(x,t) = \sum_{i=1}^m \pi_i(t) \cdot \mathfrak{N}\left(x\,\middle|\, \mu_i(t), \sigma_i^2(t) \right),
\end{equation}
where $\pi_i(t)$ are the mixture weights satisfying $\sum_i \pi_i(t) = 1$, and each $\mathfrak{N}(x\,|\,\mu_i(t), \sigma_i^2(t))$ denotes a Gaussian density with time-dependent mean $\mu_i(t)$ and standard deviation $\sigma_i(t)$. All three components---$\pi_i(t)$, $\mu_i(t)$, and $\sigma_i(t)$---are parameterized by neural networks that take time $t$ as input.

The parameters of the GMM are trained by minimizing a physics-informed loss function that enforces consistency with the governing L\'evy--Fokker--Planck equation. The loss is defined as the integrated squared residual:
\begin{equation}
    \mathcal{L}_{\mathrm{PINN}} = \int_0^T \! \mathrm{d}t \int \! \mathrm{d}x \left| \partial_t P^{\mathrm{GMM}}(x,t) - \mathcal{L}^* P^{\mathrm{GMM}}(x,t) \right|^2,
\end{equation}
where the operator $\mathcal{L}^*$ denotes the L\'evy--Fokker--Planck operator:
\begin{equation}
    \mathcal{L}^* P(x,t) = -\nabla \cdot \left[ \left( \frac{F(x)}{\Gamma} - D_{\mathrm{th}} \nabla \log P(x,t) + \int_0^1 \! \mathrm{d}\theta \int \nu(\mathrm{d}z) \frac{z P(x - \theta z,t)}{P(x,t)} \right) P(x,t) \right].
\end{equation}

Figure~\ref{fig:loss} shows the PINN loss during training for the GMM-based PINN approach. 
As seen in the figure, the loss decreases to $10^{-5}$, indicating that the numerical solution obtained from the GMM-based PINN can be regarded as an accurate approximation of the true solution of the PDE $\partial_t P=\mathcal{L}^*P$. 
We further compare this solution with that obtained from the proposed score-based particle algorithm over the time interval $[0,18]$. 
As shown in FIG.~\ref{fig:error}, the two approaches yield nearly identical results, demonstrating both the accuracy and robustness of our method.

\begin{figure*}
    \centering
    \begin{minipage}[t]{0.45\linewidth}
    \raggedright (a)\\[-0.5ex]
        \includegraphics[width=1\linewidth]{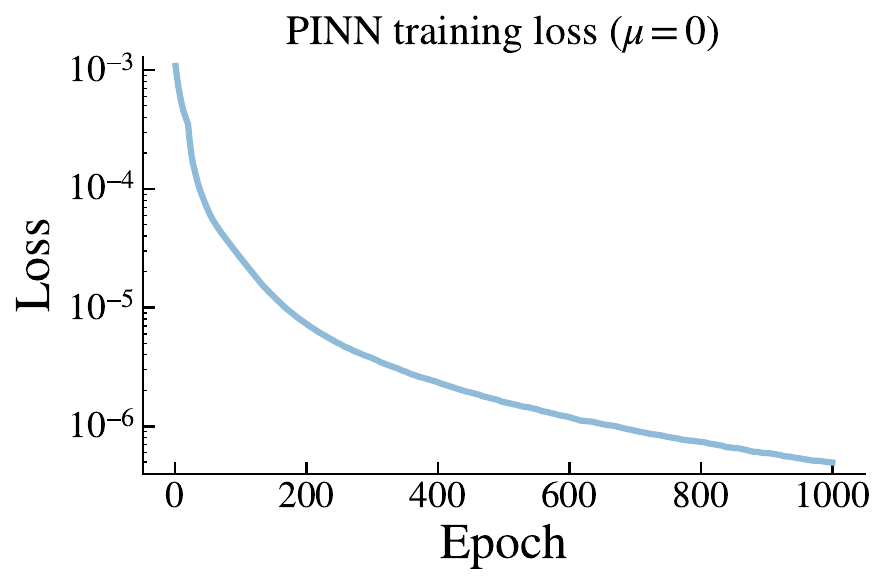}
    \end{minipage}\hfill
    \begin{minipage}[t]{0.45\linewidth}
    \raggedright (b)\\[-0.5ex]
        \includegraphics[width=1\linewidth]{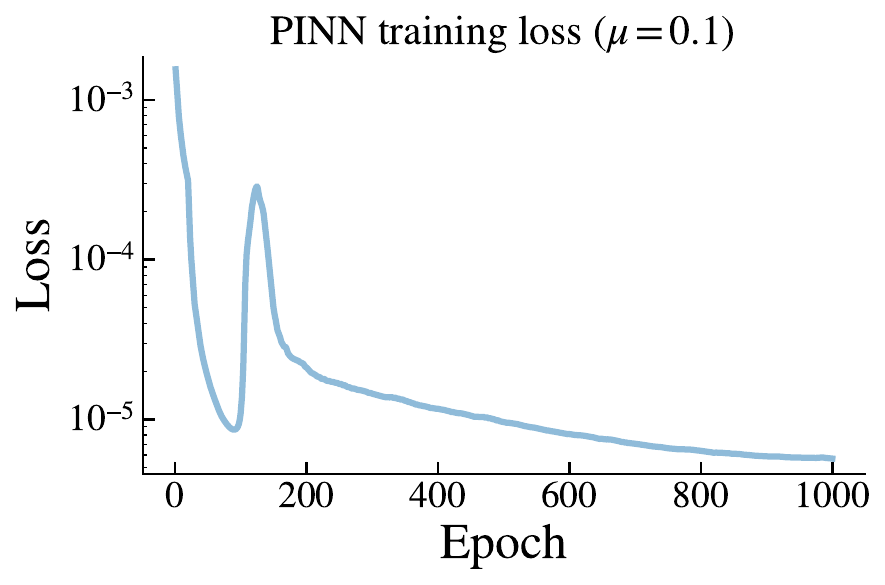}
    \end{minipage}
    \caption{
Logarithmic-scale plot of the loss function evolution during the training process of the model solving a partial differential equation. The horizontal axis represents the number of training epochs, while the vertical axis shows the loss values scaled as $\log_{10}(\text{Loss})$. The loss initially decreases rapidly, followed by transient oscillations around epoch 200, and eventually converges to a stable and minimal value, indicating effective model training and convergence.
}\label{fig:loss}
\end{figure*} 

\begin{figure*}
    \centering
    \begin{minipage}[t]{0.8\linewidth}
    \raggedright (a)\\[-0.5ex]
        \includegraphics[width=1\linewidth]{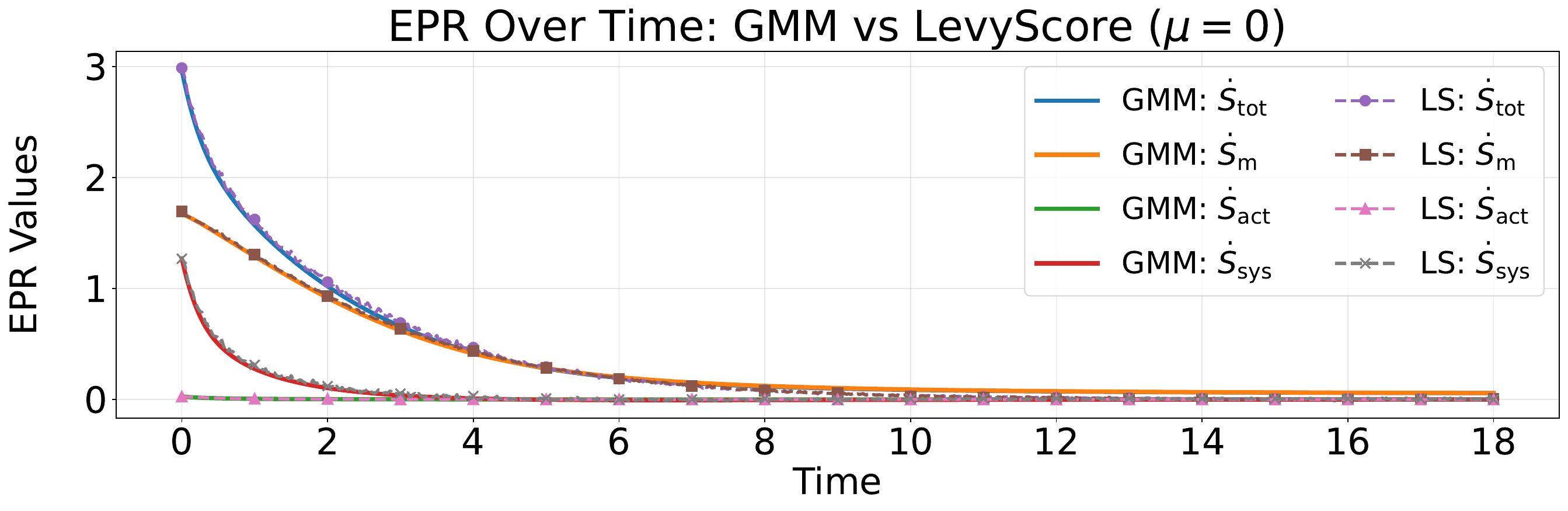}
    \end{minipage}\hfill
    \begin{minipage}[t]{0.8\linewidth}
    \raggedright (b)\\[-0.5ex]
        \includegraphics[width=1\linewidth]{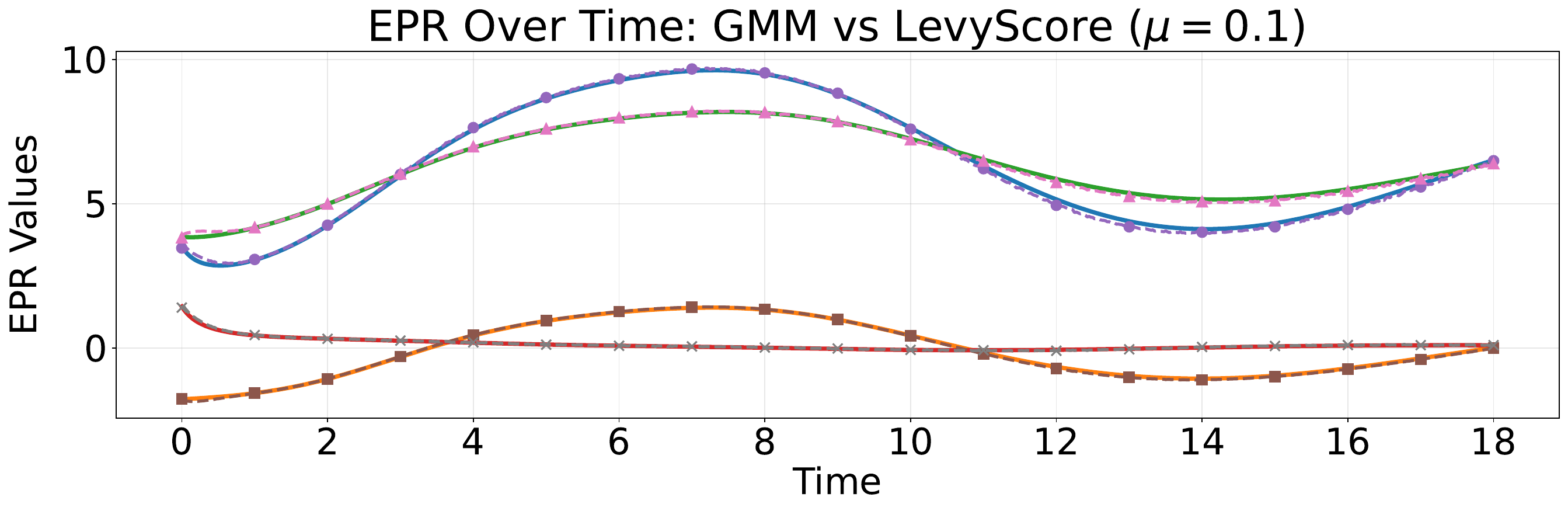}
    \end{minipage}
    \caption{Comparison between two numerical approaches: the Gaussian mixture method and our score-based particle algorithm. The solid lines represent results obtained using the Gaussian mixture method with physics-informed neural networks (PINNs), while the dashed lines correspond to results from our method.}\label{fig:error}
\end{figure*} 

\subsection{Verification of the fluctuation theorems and generalized second law}
This section is to verify the detailed fluctuation relation,
\begin{align}
    \frac{\rho_{\mathcal{R}}(\Sigma)}{\rho_{\mathcal{R}}(-\Sigma)} = e^{\Sigma}
    \quad \text{or equivalently} \quad
    \log\!\left[\frac{\rho_{\mathcal{R}}(\Sigma)}{\rho_{\mathcal{R}}(-\Sigma)}\right] = \Sigma.
\end{align}

To this end, we numerically generate $N=10^6$ trajectories for both cases ($\mu = 0$ and $\mu = 0.1$) using the discretization scheme in Eq.~\eqref{app:EMEXAMPLE1} with time step $\Delta t$ ($\Delta t=10^{-3}$ for $\mu=0$ and $\Delta t=10^{-6}$ for $\mu=0.1$), and compute the corresponding values of $\mathcal{R}$ defined in Eq.~\eqref{eqn:Rn}. The resulting histogram of $\mathcal{R}$ is shown in Fig.~\ref{fig:rhodistirbution}. Furthermore, Fig.~\ref{fig:verificationDFT}(a) and (b) in the \textbf{End Matter} present the numerical evaluation of $\rho_{\mathcal{R}}(\Sigma) / \rho_{\mathcal{R}}(-\Sigma)$, which exhibits excellent agreement with the theoretical prediction. Moreover, it is well known that the IFT can be derived from the DFT, and the same holds for the generalized second law. Therefore, the central results of the fluctuation theorems and the second law of the mian text can be explicitly verified within this example.

We note that the time intervals used in the two cases are different. For 
$\mu=0$, we use $T=1$ while  for  $\mu = 0.1$, we set $T = 0.035$.
(The larger $T$,
the larger value the distribution of  $\mathcal{R}$ tends to move to, and the numerical error of computing the ratio $\rho_{\mathcal{R}}(\Sigma) / \rho_{\mathcal{R}}(-\Sigma)$ 
is not easy to control. This is also the reason we use a much smaller time step size for $\mu=0.1$.)

\begin{figure*}
    \centering
    \begin{minipage}[t]{0.45\linewidth}
    \raggedright (a)\\[-0.5ex]
        \includegraphics[width=1\linewidth]{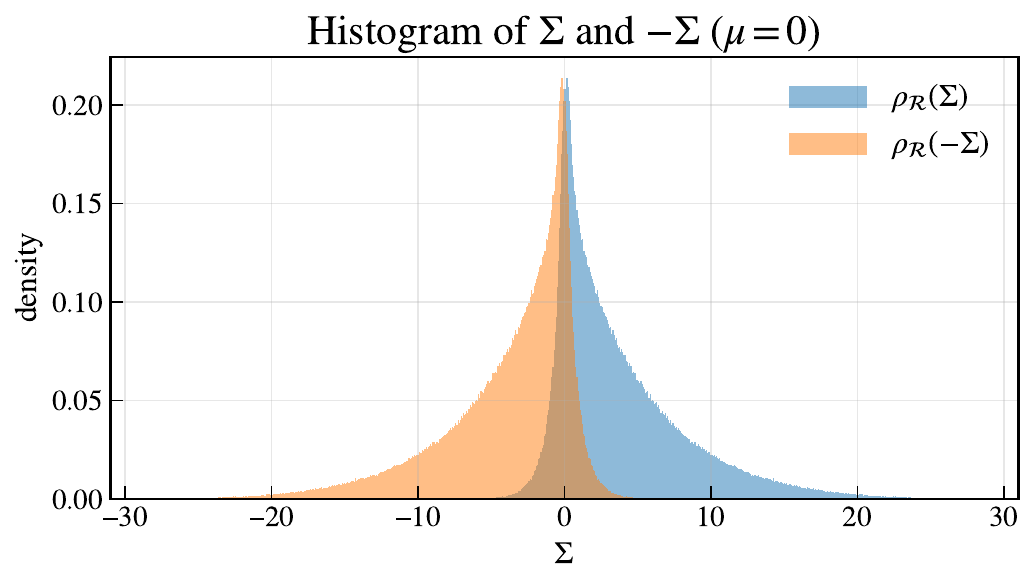}
    \end{minipage}\hfill
    \begin{minipage}[t]{0.45\linewidth}
    \raggedright (b)\\[-0.5ex]
        \includegraphics[width=1\linewidth]{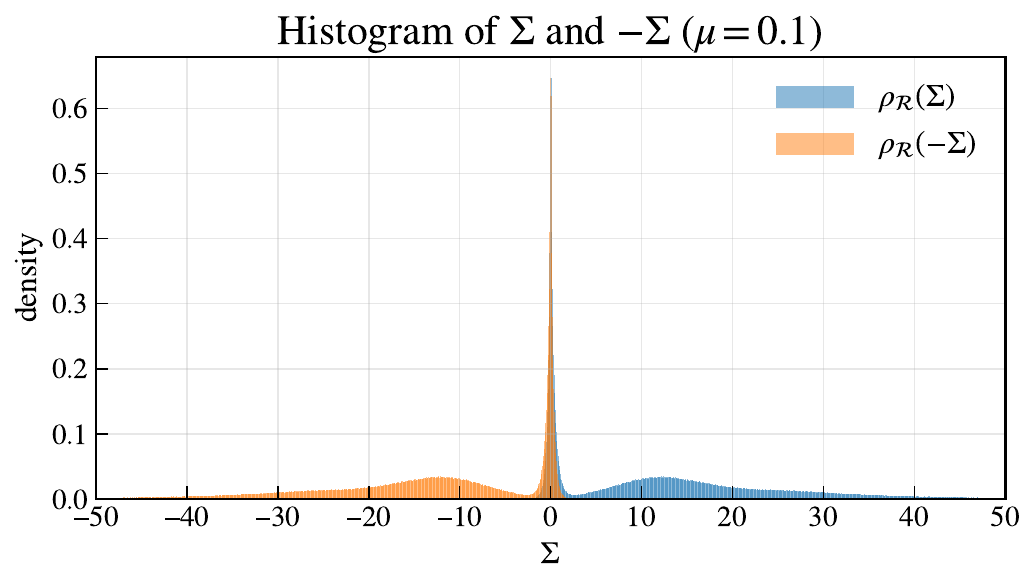}
    \end{minipage}
    \caption{
The empirical histogram distribution $\rho_{\mathcal{R}}(\Sigma)$ for a Brownian particle immersed in a periodic active bath is shown for $N=10^6$ sample trajectories and (a) $\mu = 0$ and (b) $\mu = 0.1$.}\label{fig:rhodistirbution}
\end{figure*}

\subsection{More parameters for the mean jump size of kicks}
In the main text, we examined two cases for the mean jump size of the kicks, $\mu=0$ and $\mu=0.1$, and observed that the system departs from equilibrium when $\mu$ increases from $0$ to $0.1$. 
We now explore intermediate values of $\mu$, increasing it gradually from $0$ to $0.025$, $0.05$, $0.057$, and finally $0.1$. 
As shown in FIG.~\ref{fig:moreparameter}, the active entropy production grows with $\mu$, rising from nearly zero to a finite positive constant, while both the medium and system entropy productions remain close to zero. 
This demonstrates that the departure from equilibrium is primarily driven by active fluctuations. 

\begin{figure*}
     \centering
    \includegraphics[width=\linewidth]{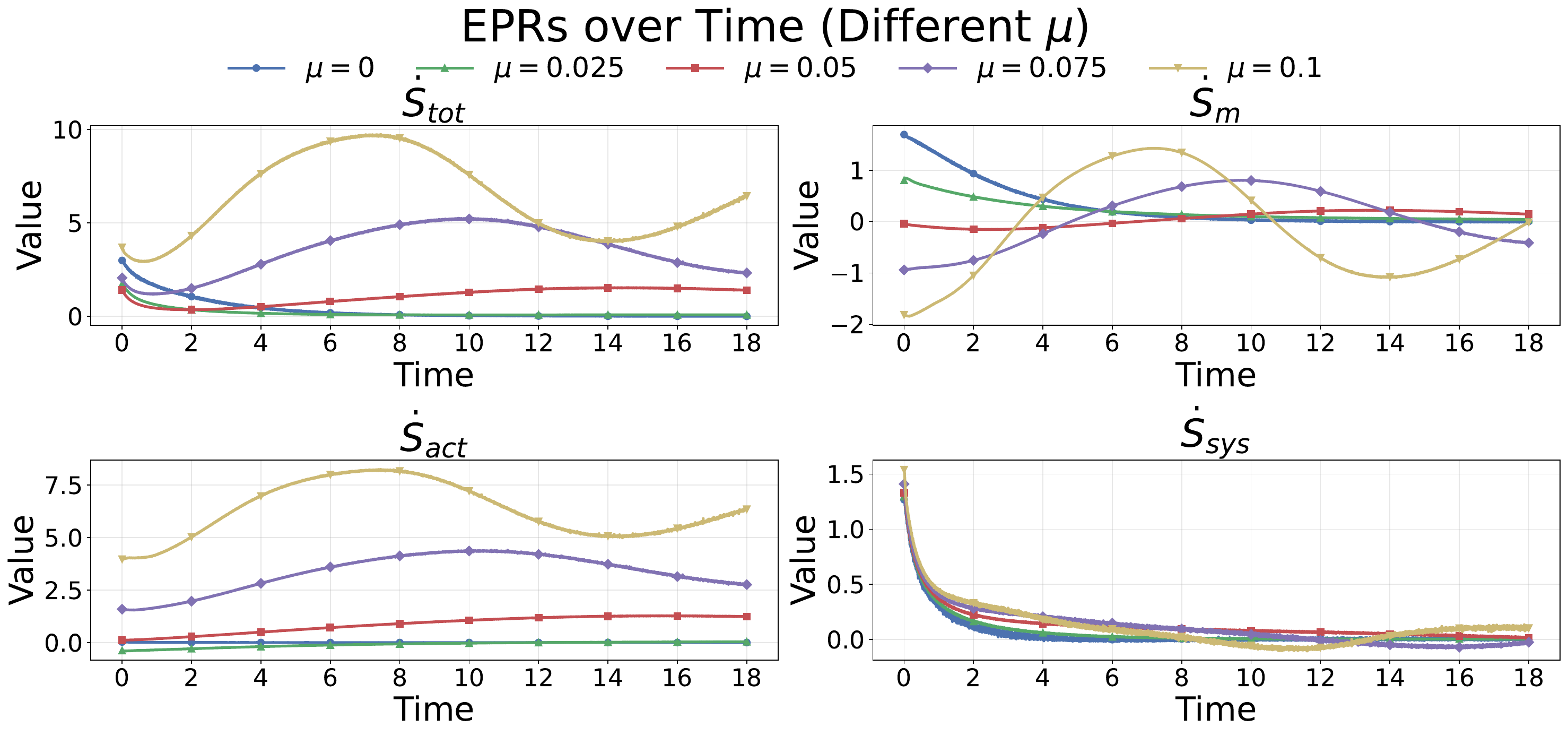} \caption{Total, active, medium, and system entropy productions for different  mean jump sizes $\mu$= 0 (blue lines with circle), 0.025 (green lines with upper triangle), 0.05 (red lines with squares), 0.075 (purple lines with prism) and 0.1 (yellow lines with lower triangle). 
The active entropy production increases from nearly zero to a finite positive value as $\mu$ grows from $0$ to $0.1$, while the medium and system entropy productions remain close to zero, indicating that the departure from equilibrium is dominated by active fluctuations.}\label{fig:moreparameter}
\end{figure*}

\subsection{Extended example for comparison: A Brownian particle in a harmonic potential with random jumps}
To compare with the example with periodic potential, we now consider a Brownian particle immersed in a harmonic active bath,  and the dynamics follows the SDE:
\begin{equation}\label{eqn:example1_extension}
    \begin{aligned}
        \mathrm{d} r(t) 
        =&\; -\frac{A}{\Gamma} r(t)\, \mathrm{d} t
        + \sqrt{2D_\mathrm{th}}\, \mathrm{d} W_t
        + \int z\mathcal{N}(\d t,\d z),
    \end{aligned}
\end{equation}
where the drift term is generated by the harmonic potential $V(r) = \tfrac{1}{2}A r^2$, the parameters used in the numerical simulations are listed in Table~\ref{tab:table1_extension}.

\begin{table*}
\caption{\label{tab:table1_extension}List of model parameters used in the simulations for the extended example.}
\begin{ruledtabular}
\begin{tabular}{cccc}
Parameter & Notation & Value & Dimension \\
\hline
Thermal diffusivity & $D_\mathrm{th}$ & 1.0 & nm$^2$/s \\
Viscous drag & $\Gamma$ & 1.0 & pN s/nm \\
Potential strength & $A$ & 1.0 & pN/nm \\
Mean of jump amplitude & $\mu$ & 0 & nm \\
Standard deviation of jump amplitude & $\sigma$ & $0.25$ & nm \\
Simulation time step & $\Delta t$ & $10^{-2}$ & s \\
Total simulation time & $T$ & 5 & s \\
\end{tabular}
\end{ruledtabular}
\end{table*}

\begin{figure*}
    \centering
    \begin{minipage}[t]{0.32\linewidth}
    \raggedright (a)\\[-0.5ex]
        \includegraphics[width=1\linewidth]{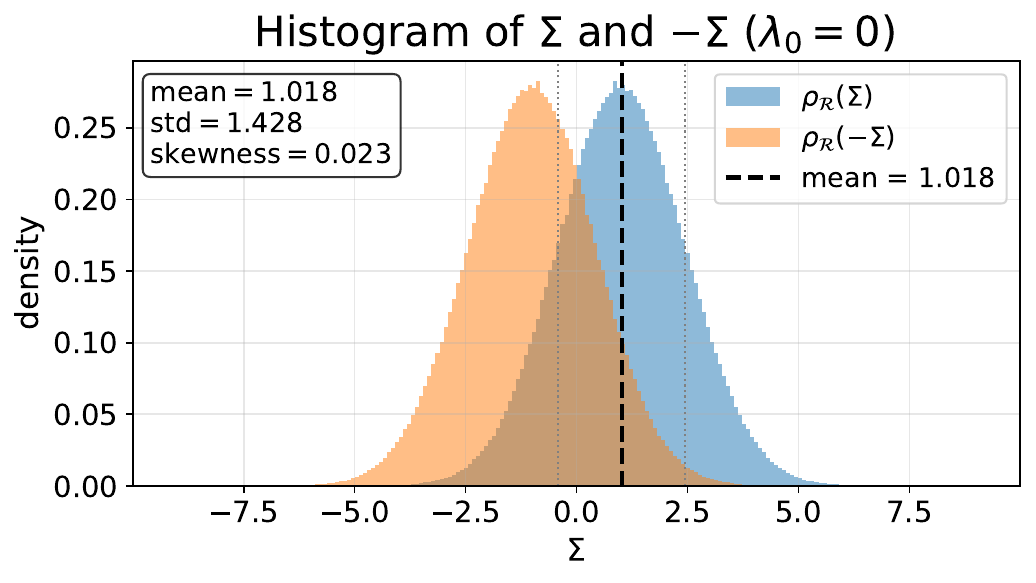}
    \end{minipage}\hfill
    \begin{minipage}[t]{0.32\linewidth}
    \raggedright (b)\\[-0.5ex]
        \includegraphics[width=1\linewidth]{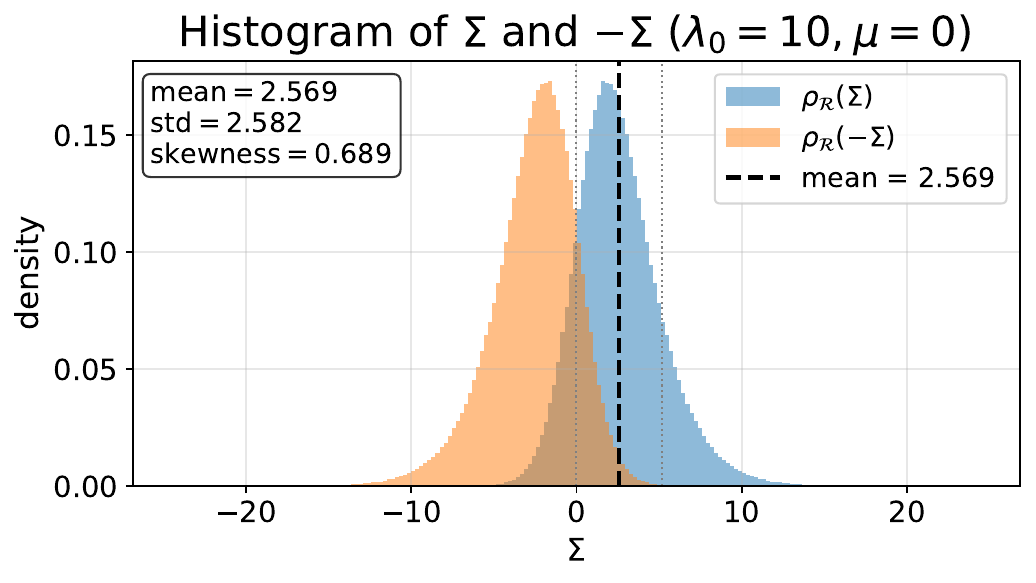}
    \end{minipage}
    \begin{minipage}[t]{0.32\linewidth}
    \raggedright (c)\\[-0.5ex]
        \includegraphics[width=1\linewidth]{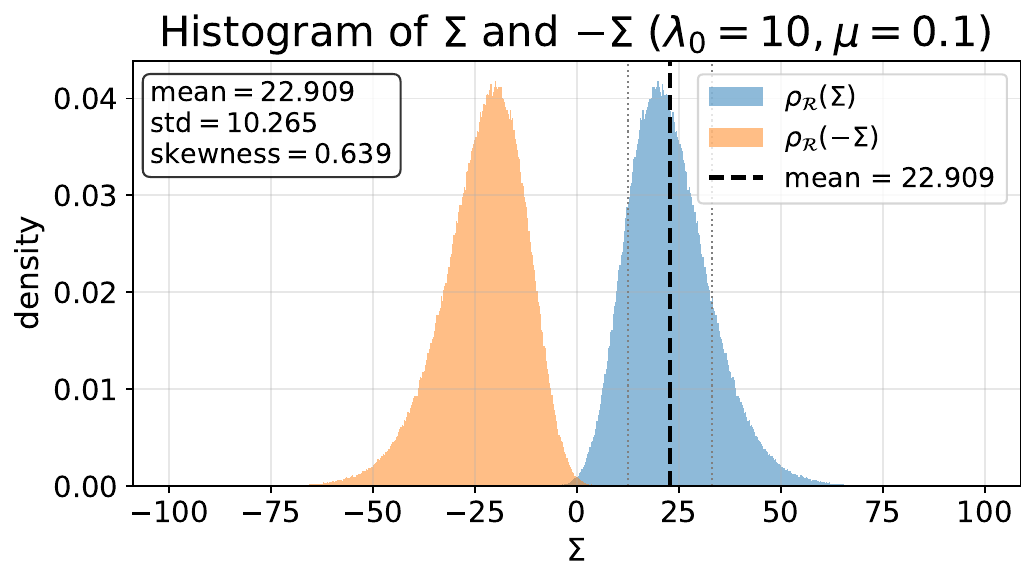}
    \end{minipage}
    \caption{
Empirical probability density $\rho_{\mathcal{R}}(\Sigma)$ of a Brownian particle confined in the harmonic potential. 
(a) $\lambda_0=0$ (no jumps), corresponding to purely thermal diffusion; 
(b) $\lambda_0=10$, $\mu=0$, corresponding to unbiased active jumps; 
(c) $\lambda_0=10$, $\mu=0.1$, corresponding to biased active jumps .
}\label{fig:quadratic_rhodistirbution}
\end{figure*} 

\begin{figure*}
    \centering
    \begin{minipage}[t]{0.32\linewidth}
    \raggedright (a)\\[-0.5ex]
        \includegraphics[width=1\linewidth]{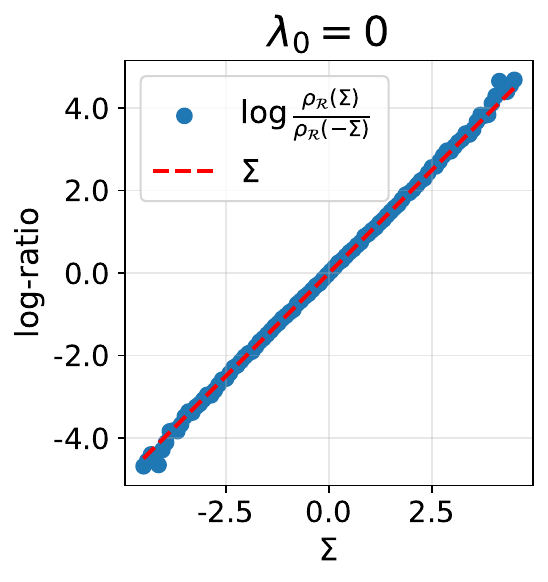}
    \end{minipage}\hfill
    \begin{minipage}[t]{0.32\linewidth}
    \raggedright (b)\\[-0.5ex]
        \includegraphics[width=1\linewidth]{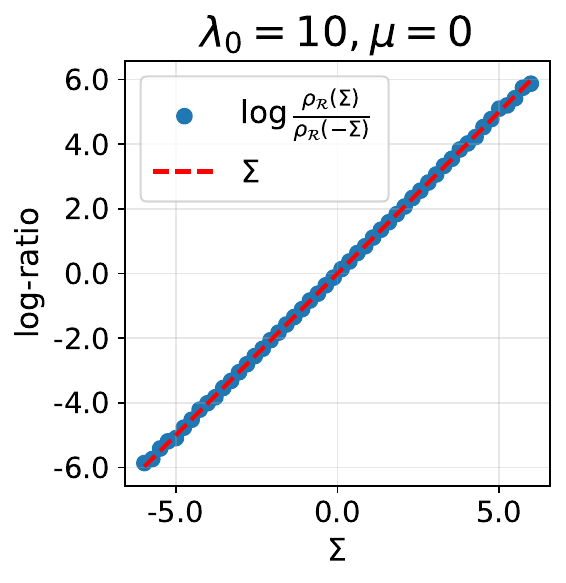}
    \end{minipage}
    \begin{minipage}[t]{0.32\linewidth}
    \raggedright (c)\\[-0.5ex]
        \includegraphics[width=1\linewidth]{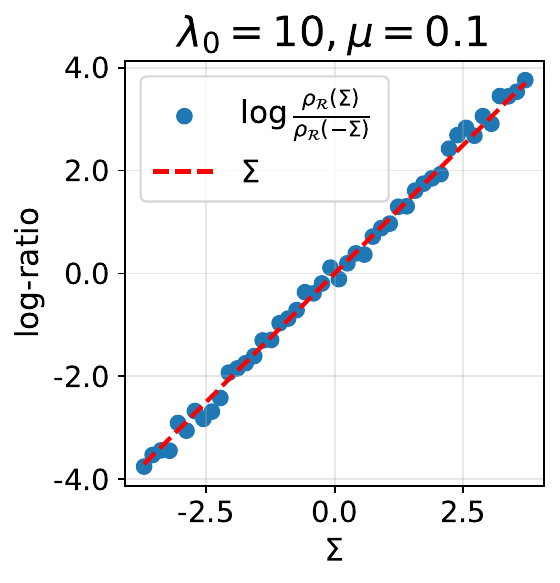}
    \end{minipage}
    \caption{Numerical verification of the detailed fluctuation theorem for the three parameter sets considered in Fig.~\ref{fig:quadratic_rhodistirbution}. 
The plots compare $\log[\rho_{\mathcal{R}}(\Sigma)/\rho_{\mathcal{R}}(-\Sigma)]$ as a function of $\Sigma$ with the theoretical prediction, showing clear agreement both in the absence and presence of L\'evy-type jump dynamics.
}\label{fig:quadratic_DFT}
\end{figure*}

To this end, we numerically generate $N = 10^6$ trajectories starting from an initial Gaussian distribution with mean $1$ and standard deviation $1$. Each trajectory is evolved up to the final time $T=5$ using the Euler–Maruyama discretization of Eq.~\eqref{eqn:example1_extension}. Three representative sets of parameters are examined: (i)~$\lambda_0 = 0$ (no jumps), (ii)~$\lambda_0 = 10$, $\mu = 0$ (unbiased active jumps), and (iii)~$\lambda_0 = 10$, $\mu = 0.1$ (biased active jumps with a positive mean). 

The resulting empirical distributions of $\mathcal{R}$ are shown in Fig.~\ref{fig:quadratic_rhodistirbution}. In the absence of jumps [Fig.~\ref{fig:quadratic_rhodistirbution}(a)], the distribution is narrow and nearly Gaussian. When L\'evy-type symmetric jumps are introduced [Fig.~\ref{fig:quadratic_rhodistirbution}(b)], the distribution broadens considerably and develops pronounced heavy tails, reflecting occasional large displacements triggered by active events. Further introducing asymmetry through a nonzero jump mean $\mu=0.1$ [Fig.~\ref{fig:quadratic_rhodistirbution}(c)] leads to a clear shift of the histogram toward positive values, indicating a net directional bias introduced by the active jump dynamics.

The verification of the detailed fluctuation relation for the three scenarios is displayed in Fig.~\ref{fig:quadratic_DFT}. The numerical results remain in excellent agreement with the theoretical linear prediction even when L\'evy-type noise is included, confirming the robustness of DFT under both symmetric and asymmetric non-Gaussian perturbations.

\section{Example 2--An active Brownian particle cross-linked to a Rouse networked polymer}

This Appendix provides additional details on the second example discussed in the main text.

The dynamics is governed by
\begin{equation}\label{app:example2}
    \begin{aligned}
       \Gamma \frac{\d r_{\mathrm{A},j}}{\d t}=& - k\sum_{l=1}^m (r_{\mathrm{A},j} - r^{(l)}_{1,j} ) +  \eta_{\mathrm{th},j}(t) +  \eta_{\mathrm{act},j}(t),\\
       \Gamma\frac{\d r^{(l)}_{i,j}}{\d t}=& - k\left(2 r^{(l)}_{i,j}  -r^{(l)}_{i+1,j} -r^{(l)}_{i-1,j} \right)+  \eta^{(l)}_{\mathrm{th},j},\quad i=1,2\cdots,n,\\
        \Gamma\frac{\d r^{(l)}_{n+1,j}}{\d t}=& 0,
    \end{aligned}
\end{equation}
for $l \in \{1,2,\cdots,m\}$, and $j \in \{1,2\}$, with fixed boundary conditions for the terminal beads ($r^{(l)}_{n+1,j}$). The parameters used in this experiment are listed in Table \ref{tab:table2}.

\begin{table*}[htb] 
\caption{\label{tab:table2}List of model parameters used in simulations for Example 2.}
\begin{ruledtabular}
\begin{tabular}{cccc}
 Parameter  & Notation & Value & Dimension 
 \\ \hline
 Thermal energy & $k_B\mathcal{T}$ & 4.114 & pN nm\\
 Viscous drag  & $\Gamma$  
 & 30 & pNs/nm \\
 Poisson parameter & $r_\mathrm{0}$ &  5 & 1\\
 Jump size & $v_0$ & 0.1 & nm\\
  Simulation time step & $\Delta t$ & $10^{-3}$ & $s$ \\
Spring constant & k & 5 & pN/nm\\
\end{tabular}
\end{ruledtabular}
\end{table*}

The active fluctuation $\bm{\eta}_\mathrm{act}$ is modeled as the compound Poisson process $\bm{\eta}_\mathrm{act}(t)= \sum_{j=1}^{N_t}v_{0}\bm{\sigma}_{\mathrm{D}}(t)\delta(t-t_j)$ where $v_{0}$ is the constant speed of self-propulsion and $\bm{\sigma}_{\mathrm{D}}(t)$ takes four possible values, corresponding to four directions along the positive and negative \( x \)-axis and \( y \)-axis, with each value having a probability \( r_i \), \( i = 1, 2, 3, 4 \), $N_t$ is a Poisson process with a fixed
parameter $\lambda_0$. In the 2D $xy$-plane, we consider compound Poisson noise in two scenarios: unbiased and biased. In the unbiased case, each jump of the active bead has four possible directions: $(v_0, 0)$, $(0, v_0)$, $(-v_0, 0)$, and $(0, -v_0)$, all with equal probability. In contrast, the biased case corresponds to different probabilities assigned to each jump direction. Specifically for the biased case, we assign a probability of 0.7 to the jump in the direction $(v_0, 0)$, while the other three directions share an equal probability of 0.1 each.

The boundary conditions for the arms in our study are the pinned arms where the last $n$-th beads in the arms are fixed in space, i.e., $\Gamma\frac{\d\bm{r}^{(l)}_{n+1}}{\d t}\equiv 0,\quad l\in\{1, \cdots,m\}$. We consider the cases where $m=3,4$ and $n=1,3,7$. We set the initial state of the system to follow a Gaussian distribution, where the mean configuration ensures a distance of 0.5 between adjacent particles, and the covariance matrix is specified as the identity matrix. FIG. \ref{fig:3-2} shows the EP for the active polymer system with $m = 3, 4$ arms, where each arm consists of $n = 1, 3, 7$ Brownian beads and a fixed end bead in a 2D plane. As the number of arms and beads increases, the time required for the system to reach the steady state becomes longer. The panels show that, regardless of the number of arms or arm lengths, the system described by \eqref{app:example2} approximates a state near equilibrium,
even in the presence of active fluctuations. This behavior
occurs because the active fluctuations are small compared
to other forces, and the fixed boundary condition introduces an additional force that counteracts these fluctuations.

\begin{figure*}[htbp]
\centering
\includegraphics[width=1\linewidth]{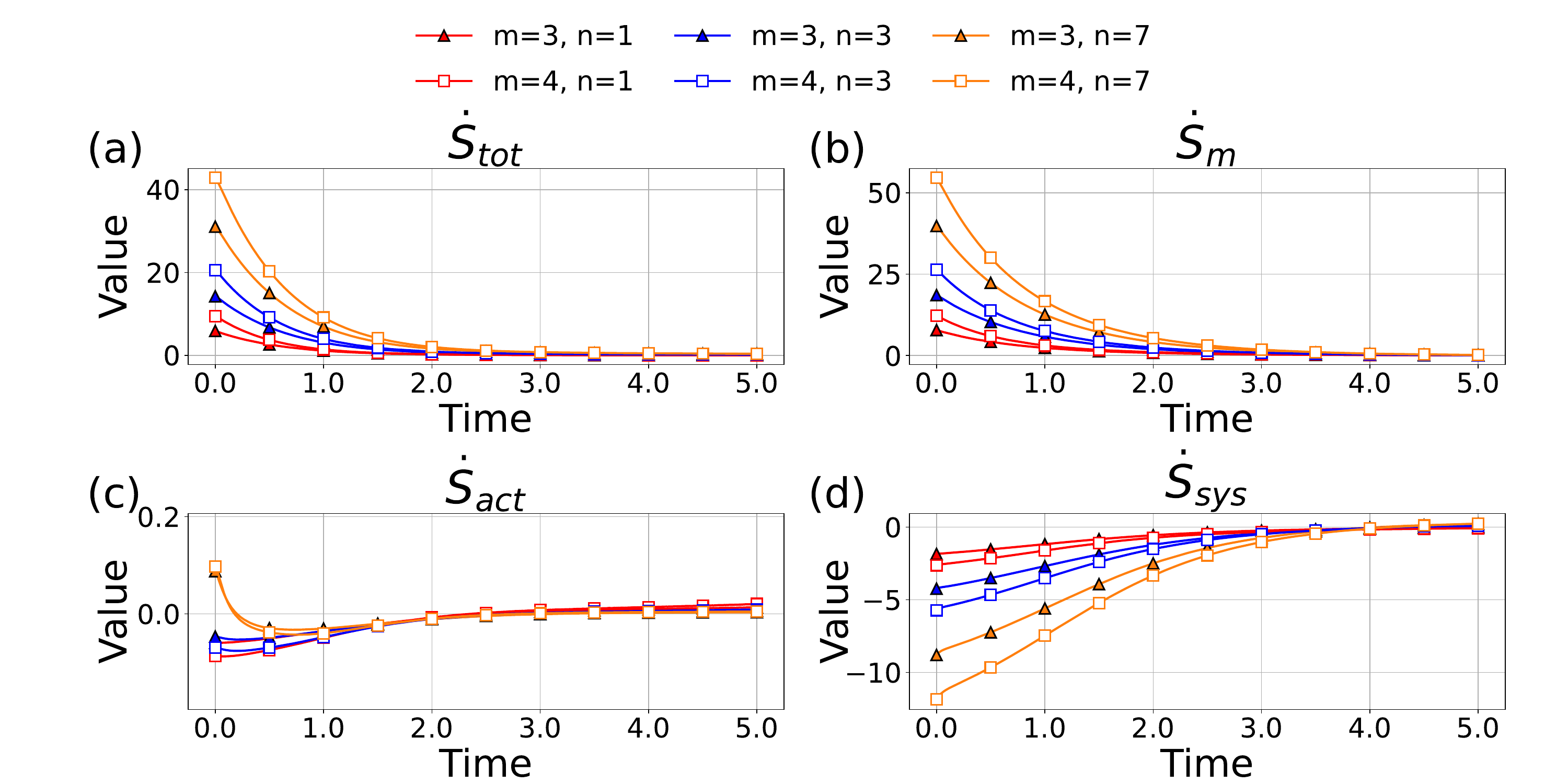} 
    \caption{EPRs of an active polymer system under different parameter combinations for the unbiased jump scenario. Each panel shows one component of the EPR: (a) $\dot{S}_\mathrm{tot}$, (b) $\dot{S}_\mathrm{m}$, (c) $\dot{S}_\mathrm{act}$, and (d) $\dot{S}_\mathrm{sys}$. Curves correspond to varying $(m, n)$, with $m=3$ (solid triangle markers) and $m=4$ (hollow square markers), and $n=1$ (red), $3$ (blue), and $7$ (orange).}
    \label{fig:3-2}
\end{figure*}
\begin{figure*}[h]
\centering
\includegraphics[width=1\linewidth]{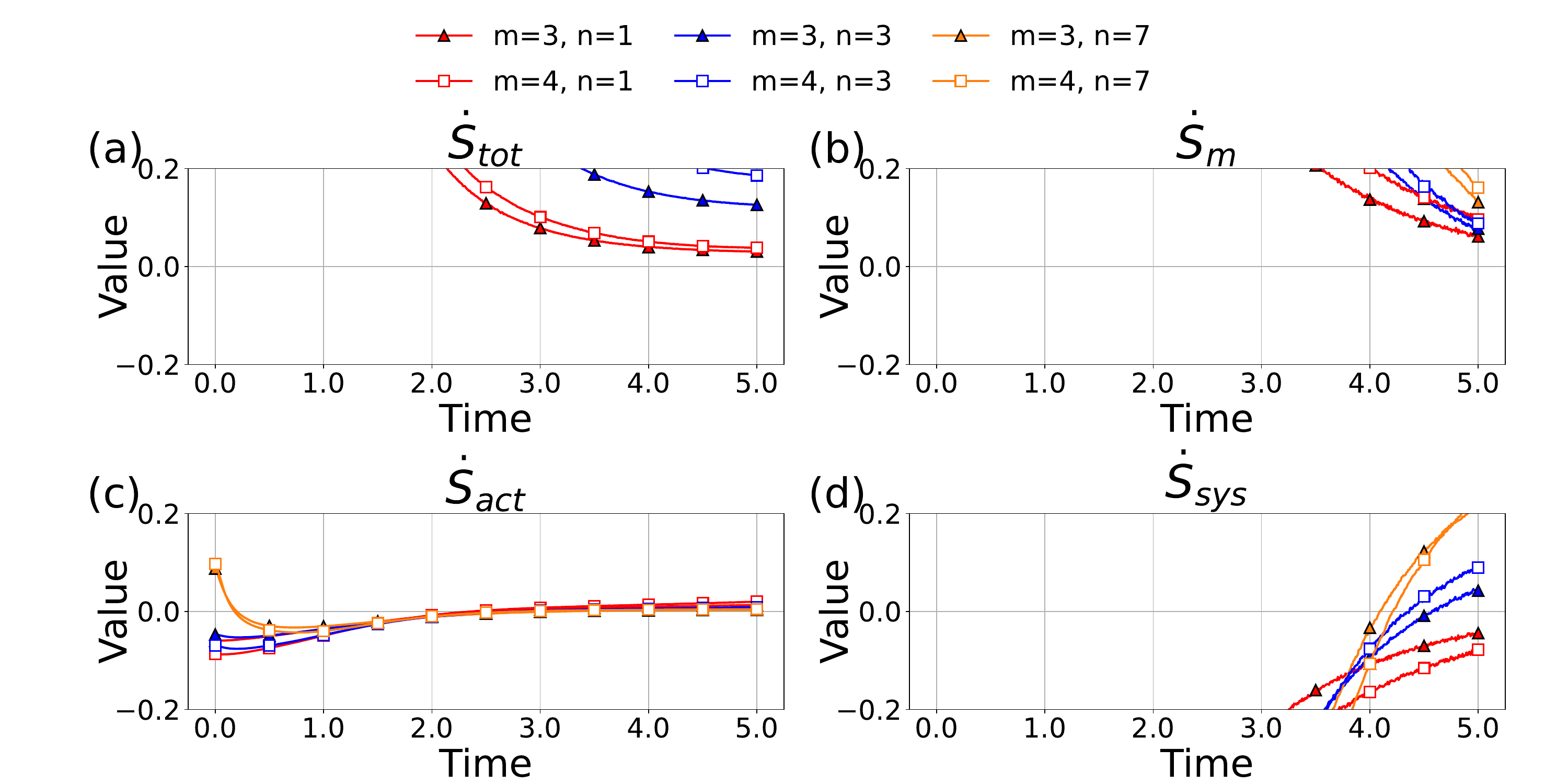} 
    \caption{Same data as in Fig. \ref{fig:3-2}, now plotted with a uniform y-axis scale across all subpanels to facilitate comparison.}
\end{figure*}

Our second example examines an active polymer system comprising an active Brownian particle (ABP) cross-linker connected to ordinary Brownian beads.

To demonstrate the validity of the algorithm, we calculate the total variation distance between $P^\mathrm{MC}$ and $P^\mathrm{NN}$ for the case \( m = 4 \) and \( n = 7 \), as well as for each dimension (with a total dimensionality of 58 in this case), as shown in FIG.~\ref{fig:example2TV}.
\begin{figure}
    \centering    \includegraphics[width=0.8\linewidth]{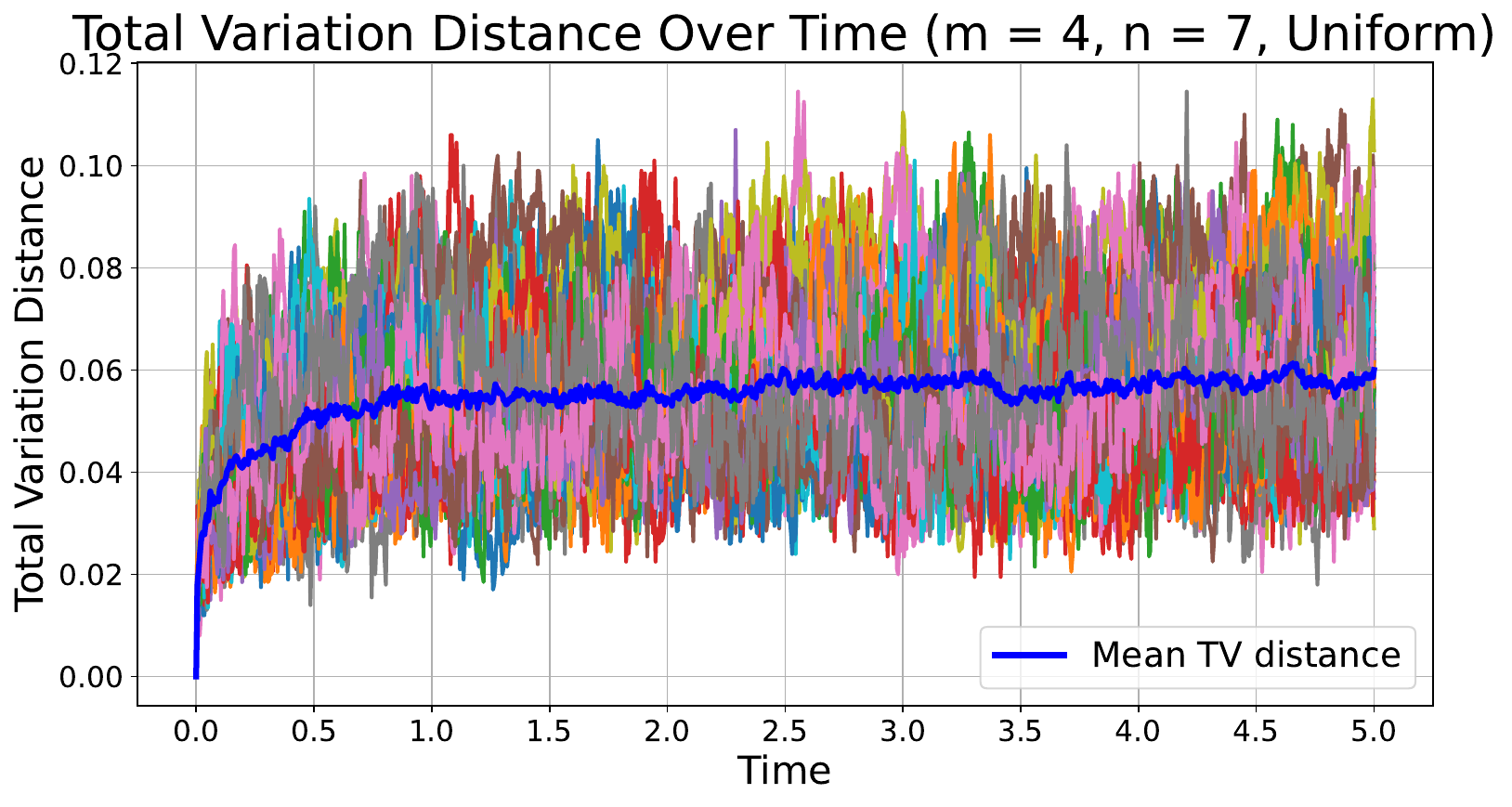}
    \caption{Total variation distances between $P^\mathrm{MC}$ and $P^\mathrm{NN}$ for the active Brownian particle cross-linker 4 arms and $7$ ordinary Brownian beads.}
    \label{fig:example2TV}
\end{figure}

FIG. \ref{fig:example2:snapshot_uniform} shows snapshots of the sample points for the case of unbiased jump noise at time $t=5$.

\begin{figure*}
\includegraphics[width=0.48\linewidth]{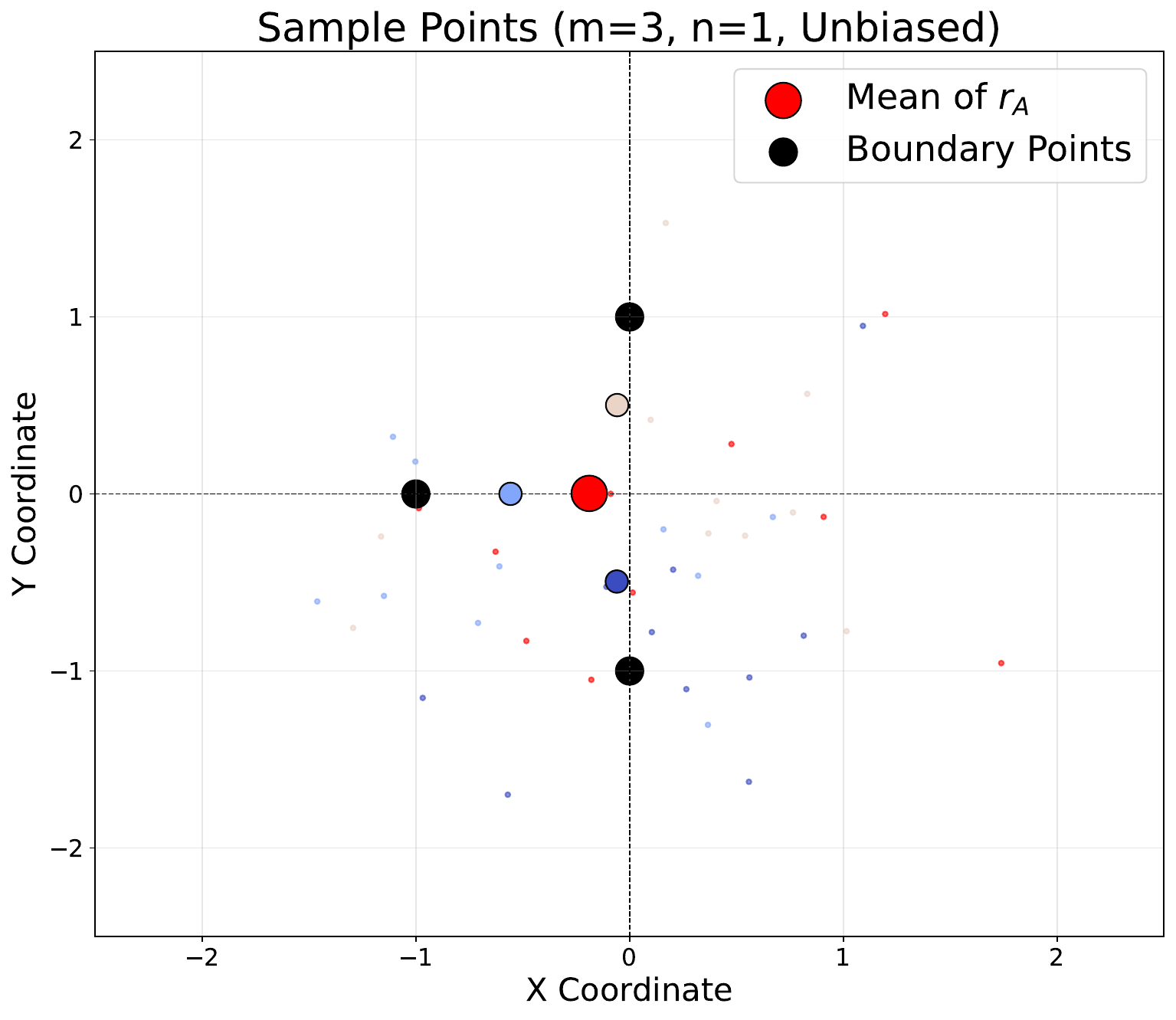}  
\includegraphics[width=0.48\linewidth]{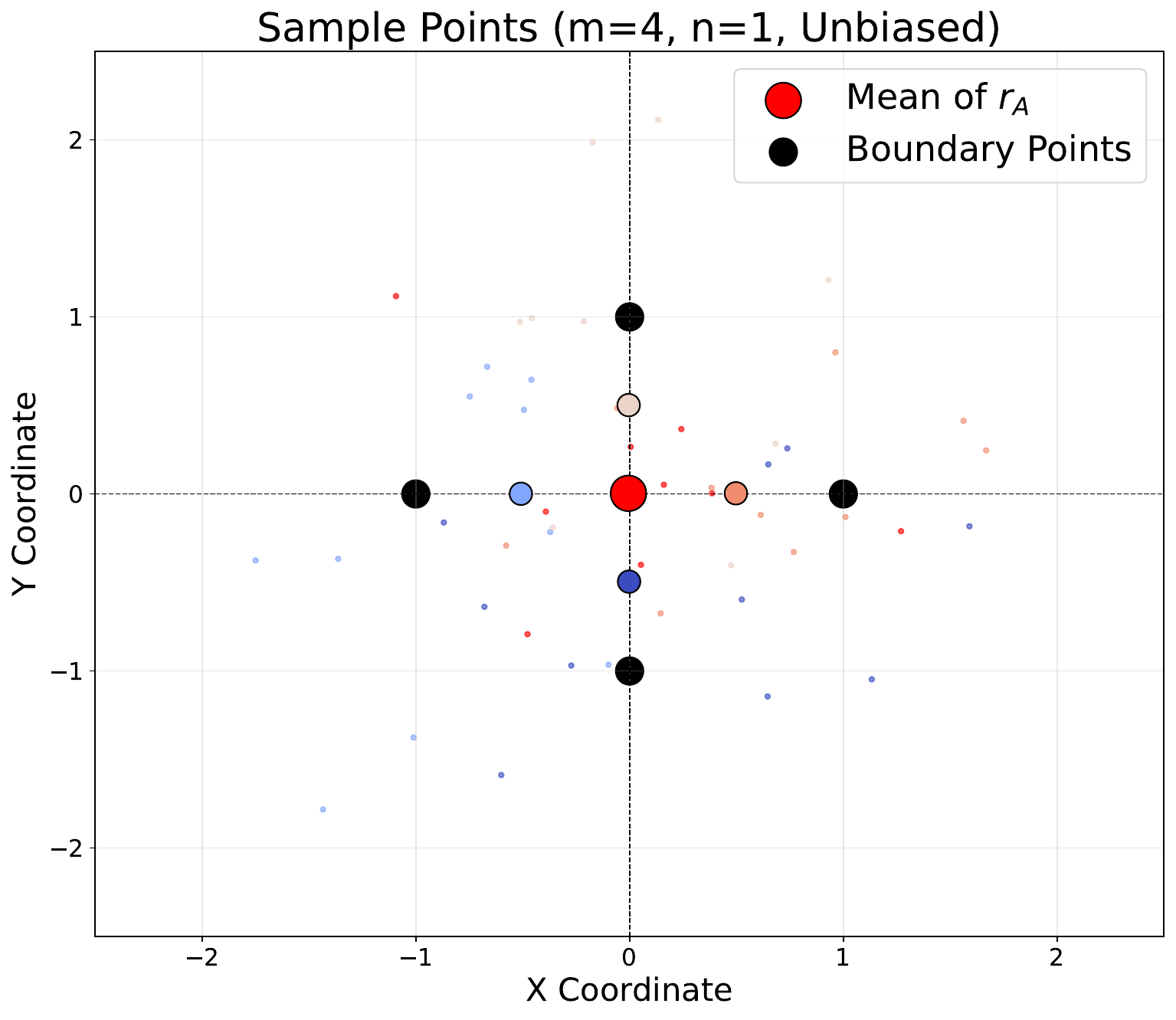}  
\includegraphics[width=0.48\linewidth]{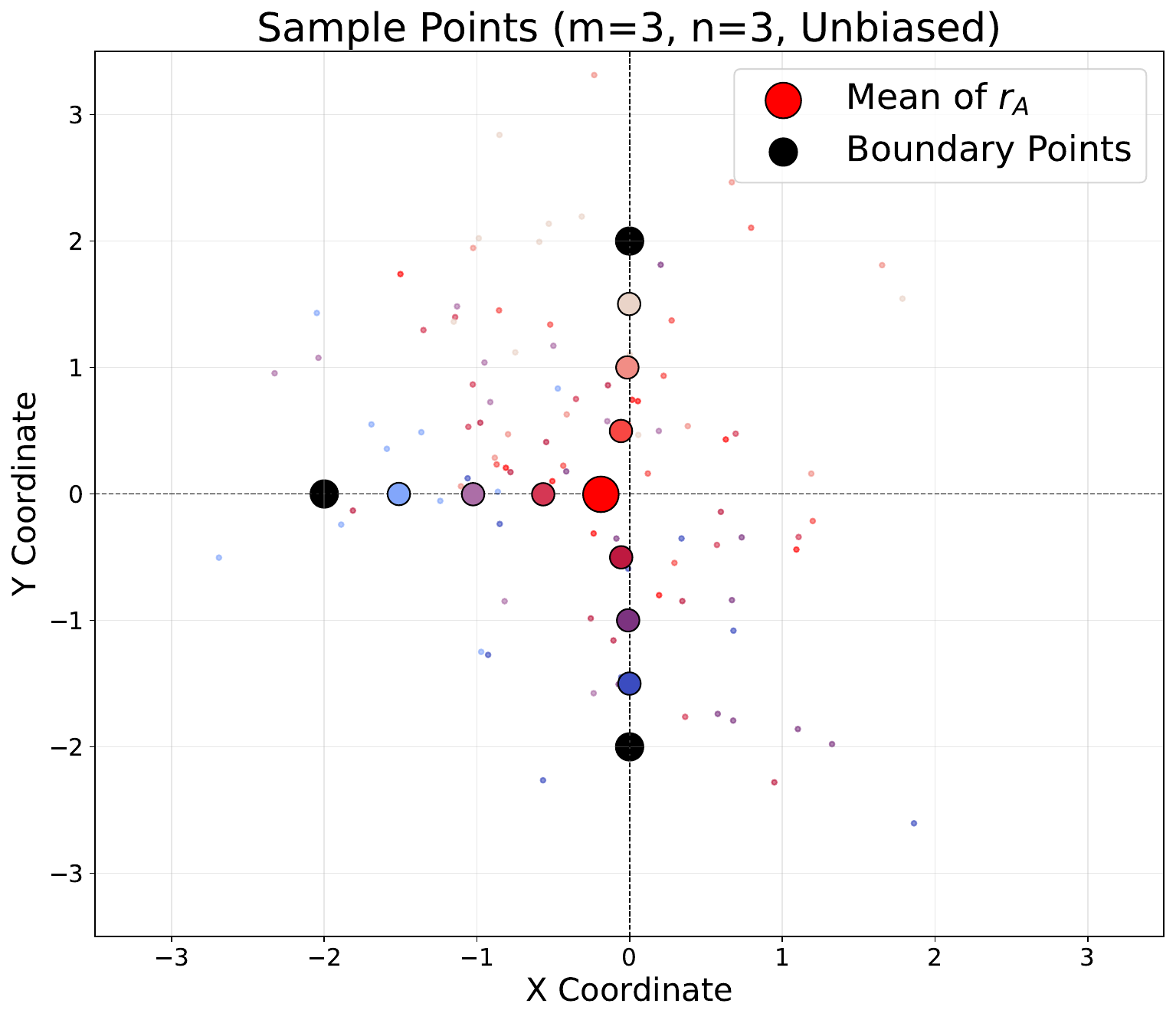} 
\includegraphics[width=0.48\linewidth]{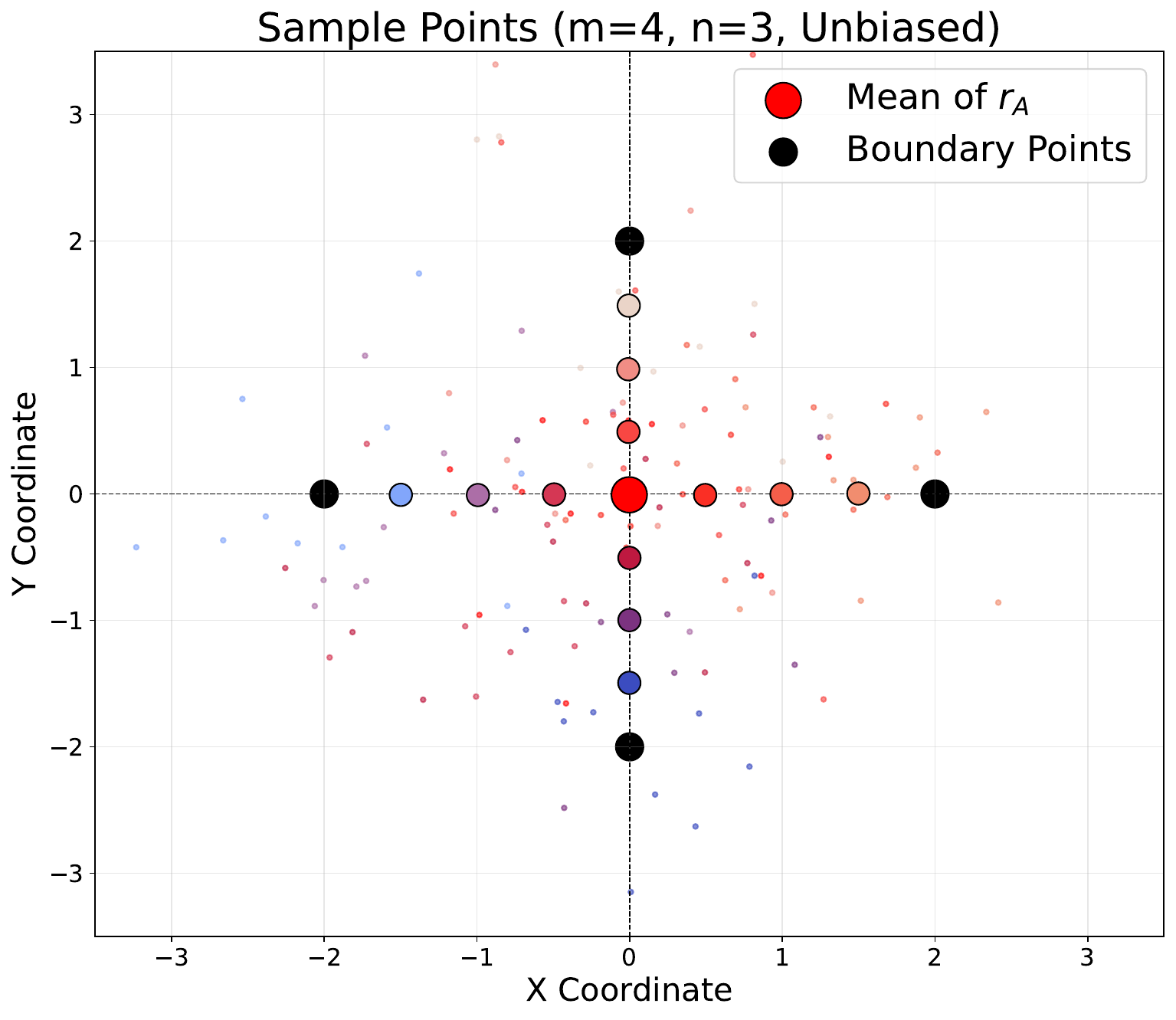} 
\includegraphics[width=0.48\linewidth]{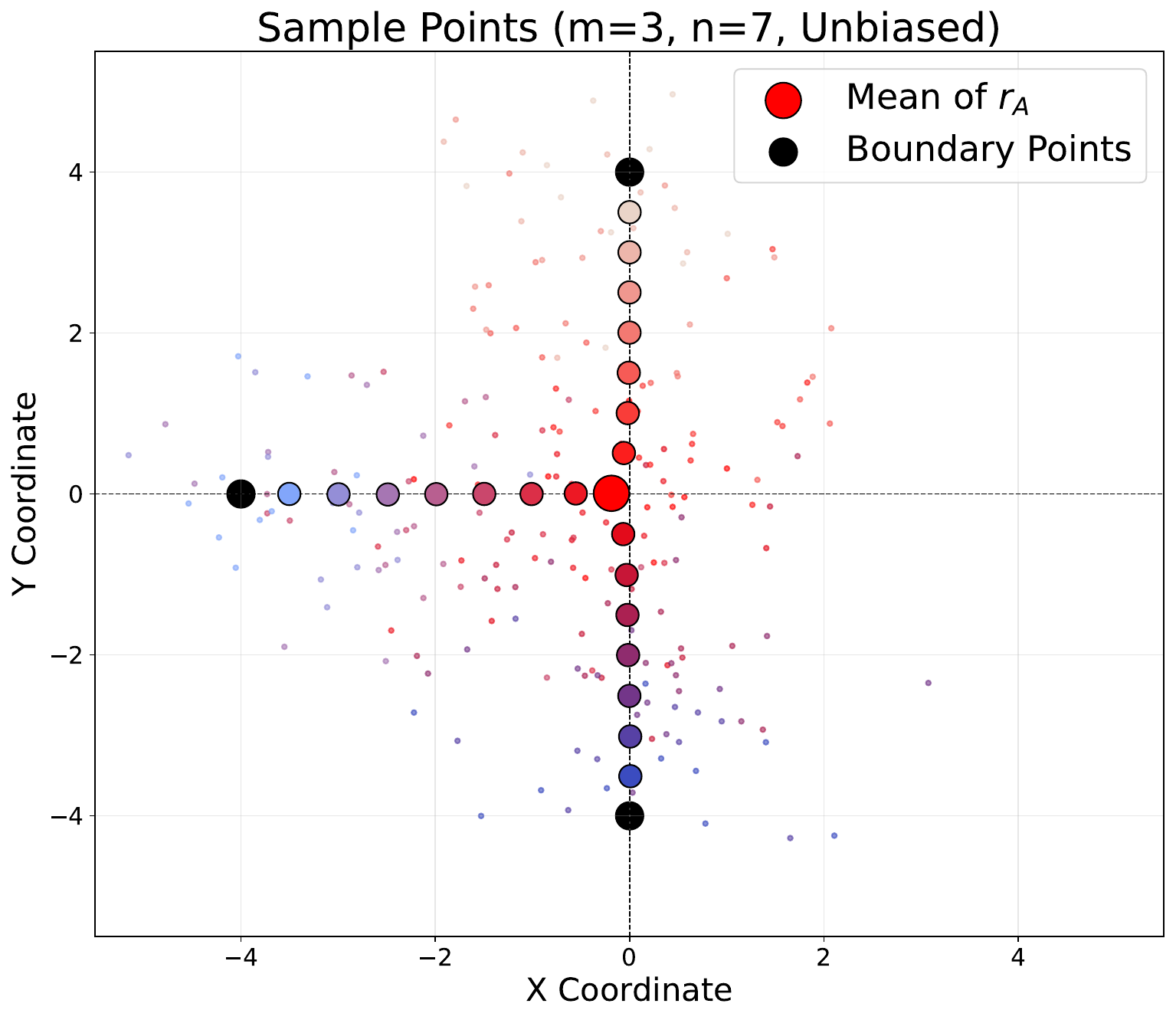} 
\includegraphics[width=0.48\linewidth]{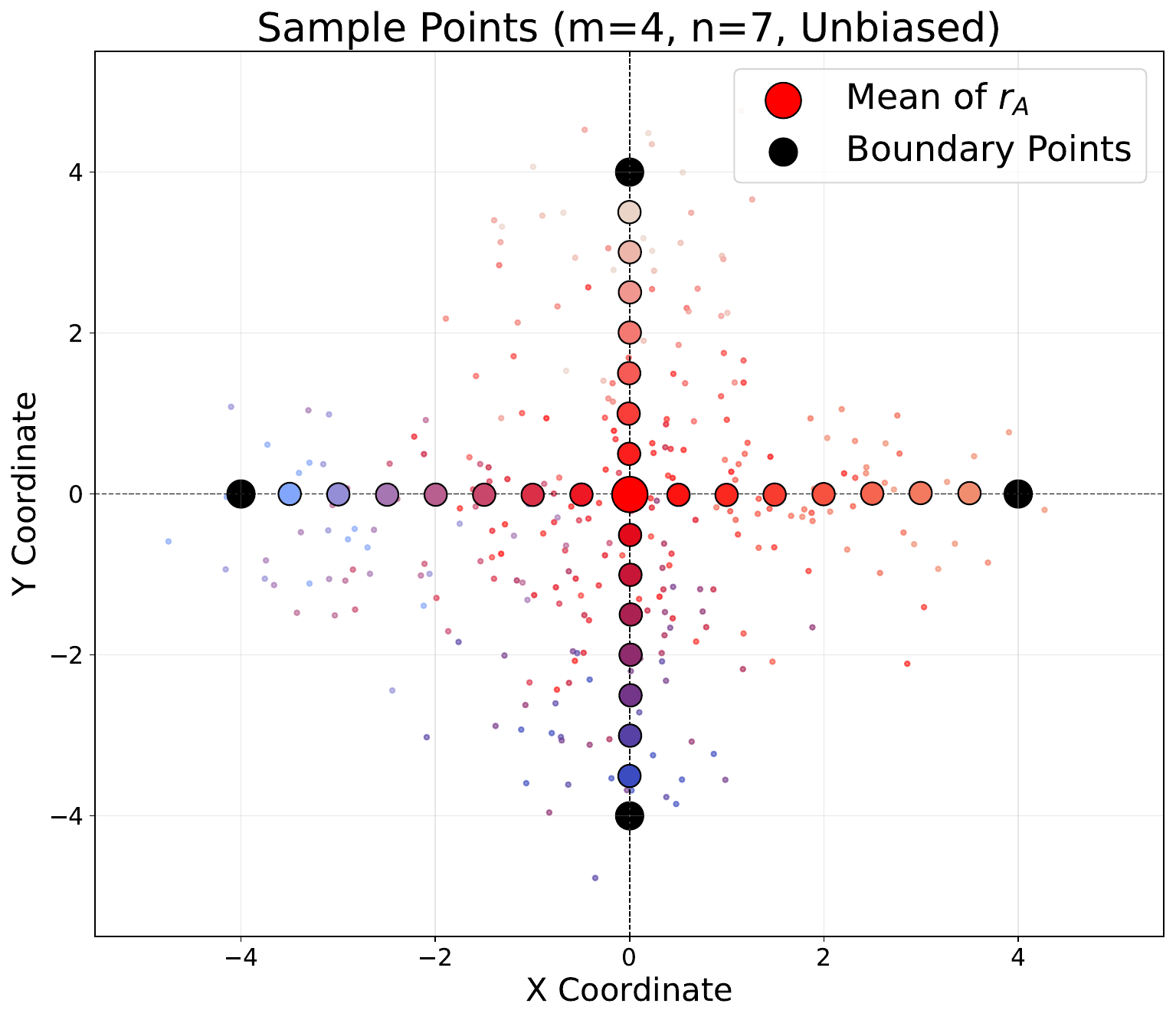}  
    \caption{Snapshot of the sample points at time $t=5$ for unbiased jump noise case.}\label{fig:example2:snapshot_uniform}
\end{figure*}

For the biased jump noise case, FIG.~\ref{fig:example2:entropy_biased} presents the EP, while FIG.~\ref{fig:example2:snapshot_nonuniform} shows the corresponding snapshots at time $t=5$.

\begin{figure*}
    \includegraphics[width=1\linewidth]{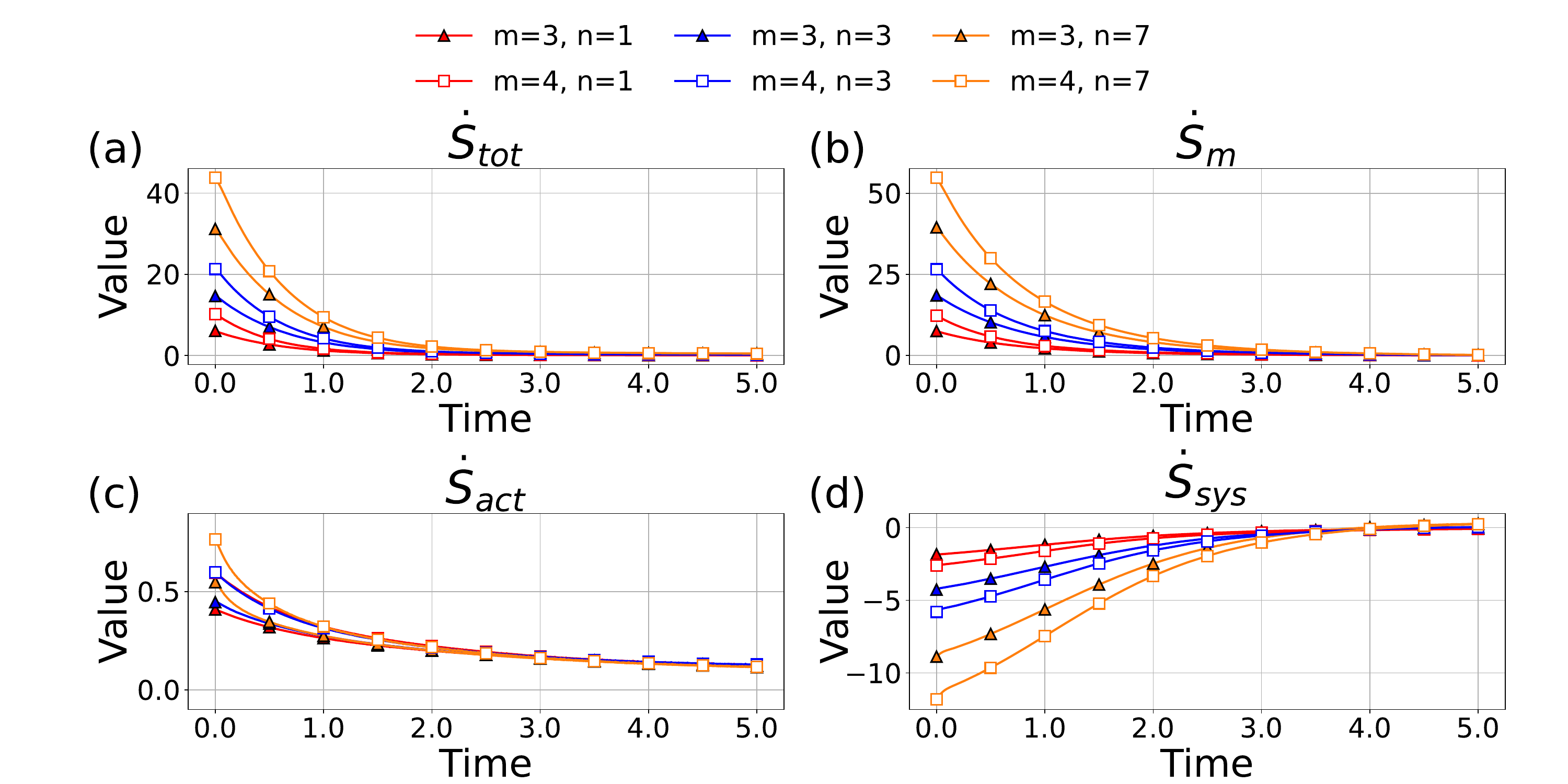} 
    \caption{EPRs of an active polymer system under different parameter combinations for the biased jump scenario. Each panel shows one component of the EPR: (a) $\dot{S}_\mathrm{tot}$, (b) $\dot{S}_\mathrm{m}$, (c) $\dot{S}_\mathrm{act}$, and (d) $\dot{S}_\mathrm{sys}$. Curves correspond to varying $(m, n)$, with $m=3$ (solid triangle markers) and $m=4$ (hollow square markers), and $n=1$ (red), $3$ (blue), and $7$ (orange).}
    \label{fig:example2:entropy_biased}
\end{figure*}

\begin{figure*}
    \includegraphics[width=1\linewidth]{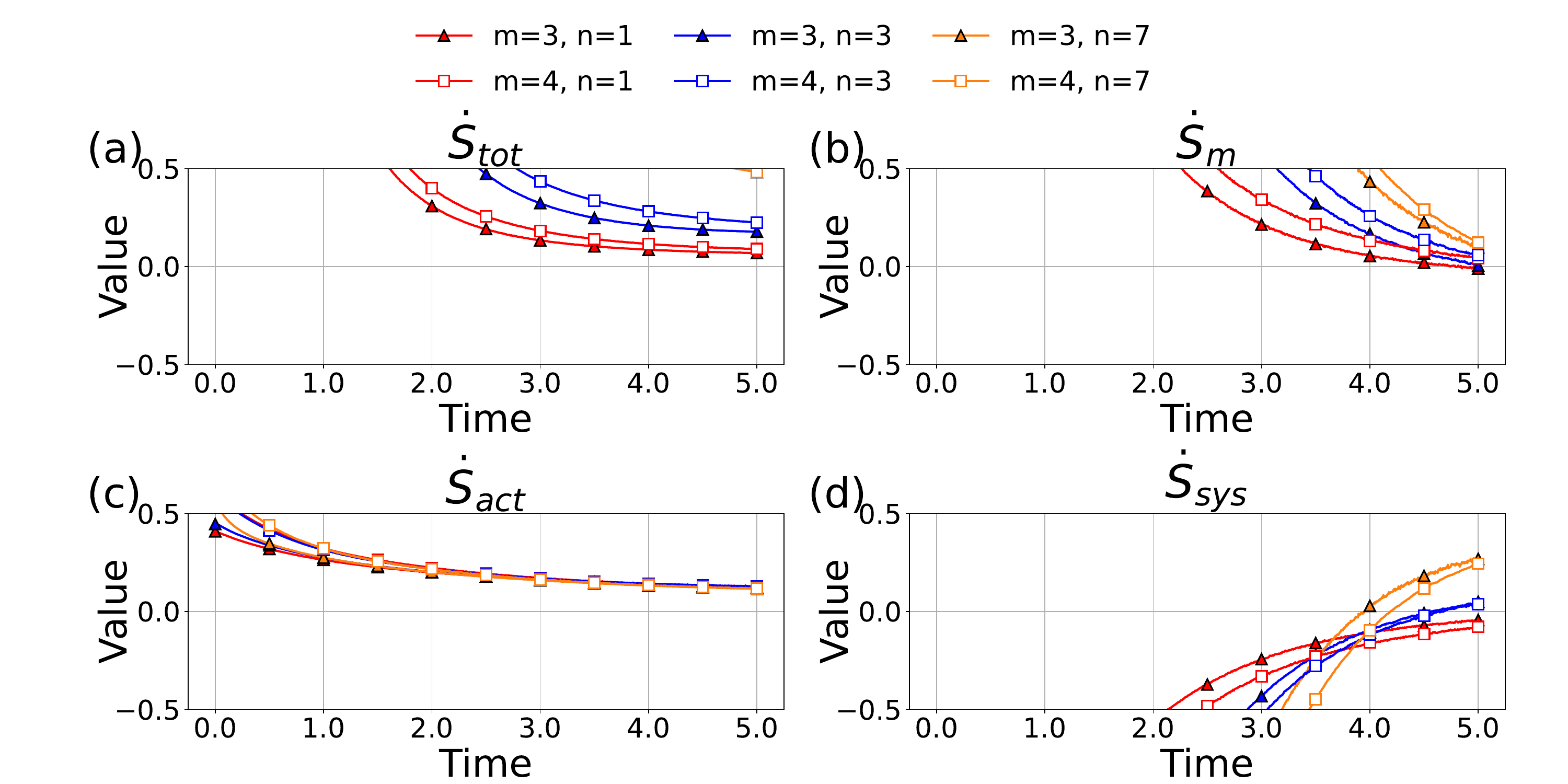} 
    \caption{Same as Fig. \ref{fig:example2:entropy_biased}, but with a consistent y-axis scale across all subplots.}
\end{figure*}
\begin{figure*}
\includegraphics[width=0.48\linewidth]{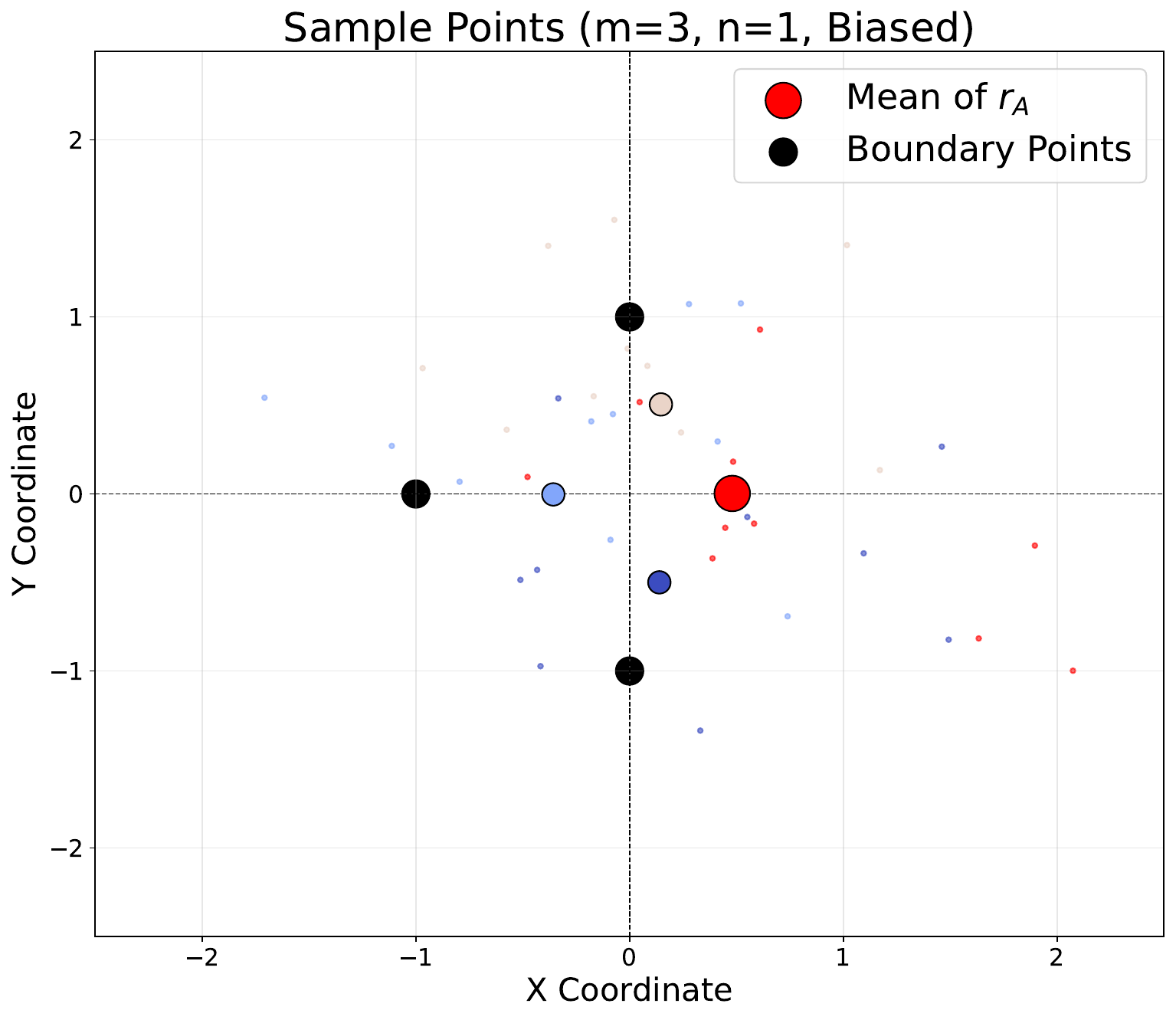}  
\includegraphics[width=0.48\linewidth]{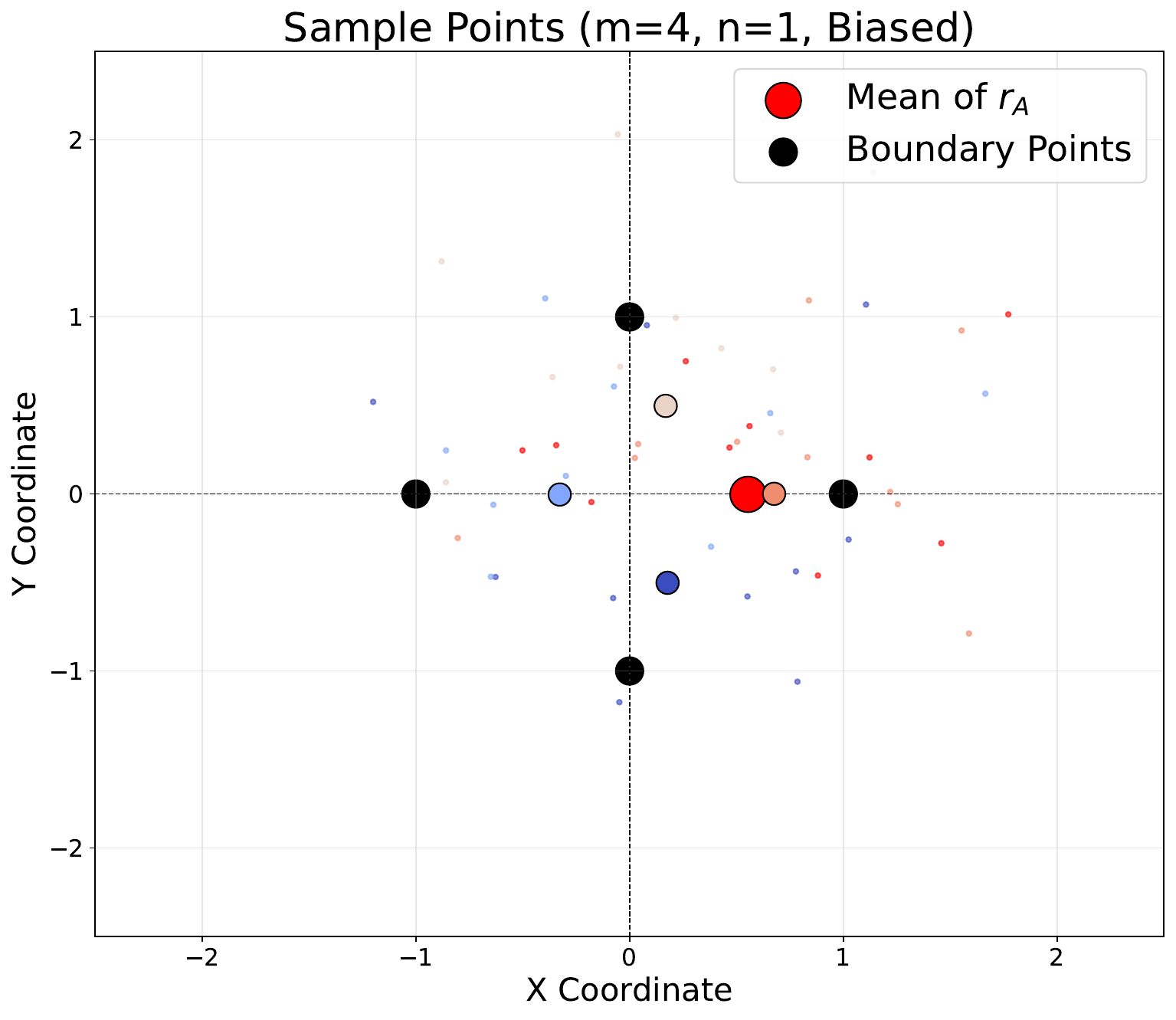}  
\includegraphics[width=0.48\linewidth]{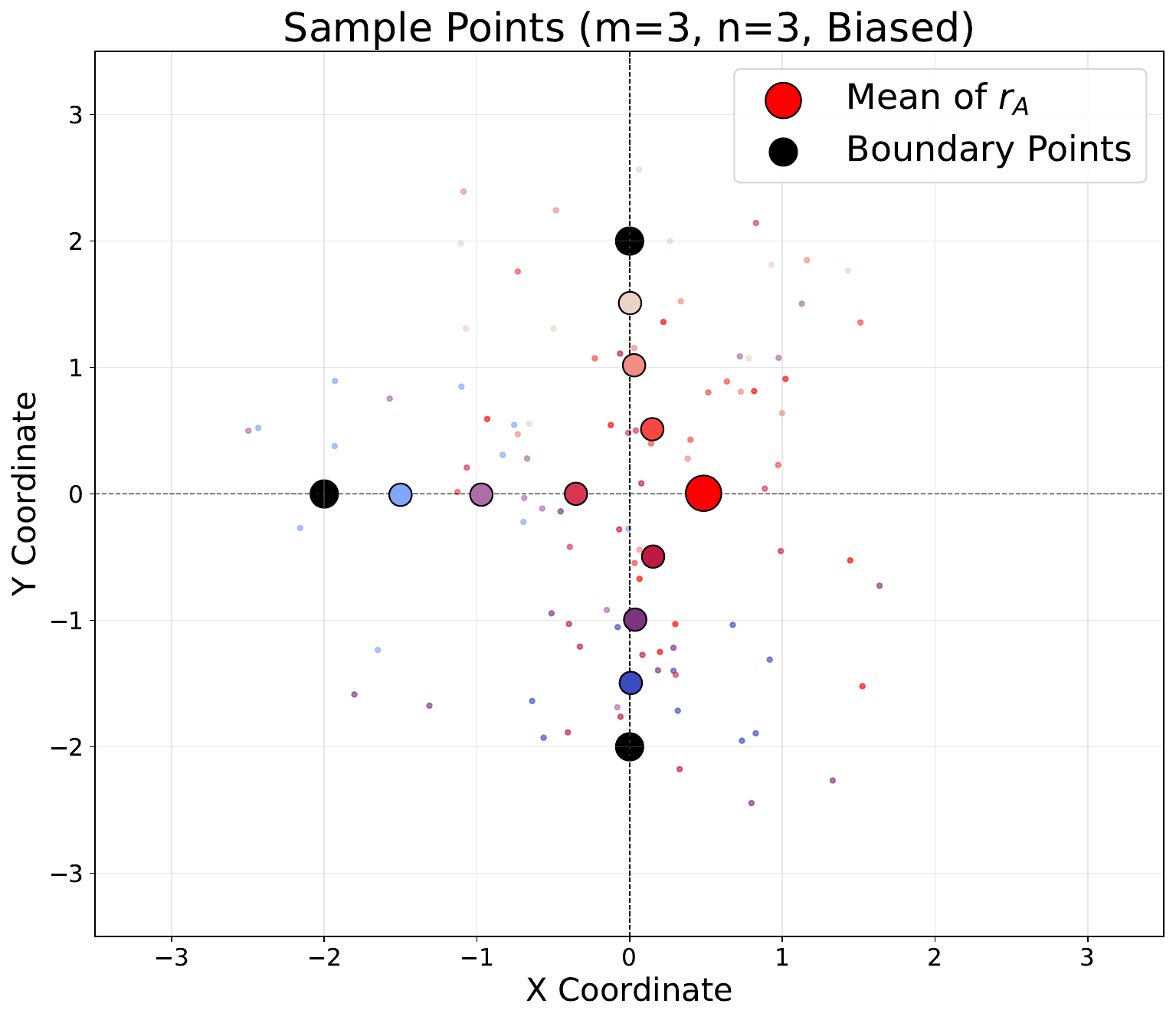} 
\includegraphics[width=0.48\linewidth]{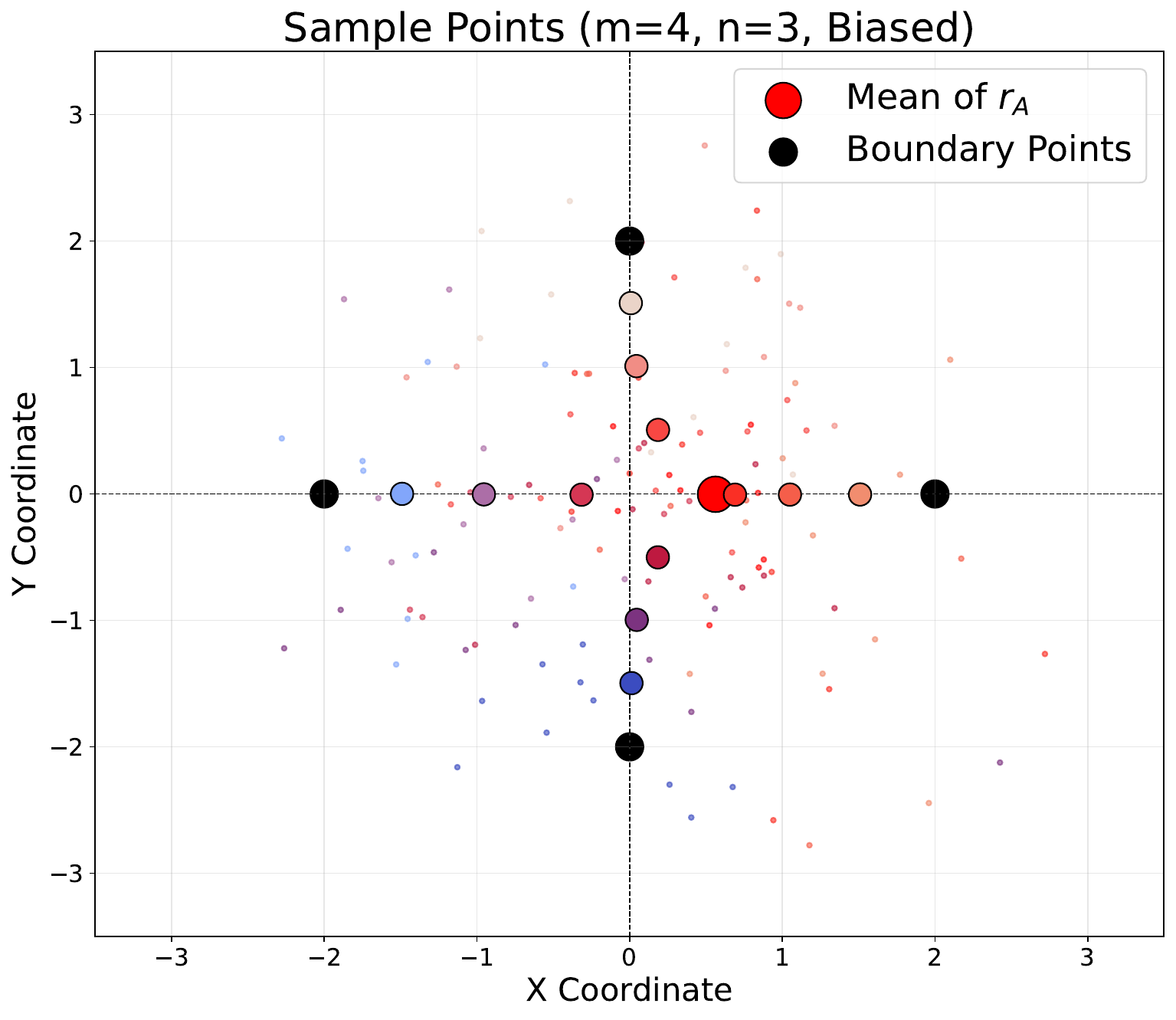} 
\includegraphics[width=0.48\linewidth]{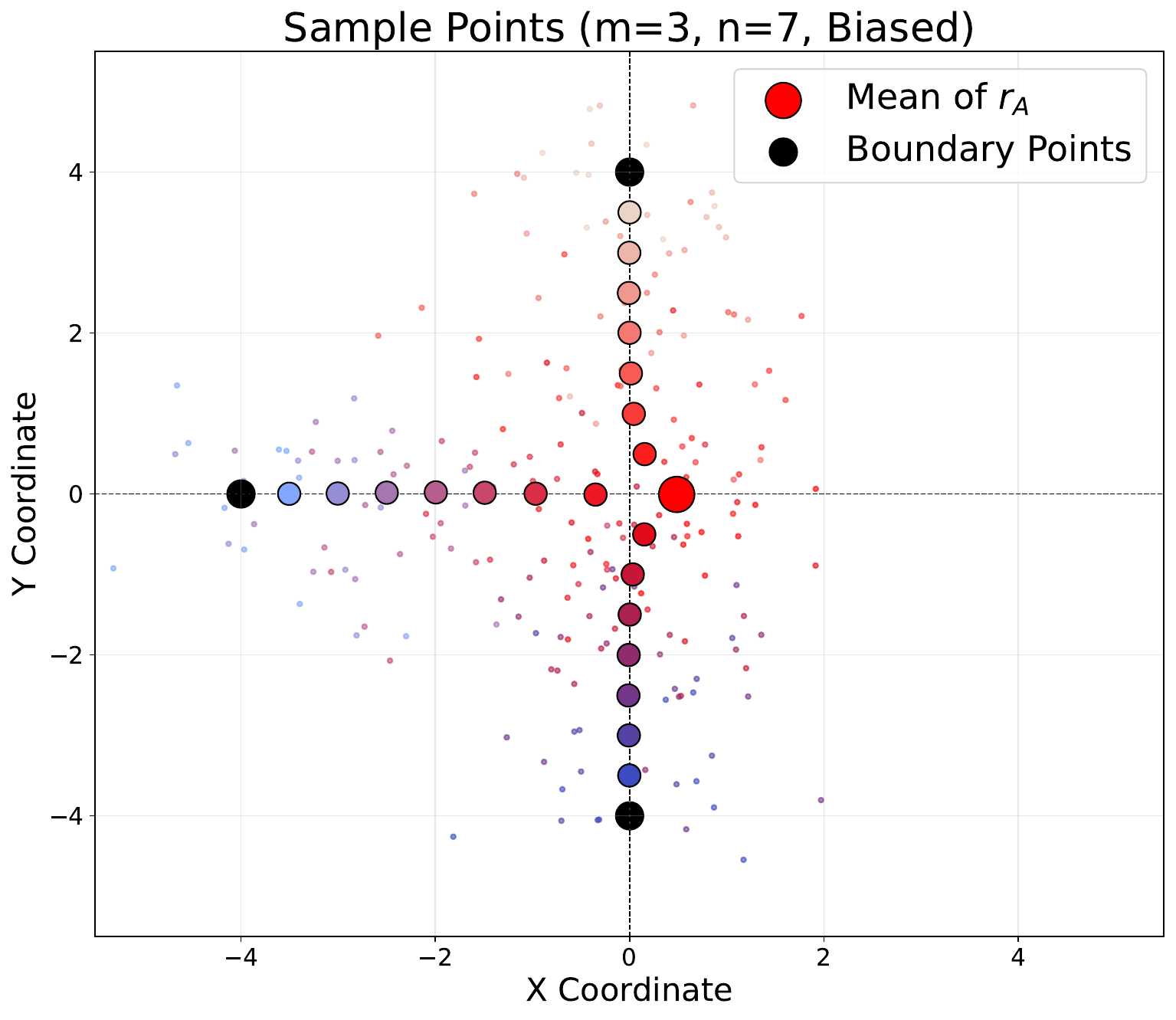} 
\includegraphics[width=0.48\linewidth]{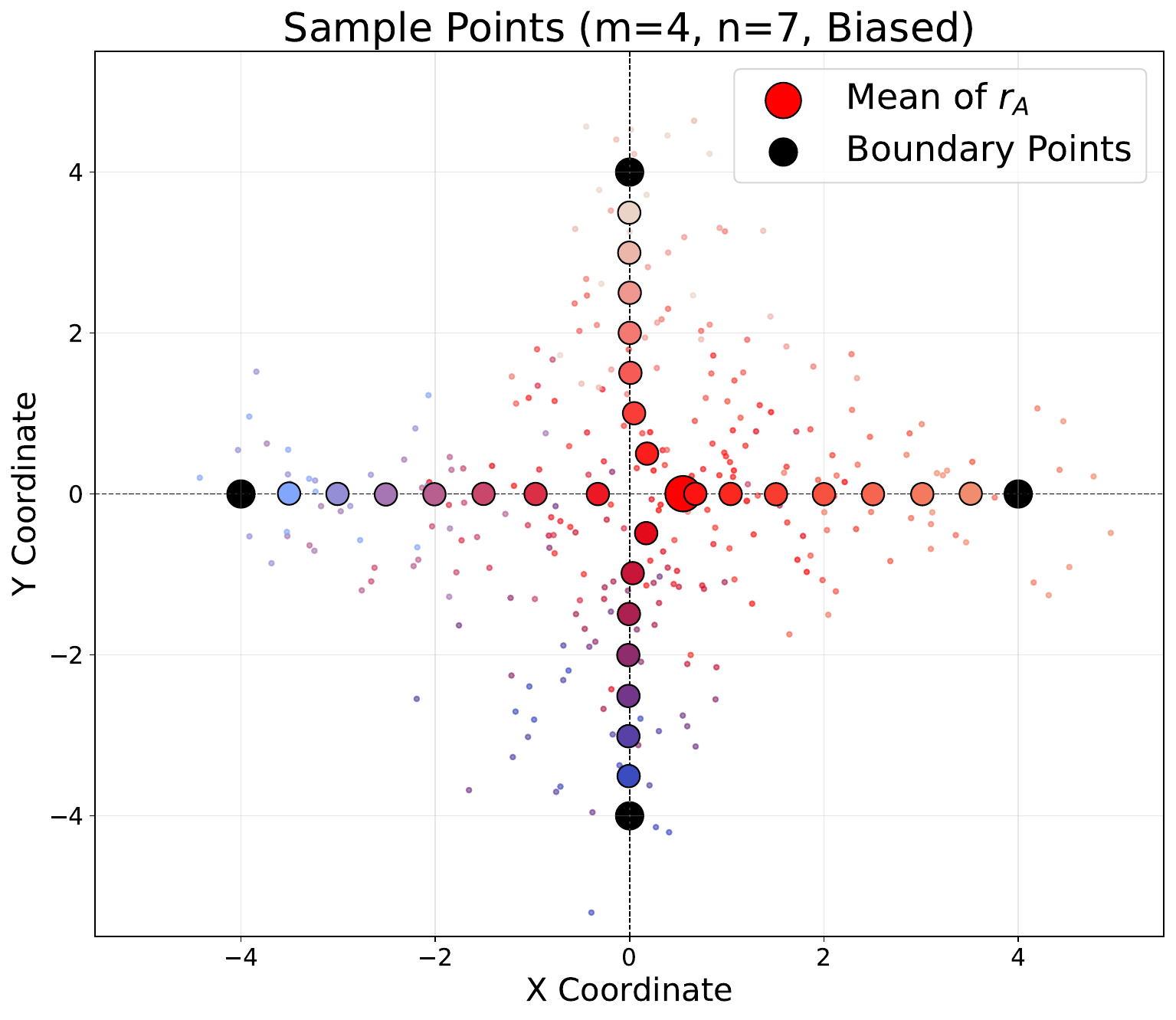} 
   \caption{Snapshot of the sample points at time $t=5$ for biased jump noise case.}\label{fig:example2:snapshot_nonuniform}
\end{figure*}

\end{document}